\documentclass[prb,twocolumn,superscriptaddress,showpacs,floatfix,longbibliography,nofootinbib]{revtex4-1}
\usepackage[utf8]{inputenc}
\usepackage{graphicx,amsfonts,amssymb,amsmath,xspace,physics}
\usepackage[utf8]{inputenc}
\usepackage[pdftex,breaklinks=true,colorlinks=true]{hyperref}
\usepackage{epstopdf}
\usepackage{tabularx}
\usepackage{color}
\usepackage{stmaryrd}
\definecolor{g}{rgb}{.1,0.4,.1} 
\definecolor{b}{rgb}{0,0.2,1}
\definecolor{rouge}{rgb}{0.82,0.,0.}
\definecolor{vert}{rgb}{0.,0.82,0.}
\definecolor{orange}{rgb}{1,0.5,0.}
\definecolor{bleu}{rgb}{0.,0.,0.82}
\definecolor{m}{rgb}{0.82,0.,0.82}
\definecolor{vert2}{rgb}{0.,0.5,0.}
\definecolor{rougeclair}{rgb}{1.0,0.7,0.7}

\newcommand{\beq}{\begin{equation}}
\newcommand{\be}{\begin{equation}}
\newcommand{\beqn}{\begin{eqnarray}}
\newcommand{\eeq}{\end{equation}}
\newcommand{\ee}{\end{equation}}
\newcommand{\eeqn}{\end{eqnarray}}
\newcommand{\nn}{\nonumber}

\newcommand{\bem}{\begin{pmatrix}}
\newcommand{\eem}{\end{pmatrix}}

\newcommand{\f}{\frac}

\newcommand{\e}{\textrm{e}}

\newlength{\ldag}
\begin{document}
\title{Two particles on a chain with disordered interaction: \\
Localization and dissociation of bound states and mapping to chaotic billiards}

\author{Hugo Perrin}
\email{hugo.perrin@sorbonne-universite.fr}
\affiliation{Sorbonne Universit\'e, CNRS, Laboratoire de Physique Th\'eorique de la Mati\`ere Condens\'ee, LPTMC, 75005 Paris, France}

\author{J\'anos K. Asb\'oth}
\email{asboth.janos@ttk.bme.hu}
\affiliation{Department of Theoretical Physics and MTA-BME  Exotic Quantum Phases "Momentum" Research Group, Budapest  University of Technology and Economics, H-1111 Budapest,  Hungary}
\affiliation{Institute for Solid State Physics and Optics, Wigner Research Centre for Physics, H-1525 Budapest P.O. Box 49, Hungary}

\author{Jean-No\"el Fuchs}
\email{fuchs@lptmc.jussieu.fr}
\affiliation{Sorbonne Universit\'e, CNRS, Laboratoire de Physique Th\'eorique de la Mati\`ere Condens\'ee, LPTMC, 75005 Paris, France}

\author{R\'emy Mosseri}
\email{remy.mosseri@upmc.fr}
\affiliation{Sorbonne Universit\'e, CNRS, Laboratoire de Physique Th\'eorique de la Mati\`ere Condens\'ee, LPTMC, 75005 Paris, France}

\date{\today}

\begin{abstract}
We consider two particles hopping on a chain with a contact interaction between them. At strong interaction, there is a molecular bound state separated by a direct gap from a continuous band of atomic states. Introducing weak disorder in the interaction, the molecular state becomes Anderson localized. At stronger disorder, part of the molecular band delocalizes and dissociates due to its hybridization to the atomic band. We characterize these different regimes by computing the density of states, the inverse participation ratio, the level-spacing statistics and the survival probability of an initially localized state. The atomic band is best described as that of a rough billiard for a single particle on a square lattice that shows signatures of quantum chaos. In addition to typical ``chaotic states'', we find states that are localized along only one direction. These ``separatrix states'' are more localized than chaotic states, and similar in this respect to scarred states, but their existence is due to the separatrix iso-energy line in the interaction-free dispersion relation, rather than to unstable periodic orbits.

\end{abstract}


\date{\today}
\maketitle
\section{Introduction}
\color{red}

\color{black}

\par The Hubbard model was first proposed in 1963, independently by Gutzwiller~\cite{Gutzwiller1963}, Kanamori~\cite{Kanamori1963} and Hubbard ~\cite{Hubbard1963}. The model was a way to understand the collective behaviour of interacting electrons in solids. In 1968, Lieb and Wu~\cite{Lieb1968} found an analytical solution for the one-dimensional (1D) case using the Bethe ansatz. Despite its simple formulation, the Hubbard problem is mathematically hard to tackle. For higher dimensional case, physicists have been able to obtain only approximate analytical or numerical results (mean-field theory, DMFT,...). An exact solution remains yet unknown. For more than 50 years, the model has attracted a lot of attention. 

\par Technical developments over the last decades made possible the experimental realisations of the Hubbard model. The first setup was proposed by Greiner et al.~\cite{Greiner2002} in 2002 using ultracold bosonic atoms trapped in optical lattices. They observed the transition from a superfluid to a Mott insulator. Many other variants of the Hubbard model have been implemented, including model for fermions~\cite{Joerdens2008} or density-dependent interaction parameter~\cite{Will2010}. There are many challenging problems to tackle for experimentalists who have to control with precision different parameters: the tunneling, the lattice geometry or the potential shape. The uncertainty on these variables can give rise to undesired effects which can affect the global quality of the results. 

In this article, we propose a detailed analysis of the effect of a disordered contact interaction for the simple 2-body Hubbard problem (of two distinguishable particles, with no internal degree of freedom) in a 1D chain. Such a random $U$ Hubbard model has already been considered at finite density (rather than for two particles) to study phase transitions in two different contexts. First, for the superconductor-insulator transition in inhomogeneous materials using a 2D attractive Hubbard model with a bimodal distribution of the interaction~\cite{Litak2000,Shenoy2008,Pradhan2018}. Second, for the many-body localization transition with a 1D repulsive Bose-Hubbard model~\cite{Sierant2017}. These authors focus study either the phase transition using thermodynamic quantities at equilibrium or the thermalization on quantities such as entanglement entropy. Here, we analyse the 2-body random $U$ Hubbard model (zero density) and concentrate on others quantities such as the different type of states or the energy spectrum. Our problem turns out to have connections to several fields -- molecular physics, disordered systems, surface physics and quantum chaos -- that we briefly review.

\par (1) The contact interaction, whether attractive or repulsive, in the Hubbard model leads to two-body bound states~\cite{Caffarel1998}. One-dimensional molecules corresponding to repulsive bound-states have been observed with bosonic atoms in an optical lattice experiment~\cite{Winkler2006}. In this context, such molecules are sometimes called doublons.

\begin{figure}[t]
\begin{tabular}{cc}
  \includegraphics[width=0.5\linewidth]{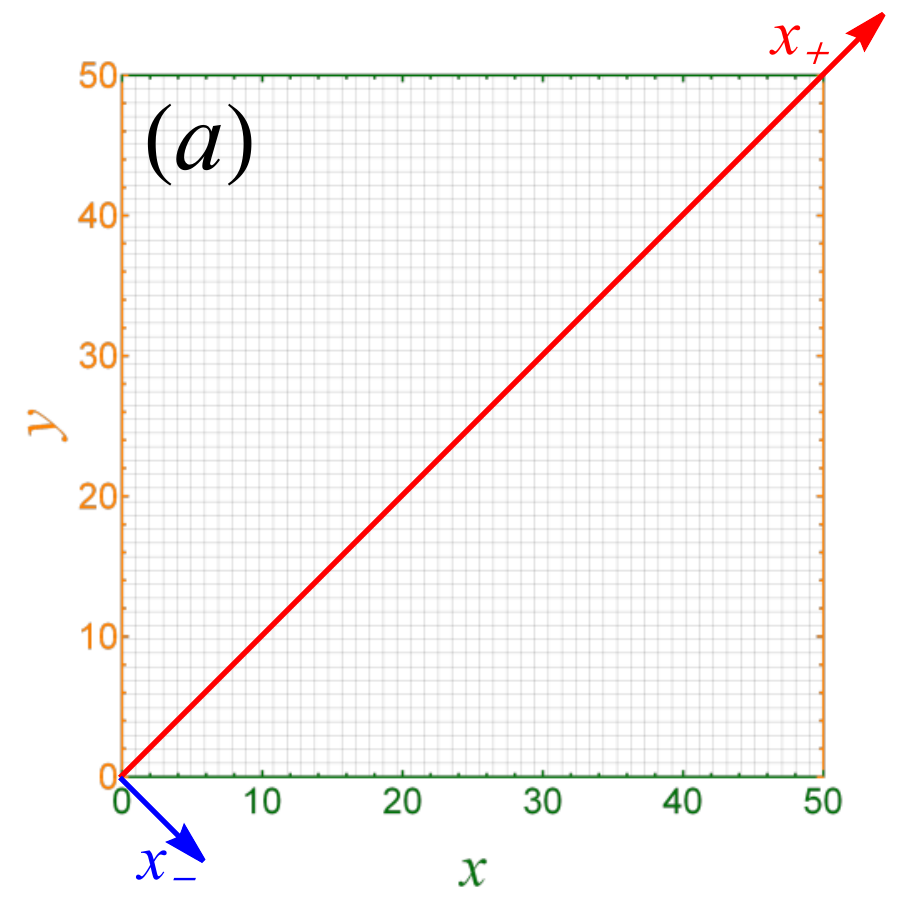}   &  \includegraphics[width=0.5\linewidth]{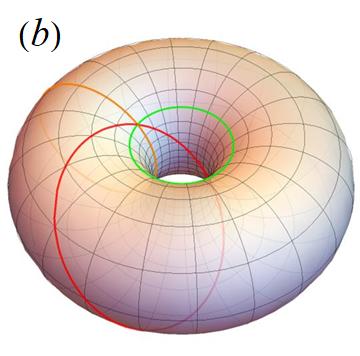} \\
    \end{tabular}
    \caption{(a): Two particles on a chain can be represented as one particle on a square lattice. The on-site contact interaction becomes a potential barrier on the diagonal of the square (red line) (b): Using periodic boundary conditions, opposite sides (green and orange lines) of the square lattice are identified. The system is topologically equivalent to a torus. }
    \label{squaretorus}
\end{figure}

\par (2) Arbitrarily weak disorder in the on-site potential in 1D and 2D quantum systems can lead to exponential localization of all energy eigenstates, by the so-called Anderson localization~\cite{Anderson1958,Abrahams1979}. Although here the disorder is in the interaction rather than in an on-site potential, the concept of Anderson localization is still useful to interpret our results, as we will see.

\par (3) The tight-binding problem of two particles on a 1D chain with contact interaction maps onto that of a single particle on a 2D square lattice with a 1D potential barrier (or impurity chain) along the diagonal (see discussion below and Fig.~\ref{squaretorus}). Our model is therefore close to the problem of a disordered surface (1D impurities) versus an ordered bulk (2D). Using this analogy to surface physics, our results can also be understood in the light of Tamm surface states~\cite{Goodwin1939,Forstmann1993} and Koster-Slater impurity states~\cite{Koster1954,Feynman1964}.

\par (4) Another 2D analogy can be made with quantum billiards. In classical billiards, one considers the dynamics of a single classical particle inside a 2D region delimited by a hard boundary. The dynamics is governed by the standard Hamilton equations with parabolic kinetic energy. Depending on the billard's shape, the system can be either (i) integrable and trajectories labeled by constants of the motion, or (ii) chaotic (non-integrable) and the trajectories are exponentially sensitive to initial conditions~\cite{Gutzwiller1990}.  To ``quantize'' the model means that the particle is now described by a vector in a Hilbert space and its dynamics is given by the Schr\"odinger equation. Despite the fact that it is problematic to talk about trajectories and sensitivity to initial conditions for quantum systems, ``quantum chaos'' has emerged as an active field of research, whose aim is to find traces of classical chaos in the usual objects of study in quantum mechanics, such as the energy levels and their eigenfunctions~\cite{Heller1993}. In these models, the ``disorder'' lies not in the potential but in the boundary's shape. Investigations of quantum chaos has been mostly done in the continuum using a parabolic dispersion relation. In our case, the model is of the tight-binding type on a square lattice, so that the dispersion relation (kinetic energy) is $E\propto \cos k_x + \cos k_y$ instead of $E\propto k_x^2 +k_y^2$. This would affect the dynamics of the particles already at the classical level. A few papers have already focused on this type of quantum billiards with lattices~\cite{Pavloff1992,Cuevas1996,Libisch2009,Wimmer2010}. Another specificity of our model is that there are no boundaries, as the system is placed on a torus, but a closed loop on the diagonal of the system that mimics a rough boundary (see Fig.~\ref{squaretorus}).

\par Two particles with a fixed interaction and in a disordered medium were intensively studied during the mid-90's. This body of work was initiated by Shepelyansky~\cite{Shepelyansky1994}, and followed by more detailed analyses~\cite{Fyodorov1995,Imry1995,Oppen1996,Ponomarev1997,Frahm2016}. Although this model has connections with superimposed band random matrices, which also appear in the context of quantum chaos, it remains quite different from ours, due to the disorder in the onsite potential.
The main result found is that the interaction is able to increase the localization length of some of the two-particle states. In our model, in the absence of interactions, there are no localized states and thus there is no delocalizing effect of the interaction.

\par Two interacting particles  with a contact interaction have also been recently discussed, either in the context of a richer one-dimensional two-band tight-binding model without disorder~\cite{DiLiberto2016} or for a disordered interacting quantum walks in~\cite{Toikka2020}, but the disorder was introduced dynamically.


\par The paper is organized as follows: first, in section~\ref{sec_hubbard}, we review the main results of the standard 2-body Hubbard problem and then introduce the disordered interaction. Then, in Sec.~\ref{sec_bulk}, we focus on the analysis of the atomic band under the effect of disorder. In Sec.~\ref{sec_mol}, we study the effect of disorder on the molecular band, when it is clearly separated from the atomic band (weak disorder regime). Next, Sec.~\ref{sec_overlap} is dedicated to the regime of overlap between the molecular and atomic bands (strong disorder regime). Eventually, Sec.~\ref{sepstates} is devoted to ``separatrix states", which, to the best of our knowledge, were not discussed before in the literature. In a last section (Sec.~\ref{sec_conclusion}), we conclude, propose several experiments and give perspectives for future studies. Several appendices give details on various parts of the work.

\section{The random \textit{U} Hubbard model}

\label{sec_hubbard}

In this section, we define the model that we study in the following sections. Consider two distinguishable particles hopping on a chain of $N$ sites, with a contact interaction between them that depends on the position. We denote the position of the first and second particle by integers $x$ and $y$, respectively, with $1 \le x,y \le N$, and use periodic boundary conditions, identifying $N$ with $0$ (see Fig.~\ref{squaretorus}). 
The Hamiltonian reads, 
\begin{eqnarray}
H &=&  -\sum_{x,y=0}^{N-1} 
\ket{x+1,y}\bra{x,y}+\ket{x,y+1}\bra{x,y} + \mathrm{h.c.} \nonumber \\
 &\quad& + \sum_{x=0}^{N-1} U_x \ket{x,x}\bra{x,x} = H_0 + U\, 
 \label{eq:hamiltonian}
\end{eqnarray}
where $H_0$ is the hopping Hamiltonian and $U$ the interaction potential. Here and in the rest of the paper we set $\hbar =1$ and measure energy in units of the hopping amplitude and length in units of the lattice constant. The interaction energy $U_x$ is a position-dependent random variable, uniformly distributed between $\bar{U}-W$ and $\bar{U}+W$. 
The random $U$ Hubbard model is a modification of the Hubbard model (here restricted to 2 particles in 1D) that depends on two parameters: the average interaction $\bar{U}$ and the fluctuations (or disorder) in the interaction $W$.
  
\par  We briefly review the trivial case without interaction $\bar{U}=W=0$. The Bloch theorem applies, and one has two particles with quasi-momentum $k_x=2\pi i/N$ and $k_y=2\pi j/N $ where $i,j\in\llbracket-N/2,N/2-1\rrbracket$, and total energy $E(k_x,k_y)=-2(\cos k_x+\cos k_y)$ (the notation $\llbracket.,.\rrbracket$ indicates that only discrete values, in unit step, are taken in the interval). Eigenvectors of the system of two particles are plane waves delocalized over the whole system. This is identical to a single particle on a square lattice with periodic boundary conditions (PBC).
\par Before considering the effect of a disordered interaction, we set a baseline by considering the translation-invariant interacting case ($W=0$ and $\bar{U}\neq 0$) in the following section.

\subsection{Translation-invariant interaction: bound and scattering states}

We start the analysis of the system with the case where there is no disorder, $W=0$, and hence, 
we have two distinguishable particles with a contact interaction $U_x = \bar{U}$, independent of  $x$. 

As we recall below, here the Hamiltonian can be solved exactly, and its eigenstates are either bound states of the two particles, or scattering states.

To obtain explicit formulas for the bound and scattering states, we change to the center-of-mass reference frame (see Appendix~\ref{ap:comframe} for details). The center-of-mass and relative coordinates, $x_\pm$, read 
\begin{align}
    x_+ &= \frac{x + y }{2};&  
    x_- &= {x - y }
\end{align}
and the center-of-mass wavenumber, $k_+$, is defined as
\begin{align}
    \ket{k_+} =\frac{1}{\sqrt{N}} \sum_{x_+} e^{i k_+ x_+} \ket{x_+}, \, \text{with } k_+=k_x+k_y = 2\pi K/N
\end{align} where $K\in\llbracket -N/2,N/2-1\rrbracket$, so that $-\pi\leq k_+ <\pi$. The Hamiltonian now reads
\begin{widetext}
\begin{eqnarray}
H&=&-\sum_{x_+,x_-\in\mathcal{L}}\ket{x_++1/2,x_-+1}\bra{x_+,x_-}+\ket{x_++1/2,x_--1}\bra{x_+,x_-}
+\mathrm{h.c.}+\bar{U}\sum_{x_+=0}^{N-1}\ket{x_+,0}\bra{x_+,0}\, ,
 \label{eq:hamiltonian2}
\end{eqnarray}
\end{widetext}
where $\mathcal{L}$ is the original square lattice expressed in the center-of-mass reference frame i.e $x_-\in\llbracket -N+1,N-1\rrbracket$ and $x_+\in\llbracket|x_-|/2,N-1-|x_-|/2\rrbracket$, $x_+$ taking integer values when $x_-$ is even and half-integer values when $x_-$ is odd.
Because of translation invariance along $x_+$, the center-of-mass momentum $k_+$ is conserved. The Hamiltonian, Eq.~\eqref{eq:hamiltonian2}, written in the $k_+$ basis, separates into $N$ decoupled chains indexed by $k_+$:
\begin{align}
H &= \sum_{k_+} \ket{k_+}\bra{k_+} \otimes h(k_+), 
\end{align}
where the Hamiltonian $h(k_+)$ of a single chain reads,
\begin{align}
h(k_+) &= -2\cos \frac{k_+}{2} \sum_{x_-}\ket{x_-+1} \bra{x_-}+\mathrm{h.c.} \nonumber \\
&\quad +\bar{U}\ket{0}\bra{0} \, .
\label{ham1}
\end{align}

\begin{figure}[h]
\begin{tabular}{c}
  \includegraphics[width=\linewidth]{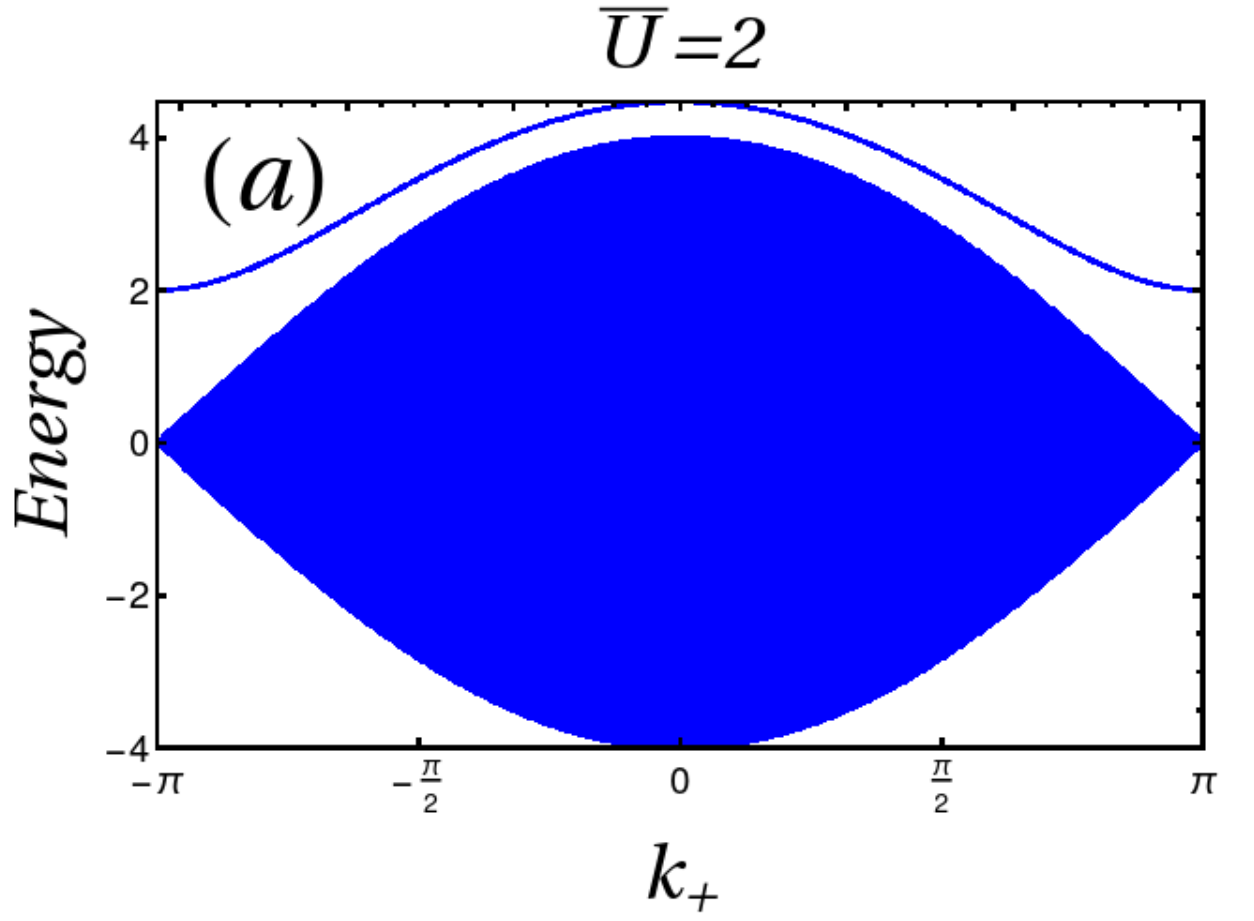}  \\ \includegraphics[width=\linewidth]{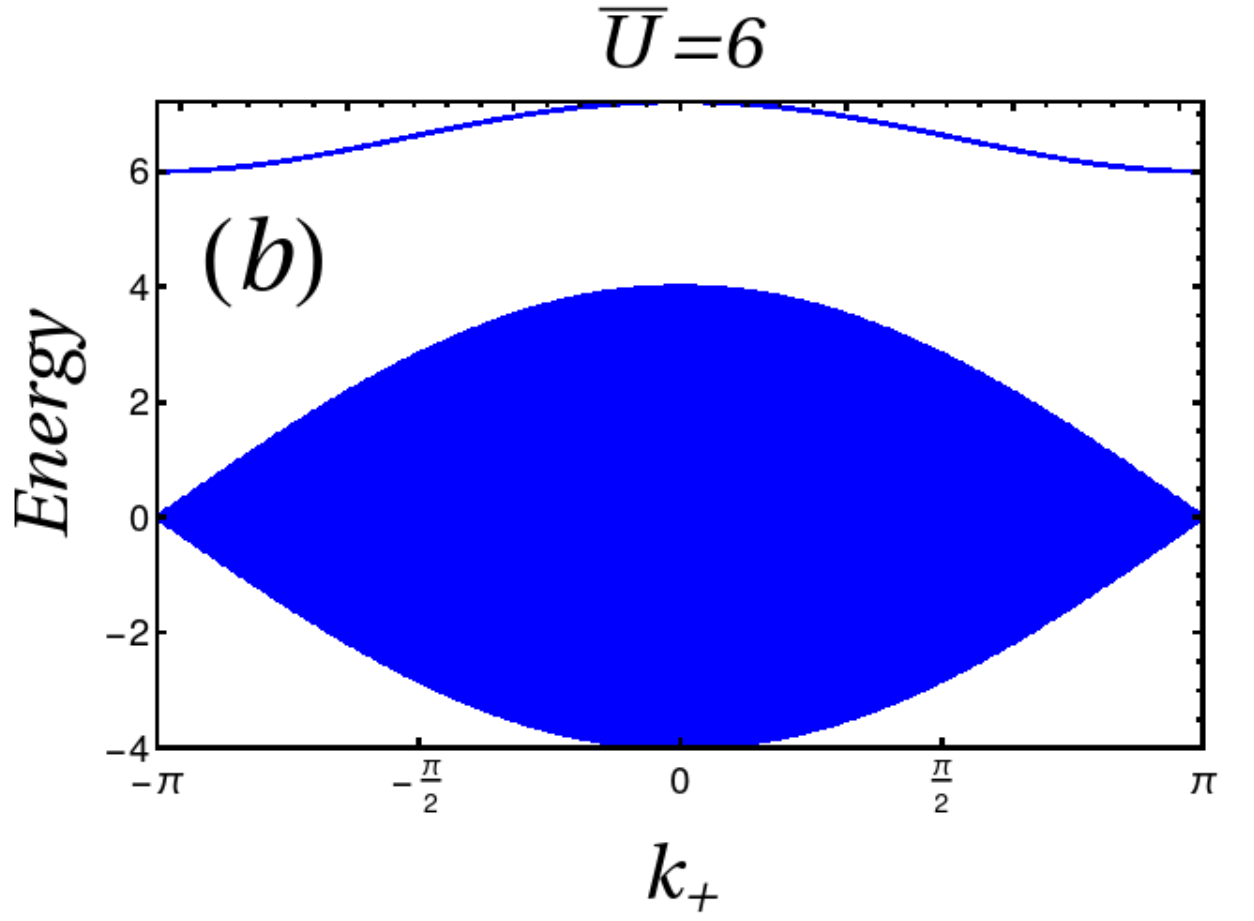}  

\end{tabular}
  \caption{Energy spectra of the standard two-particle Hubbard model (translation-invariant interaction) as a function of the center-of-mass quasimomentum $k_+$. (a) For weak interaction, $\bar{U}=2$, the band of bound two-particle states overlaps in energy with the band of scattering states, where the two particles propagate almost independently. (b) For stronger interaction, $\bar{U}=6$, a gap $\bar{U}-4=2$ separates these bands.}
    \label{nondisorderedspectrum}
\end{figure}

For each of the $N$ different values of the center-of-mass momenta $k_+$, the Hamiltonian $h(k_+)$ describes a single particle hopping on an effective chain of $N$ sites with periodic boundary conditions and with hopping amplitude $t_\text{eff}=-2\cos(k_+/2)$ in the presence of an impurity of magnitude $\bar{U}$ at the origin $x_-=0$. When $\bar{U}\neq 0$, its spectrum consists of a single bound state and a band composed of $N-1$ delocalized scattering states (see Fig.~\ref{nondisorderedspectrum}). In contrast to Tamm states~\cite{Goodwin1939,Forstmann1993}, that appear for a sufficiently strong impurity at an edge of an open chain, there is no threshold for a bound state to exist for an impurity in a periodic chain. In terms of the full Hamiltonian describing two particles, these correspond to states where the two particles move together, as a bound pair, and scattering states where they move almost independently. The eigenproblem corresponding to Hamiltonian (\ref{ham1}) was solved by Koster and Slater~\cite{Koster1954,Feynman1964}. It can also be obtained by Bethe ansatz~\cite{Caffarel1998}. Here, we summarize the main results first for bound states and then for scattering states.

\begin{figure}[h]
\begin{tabular}{c}
   \includegraphics[width=\linewidth]{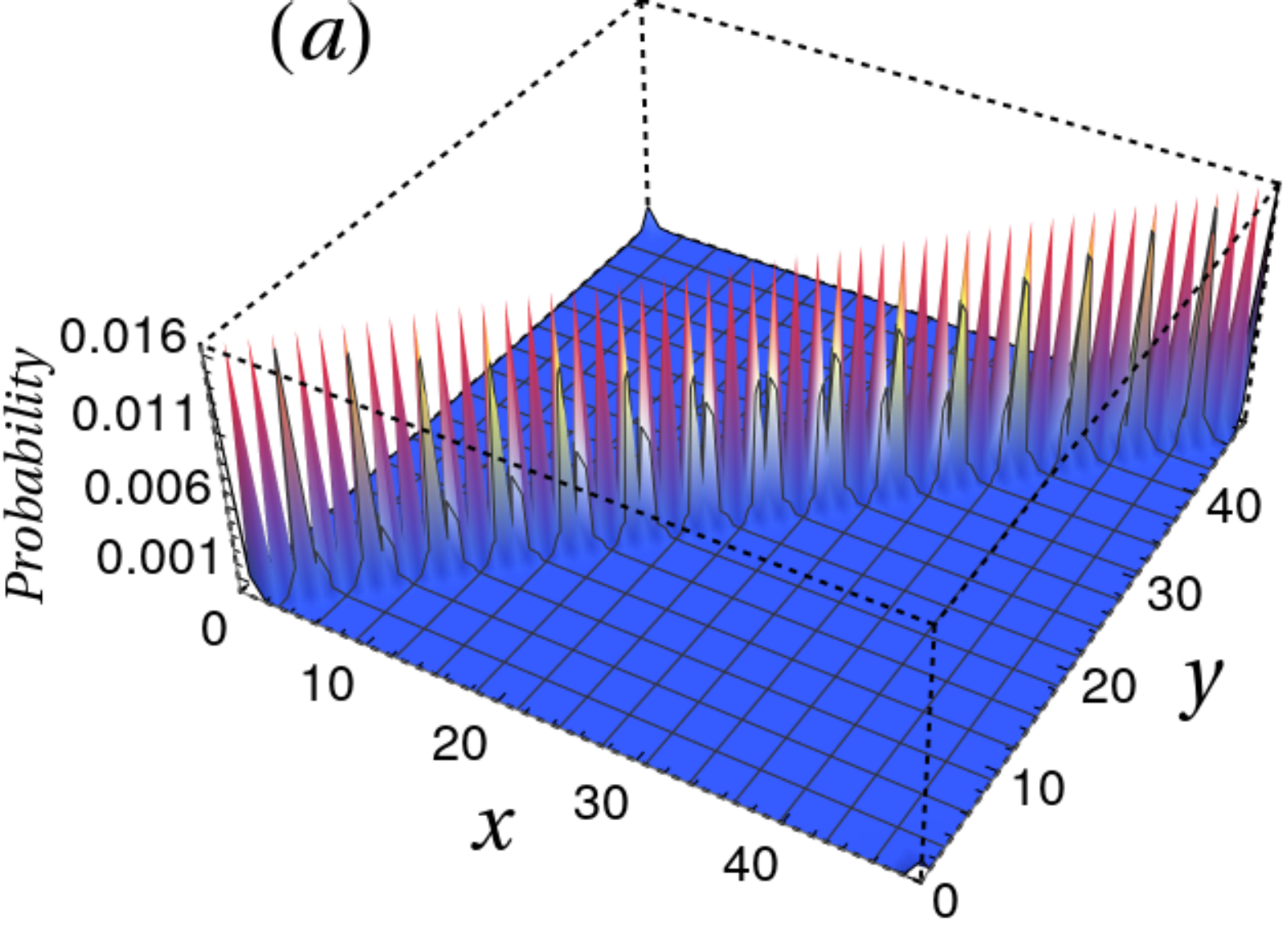}\\
  \includegraphics[width=\linewidth]{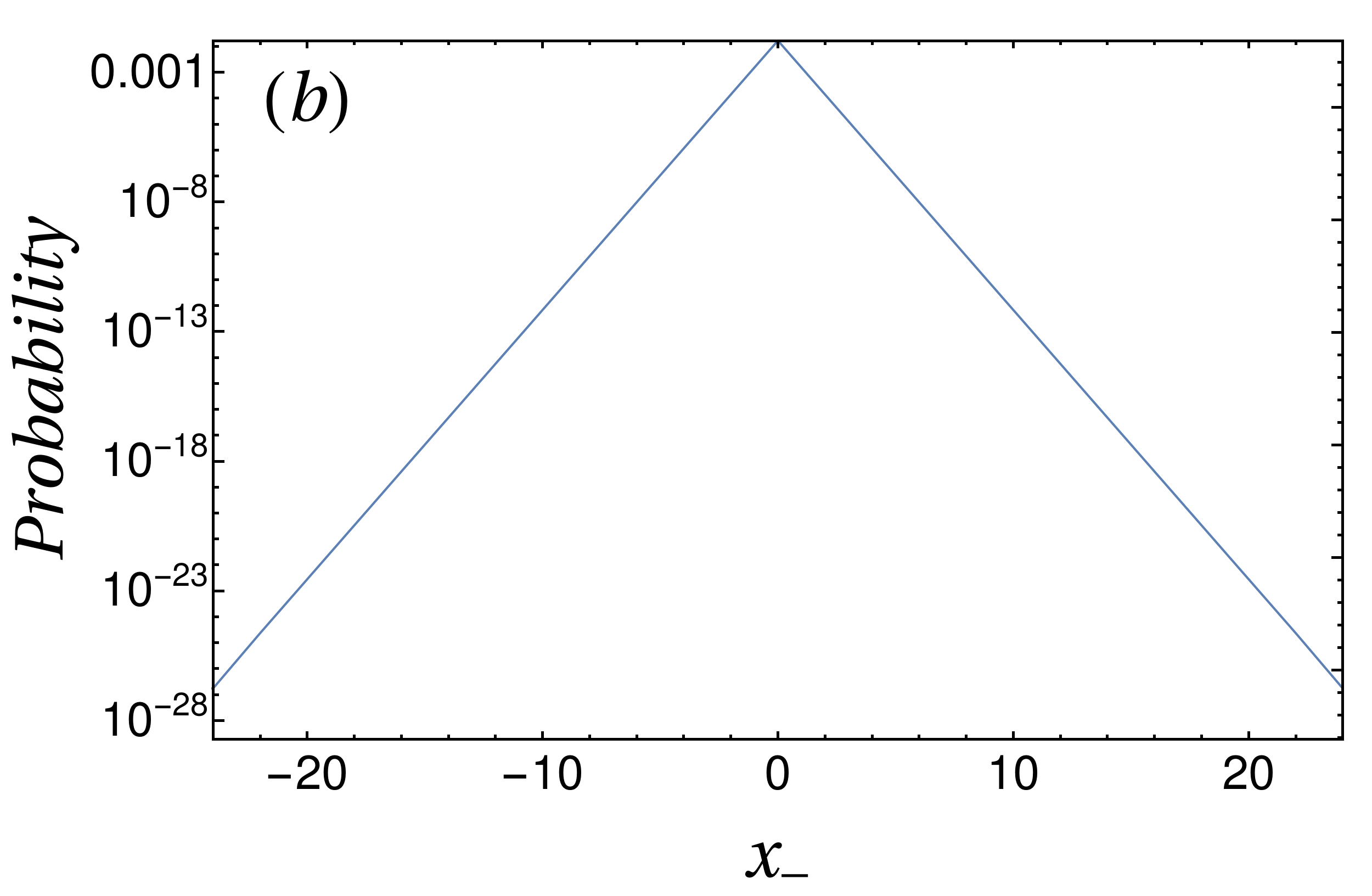}
\end{tabular}
 
  \caption{Typical bound state for a translation-invariant interaction $\bar{U}=6$ on a chain of $N=50$ sites. (a): Position distribution of an eigenmode in the coordinate $x$ and $y$ denoting the position of the first and second particle. (b): Cross-sectional cut of the position distribution, that reveals the exponential decay of the wavefunction in the antidiagonal direction $x_-$ (note the vertical logarithmic scale).}
    \label{molecule}
\end{figure}

The bound states labelled by the quantum number $-\pi \leq k_+< \pi$ have energies and wavefunctions given by:
\begin{align}
E_\text{bd}(k_+) &= \mathrm{sgn}(\bar{U})
\sqrt{\left(4\cos \frac{k_+}{2}\right)^2 + \bar{U}^2}; \label{eq:mol1}\\
\psi_\text{bd} (x_+,x_-) &= \f{e^{i k_+x_+}}{\sqrt{N}}\sqrt{\tanh{\kappa}}\, (-1)^{x_-} e^{-\kappa |x_-|},
\label{eq:mol2}
\end{align} 
when $N\gg1/\kappa$, so that finite-size effects can be neglected. Here the inverse decay length $\kappa>0$ is a function of $k_+$ given by the solution of 
\begin{equation}
\sinh{\kappa}=\frac{\bar{U}}{4|\cos(k_+/2)|}\, .
\label{eq:kappa}
\end{equation}
The inverse decay length $\kappa$ defined above is also the pure imaginary solution of the Bethe ansatz equation.
From Eq.~\eqref{eq:mol2} the bound states are plane waves along the $x_+$ direction but exponentially localized along the $x_-$ direction, see Fig.~\ref{molecule}. For simplicity and without loss of generality, we restrict to $\bar{U}>0$, in which case they correspond to repulsively bound states which are only possible when the kinetic energy is also bounded from above~\cite{Caffarel1998,Winkler2006}. The minimal and maximal bound-state energies are then $\bar{U}$ and $\sqrt{\bar{U}^2+16}$. The number of bound states is $N$ compared to a total number of states $N^2$.

If the interaction is weak, $\bar{U}<4$, the bound-state energy band and that of the scattering states overlap, while for stronger interaction, $\bar{U}>4$, there is a gap equal to $\bar{U}-4$ between these bands. In the limit of very strong interaction, the bound-state dispersion relation can be approximated as 
\begin{equation}
E_\text{bd}(k_+)\simeq \bar{U}+\frac{4}{\bar{U}}+\frac{2}{\bar{U}} 2\cos k_+ \, .
\end{equation}
This is the dispersion relation of a one-dimensional tight-binding model with hopping amplitude $2/\bar{U}$ and on-site energy $\bar{U}+4/\bar{U}$ decoupled from the bulk of the energy spectrum.


The scattering states are modified plane waves labeled by $k_+$ and $\kappa_-$. The wavefunctions are
\begin{eqnarray}
    \psi_\text{sc}(x_+,x_-)&=&Ce^{ik_+x_+}\Big[\sin (\kappa_- x_-)\nn \\
    &-&\frac{4}{ \bar{U}}\cos \frac{k_+}{2}\sin \kappa_-\cos(\kappa_- x_-)\Big] \, ,
    \label{eq:bethe}
\end{eqnarray}
where $C$ is a normalization constant and the ``wavevectors'' $\kappa_-$ are the $N-1$ real solutions of the Bethe ansatz equation given in Appendix~\ref{ap:energypertu} along with calculation details. The number of such states is $N(N-1)$. The corresponding energies are
\begin{equation}
E_\text{sc}(k_+,\kappa_-)=-4\cos \frac{k_+}{2} \cos \kappa_- \, ,
\end{equation}
with $-\pi \leq k_+< \pi$ and $-\pi \leq \kappa_- < \pi$ and are responsible for the continuum of states between $-4$ and $+4$ shown in Fig.~\ref{nondisorderedspectrum}. 

In the following, we will use ``bound states" and ``scattering states'' to refer specifically to the situation in the absence of disorder $W=0$.

\subsection{Mapping to a rough quantum billiard}
\par 

It is useful to draw an analogy between the motion of two particles on the one-dimensional lattice (chain) and the motion of a single particle on a two-dimensional square lattice. Indeed, the Hamiltonian~\eqref{eq:hamiltonian} can be interpreted in this way, with periodic boundary conditions both along $x$ and $y$ and the interaction term with $U_x$ appearing as a potential barrier on the diagonal. 

More precisely, because of the periodic boundary conditions, the disorder potential is along a closed line that wraps once around the two non-contractible loops on the torus, see Fig.~\ref{squaretorus}.  

Seen as describing two-dimensional motion, our problem is close to a quantum model of a rough billiard, as that studied in Ref.~\cite{Cuevas1996}. There, a single particle hopping on a square lattice inside a square box was considered in the presence of disorder on the boundaries in the form of random on-site potential, uniformly distributed between $-W$ and $W$. Our model differs from~\cite{Cuevas1996} in three major ways: we have (1) periodic rather than open boundary conditions; (2) disorder on a diagonal line instead of the four edges of a square; and (3) the average defect potential $\bar{U}$ can be nonzero. Despite these differences, we expect qualitatively the same kind of results as both cases may be described as a 1D chain of impurities (disorder) embedded into a clean 2D system.
\par Another type of rough boundary has been studied in~\cite{Frahm1997}. The smooth circular shape of a continuous billiard is randomly deformed on each point of the border. The authors distinguish two regimes: (1) at large roughness or high energy, they observe ergodic dynamics with a GOE distribution for the level-spacing statistics as expected for chaotic wavefunction. (2): at smaller roughness and low energy (but still in the classical chaotic regime), an exponential localization of the wavefunction is shown but in the angular momentum space. In this situation, the level-spacing statistics is a Poisson distribution with a Shnirelman peak at small spacing because of quasi-degenerate states due to time reversal symmetry. This exponential localisation remains quite different from the Anderson localisation of the molecular states in real space that we find and discuss in Sec.~\ref{sec_mol}. In particular, this localization occurs in action space and does not correspond to an exponential localization along the boundary. In addition, in our case the Anderson localized states come from bound states at vanishing disorder which are 1D localized, while such a 1D state does not exist in the integrable circular billiard: all the states are 2D delocalized. Eventually, we do not observe Schnirelman peak at the origin of the Poisson distribution in the level-spacing statistics. 
\subsection{Atomic and molecular states}
The model of two particles on a chain interacting via a disordered contact interaction is the main focus of the present article. In the following sections, we discuss different regimes depending on the parameters $ \bar{U}$ and $W$. 

In the presence of disorder $W\neq 0$, $k_+$ is no longer a conserved quantity, and one cannot resolve the energy spectrum as a function of $k_+$. In this case, and because of the possibility of overlapping bands, the distinction between scattering states and bound states is no longer pertinent. Actually, the energy spectrum separates in a bulk band with a large density of states and energy $E$ between $-4$ and $+4$, and an impurity band with a low density of states and energy such that $|E|>4$ (see Fig.~\ref{fig:dos}). We will refer to the eigenstates with energy $|E|<4$ as ``atomic states'' and to those with $|E|>4$ as ``molecular states''. The corresponding bands will be called atomic band and molecular band. When the two bands are separated by a gap, this distinction between atomic and molecular states coincides with that introduced when $W=0$ between scattering and bound states. However, when the two bands overlap [see Fig.~\ref{nondisorderedspectrum}(a)], the distinction between atomic and molecular states does not match that between scattering and bound states. The reason is that the disorder couples the bound states that overlap in energy with the scattering states. As a result, the bound states dissociate and do not remain localized: in such a case, we consider that the states in the overlapping region also belong to the ``atomic band''. In the rest of the article, we discuss in turn atomic states (Sec.~\ref{sec_bulk}), molecular states (Sec.~\ref{sec_mol}), the specific situation in which the two bands overlap (Sec.~\ref{sec_overlap}) and eventually separatrix states that exist near the center of the atomic band (Sec.~\ref{sepstates}).

\section{Atomic states}
\label{sec_bulk}
\par 
In this section, we analyze the ``atomic states'', i.e. eigenstates with energy between $-4$ and $+4$. 

Because of the analogy to the two-dimensional quantum billiard~\cite{Cuevas1996}, we expect the spectrum to show features familiar from quantum chaos theory.

\subsection{Density of states and inverse participation ratio}
\par
\begin{figure}[h]
   \includegraphics[width=\linewidth]{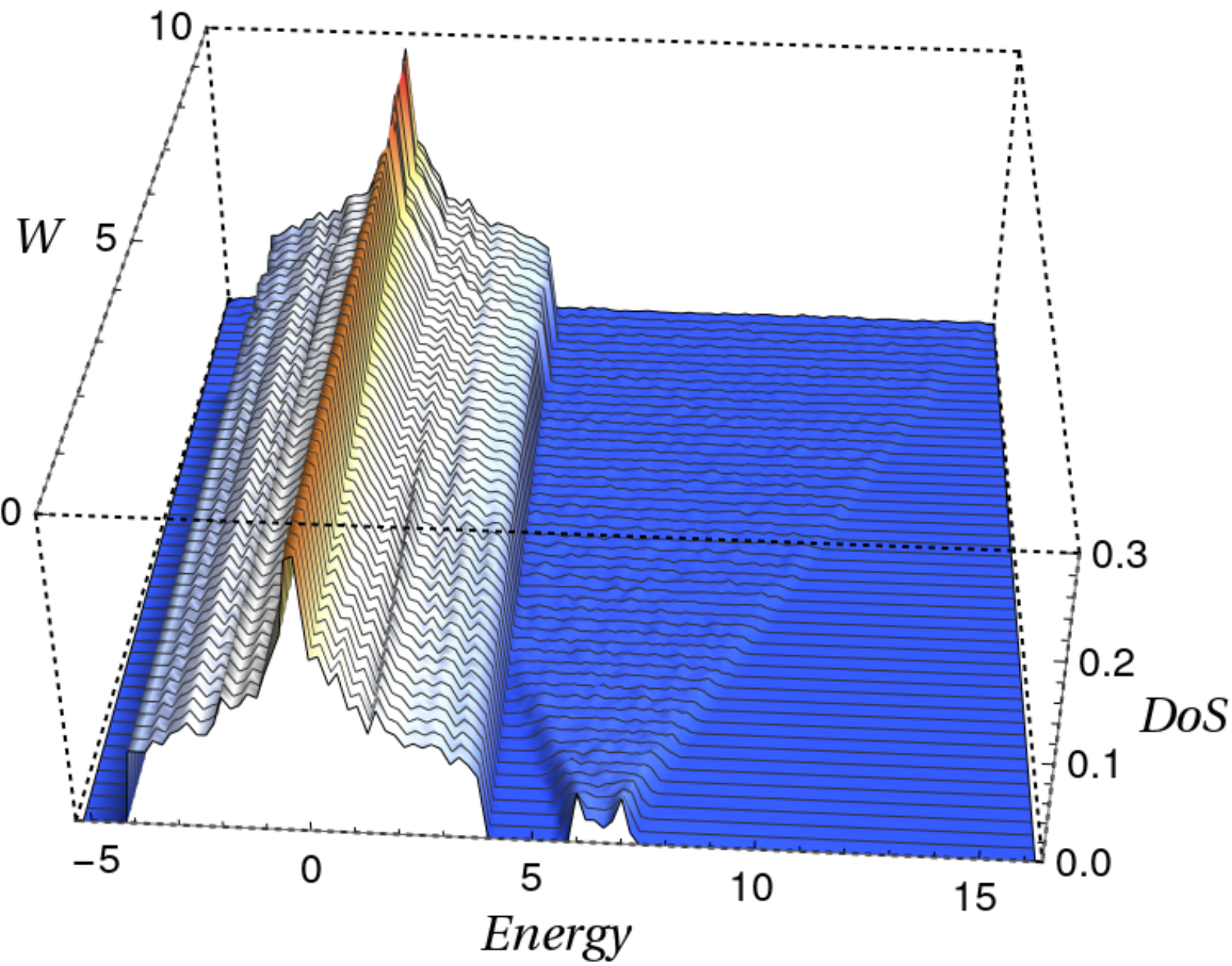}
  \caption{Density of states of disordered interacting particles for a chain of $N=50$ sites, an average interaction $ \bar{U}=6$ and 10 realisations over the disorder. The atomic band has a DoS close to that of the square lattice while the molecular band is similar to that of a 1D Anderson model.}
    \label{fig:dos}
\end{figure}

 The density of atomic states is similar to the density of states (DoS) of a clean 2D square lattice, with a van Hove singularity at the center of the band ($E=0$) and a constant density at the band edges  ($E=\pm4$), as shown in Fig.~\ref{fig:dos}. Since, in these eigenstates, the particles are mostly far away from each other, changing the contact interaction strength, $\bar{U}$, or increasing its disorder, $W$, does not significantly affect the DoS.

To quantify the localization of the atomic states and track the effects of the disorder, we use the inverse participation ratio (IPR), $I_2$. For normalized eigenfunctions $\Psi(x,y)$, it is defined by:
\begin{align}
    I_2 = \sum_{x,y}|\Psi(x,y)|^4\, .
    \label{eq:ipr}
\end{align} 
The participation ratio $P_2=1/(N^2 I_2)$ represents the fraction of sites that are occupied by the wavefunction (see Appendix~\ref{ap:iprsize}). Note that, because of the $N^2$ factor, the IPR is not simply the inverse of the participation ratio~\cite{Wegner1980}, and hence is sometimes called inverse participation or inverse participation number. As we increase the system size $N$, wavefunctions that are completely delocalized over the whole system are expected to have $I_2 \sim 1/N^2$; those localized along one direction and delocalized along the other (such as bound states in the absence of disorder) should have $I_2 \sim 1/N$; while completely localized wavefunctions should have $I_2 \sim N^0$. 
\begin{figure}[!h]
\begin{tabular}{cc}
  \includegraphics[width=0.5\linewidth]{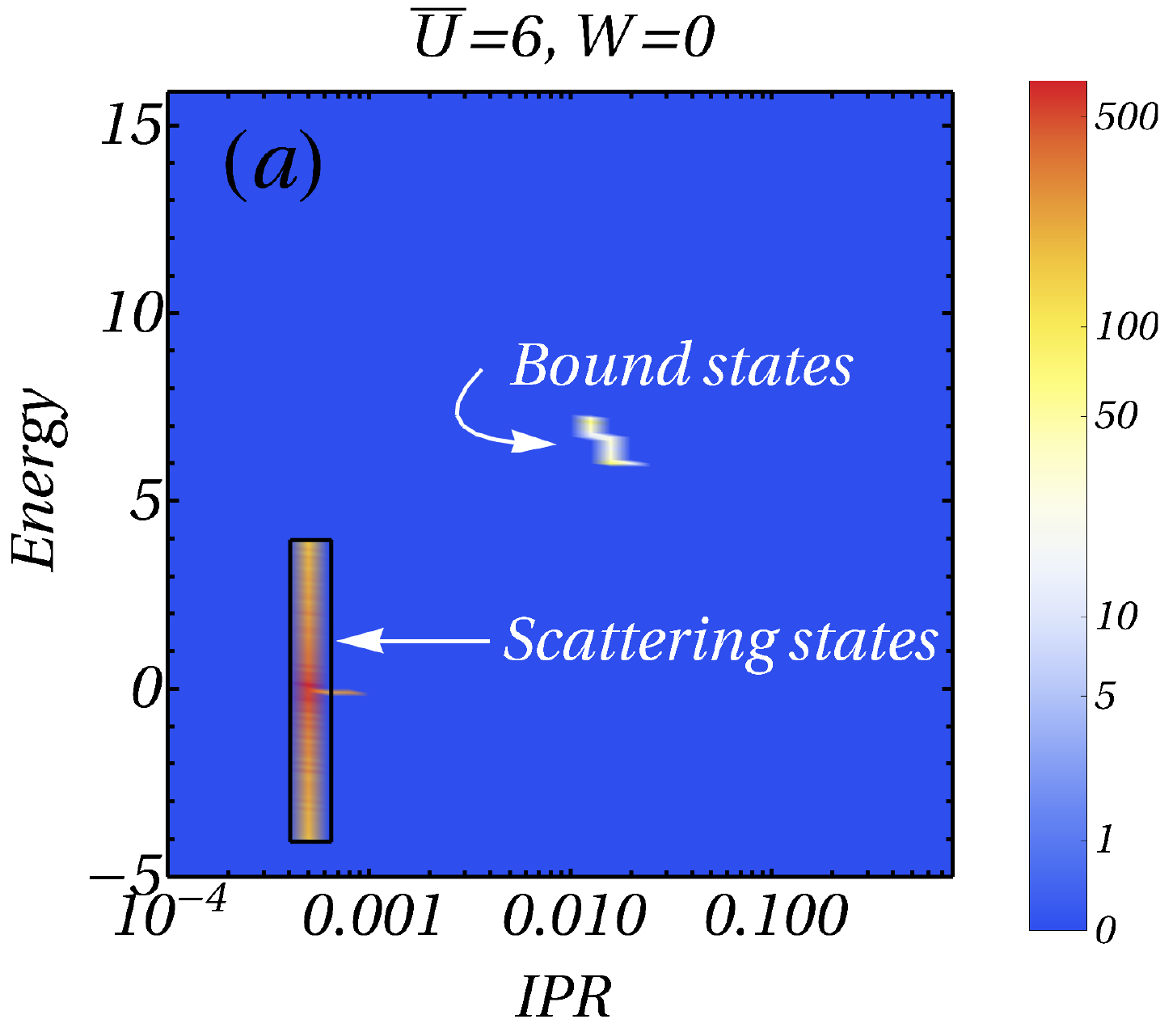}   &   \includegraphics[width=0.5\linewidth]{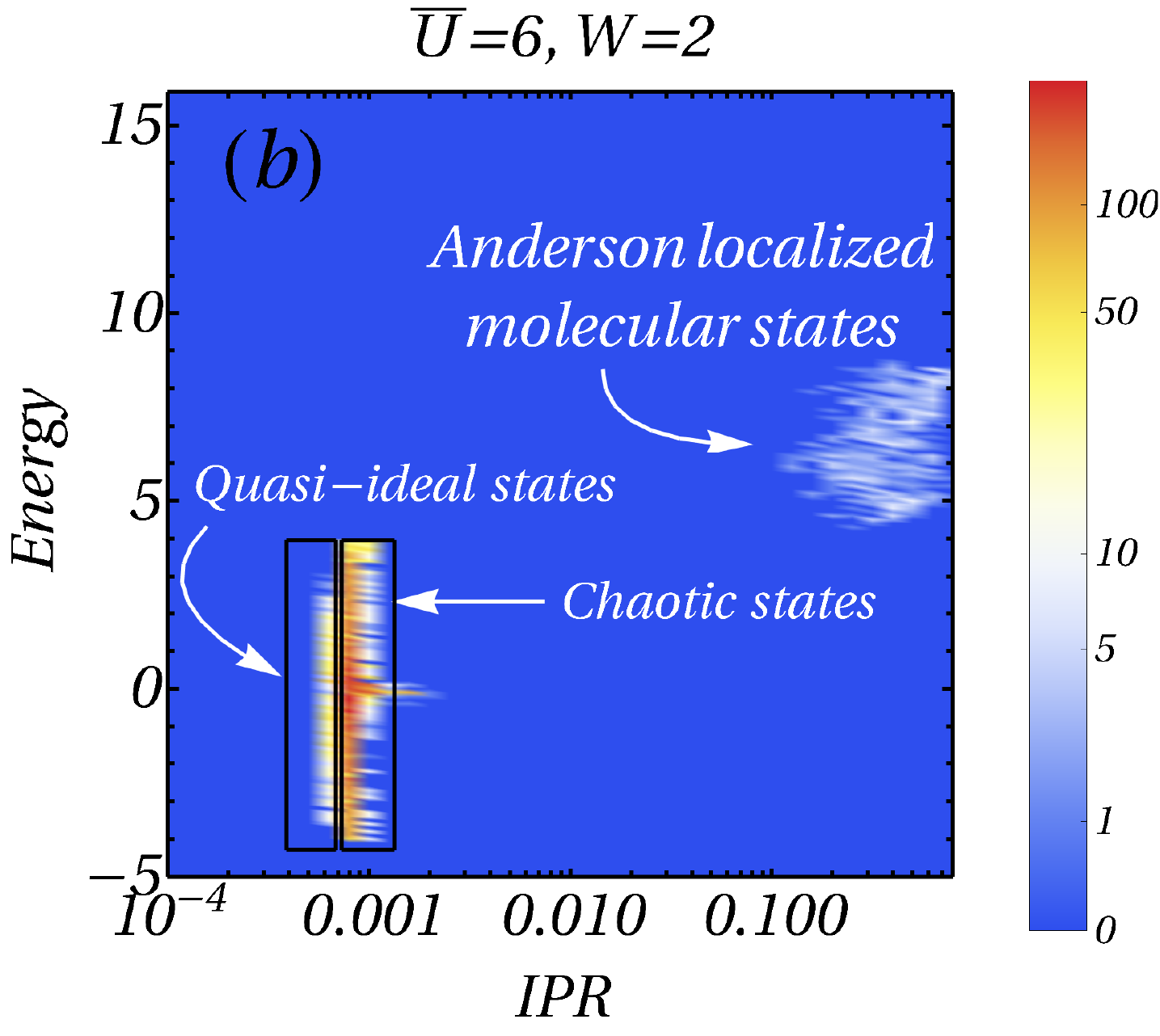}  \\
    \includegraphics[width=0.5\linewidth]{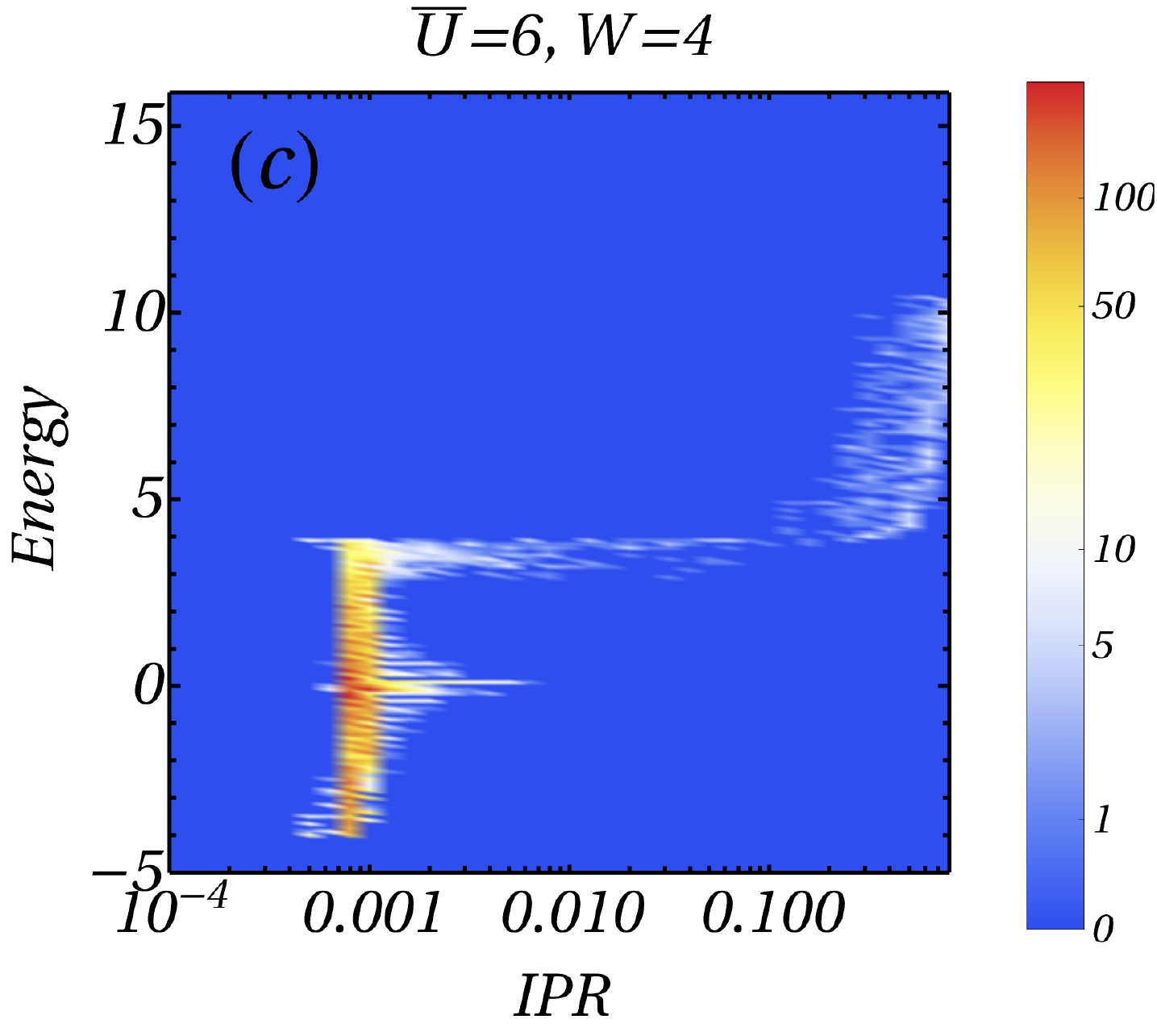}    &     \includegraphics[width=0.5\linewidth]{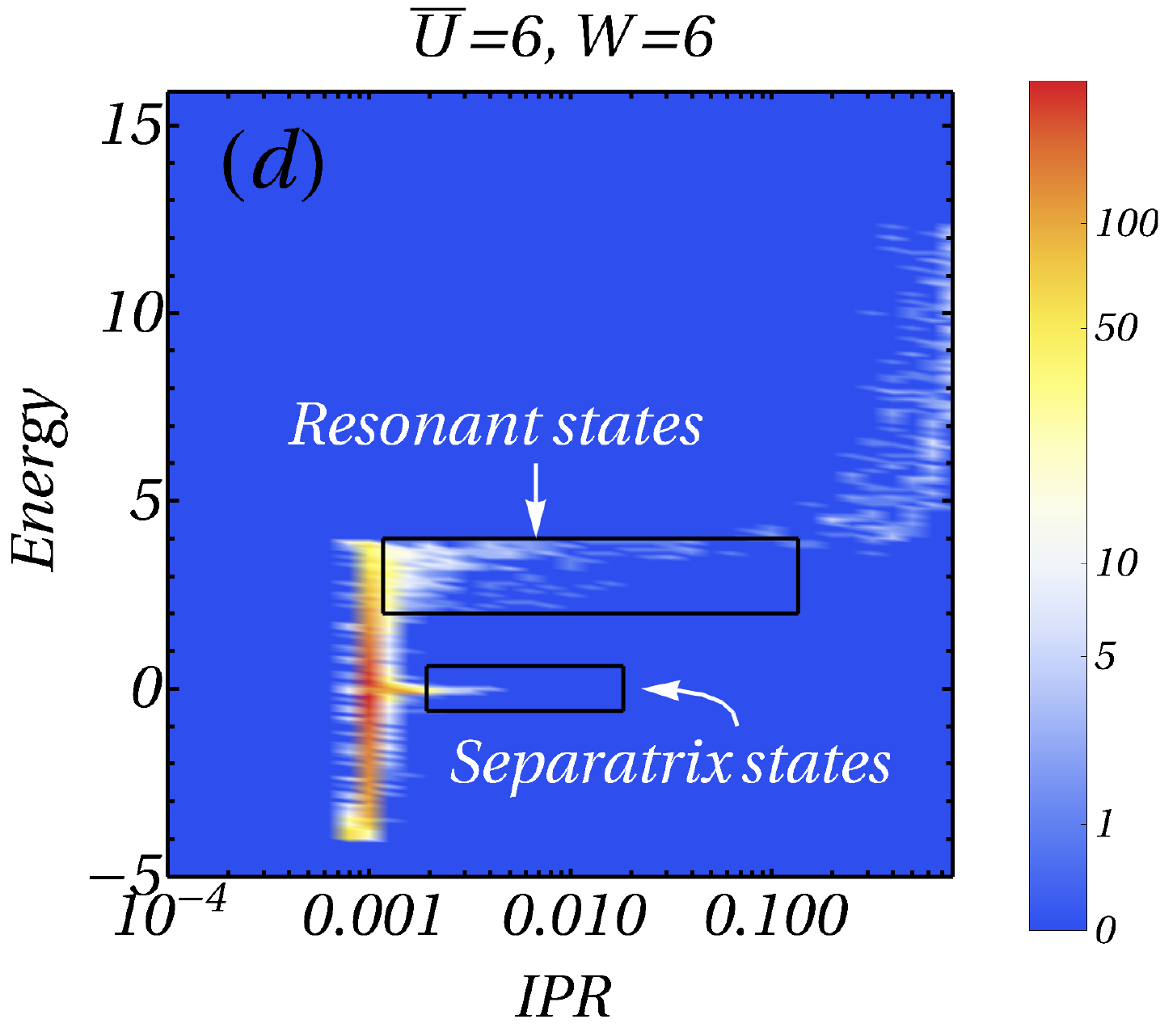} \\
        \includegraphics[width=0.5\linewidth]{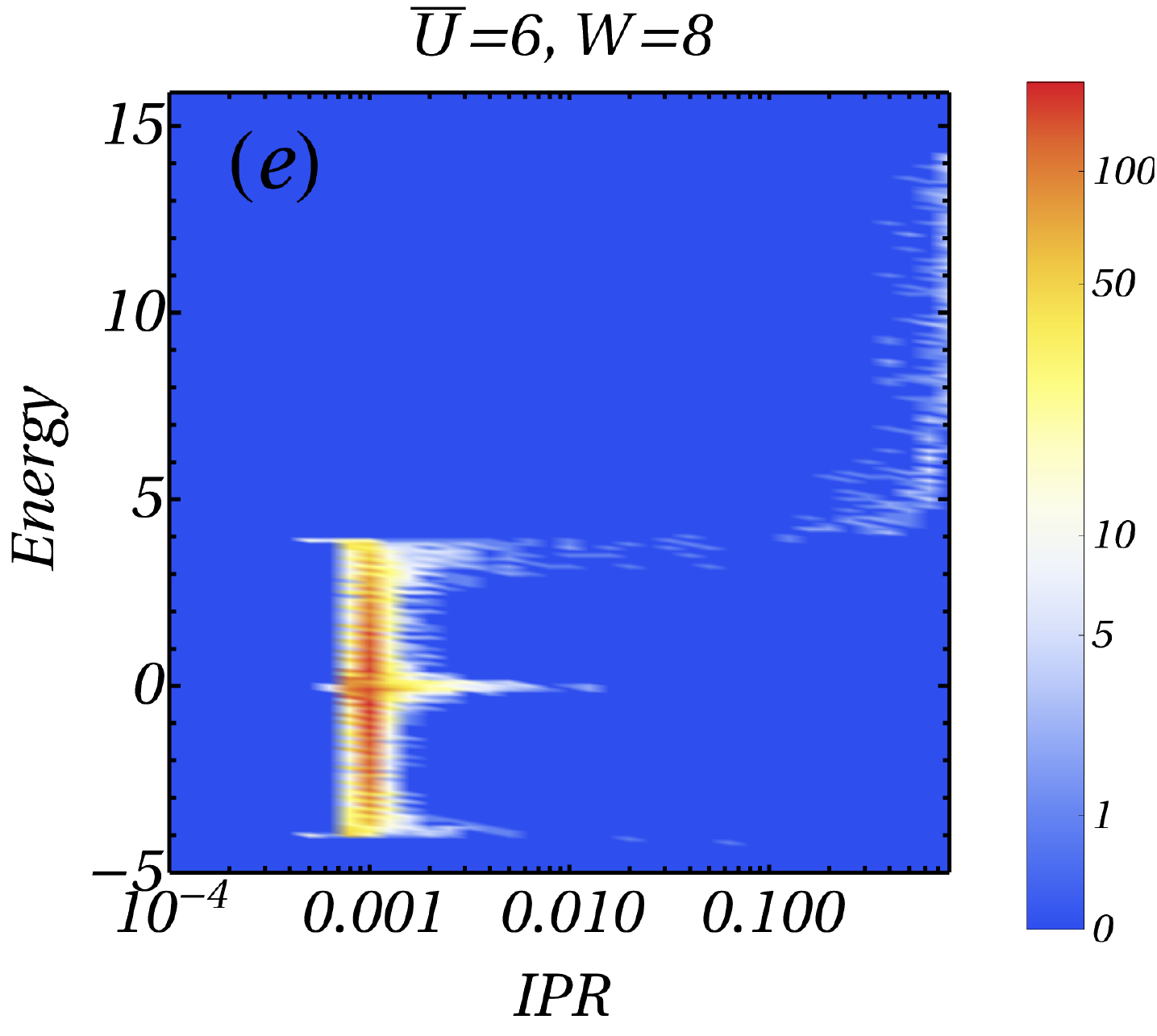} &    \includegraphics[width=0.5\linewidth]{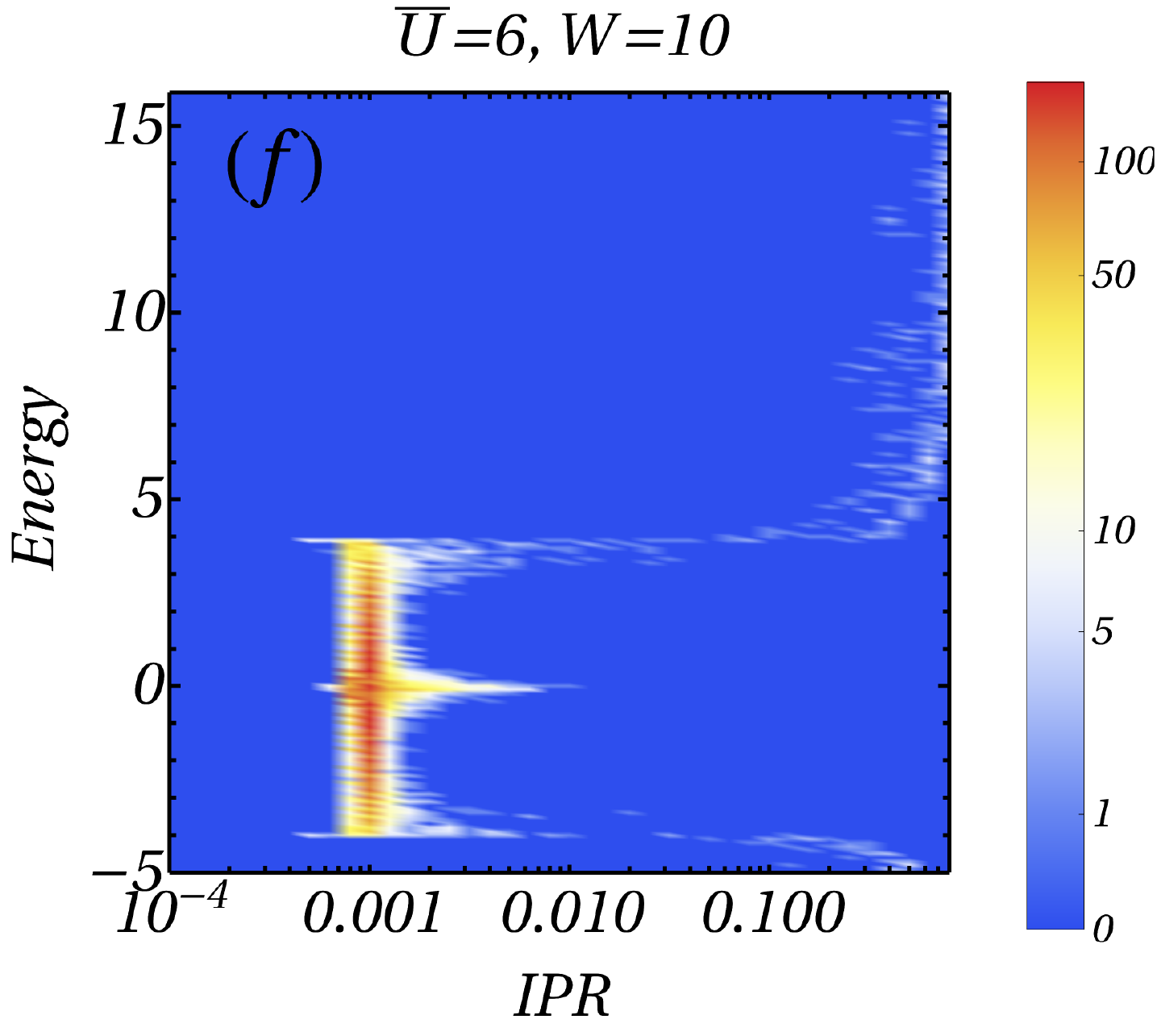} 
\end{tabular}
   
  \caption{Density of states (arbitrary units) as a function of energy and IPR for a chain of $N=50$ sites and $n_d=10$ different realisations over the disorder. The average on-site potential is $ \bar{U}=6$ and different values of $W$ between $0$ and $10$ are considered. Roughly speaking, here, an IPR in the $10^{-3}$ range means 2D-delocalized, in the $10^{-2}$ range means 1D-delocalized and in the $10^{-1}$ range means localized. Different type of states are indicated in white.}
   \label{spectrum}
\end{figure}

\par Through full diagonalization of the Hamiltonian of systems with $N=50$, as shown in Fig.~\ref{spectrum}, we observe different types of eigenstates with different localization properties. We focus here on delocalized atomic states with energy between $-4$ and $4$ and analyze the molecular states later. Depending on the energy range and on the parameters $\bar{U}$ and $W$, we identify four broad categories of such atomic states -- (a) quasi-ideal states, (b) chaotic states, (c) resonant states, and (d) separatrix states -- that we discuss in turn. States (a), (c) and (d) are finite-size effects that are expected to become negligible in the thermodynamic limit compared to states (b) that form the majority of the atomic band.

\medskip

\par (a) In Fig.~\ref{spectrum}(a), at vanishing disorder $W=0$, and interaction strength $\bar{U}=6$, we observe an IPR of $I_2 \approx 6.10^{-4}\simeq 1.5/N^2$ for $N=50$, which corresponds to a participation ratio $P_2 \sim 67\%$. This is compatible with the analytical solution of the problem at $W=0$ (see Eq.~\ref{eq:bethe}), which gives modified planes waves that we called ``scattering states". 

Their participation ratio varies continuously as a function of $\bar{U}$ between $2/3\simeq 67\%$ when $\bar{U}\to \infty$ (similar to standing waves in a box with open boundary conditions) and $100\%$ for plane waves when $\bar{U}\to 0$. When numerically computing the IPR, we took care in removing the possible degeneracy of scattering states by slightly twisting the boundary conditions (i.e. adding a small random magnetic flux across the two non-contractible loops of the real-space torus).

\begin{figure}[!h]
\begin{tabular}{c}
    \includegraphics[width=0.8\linewidth]{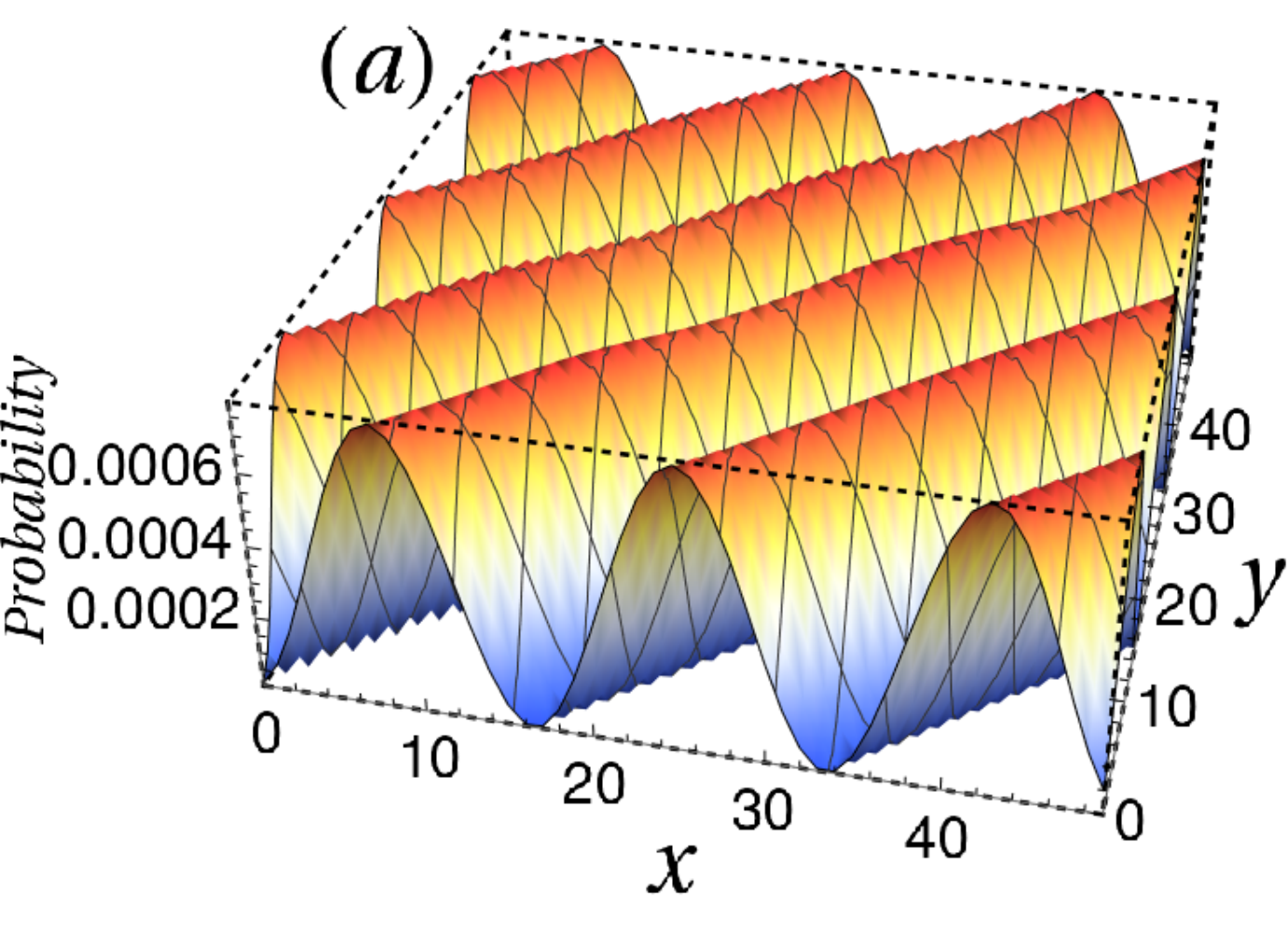}      \\
      \includegraphics[width=0.8\linewidth]{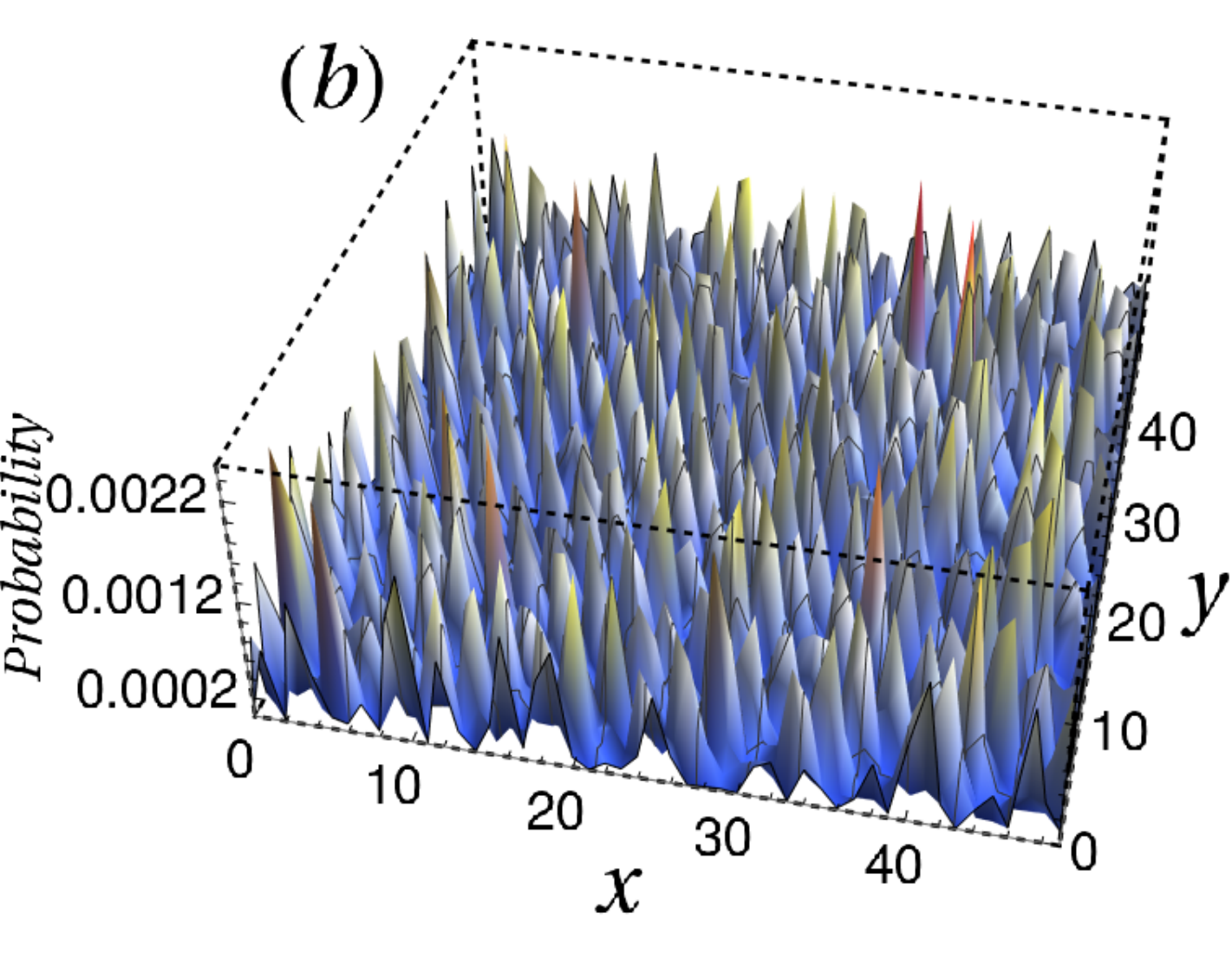}      \\
       \includegraphics[width=0.8\linewidth]{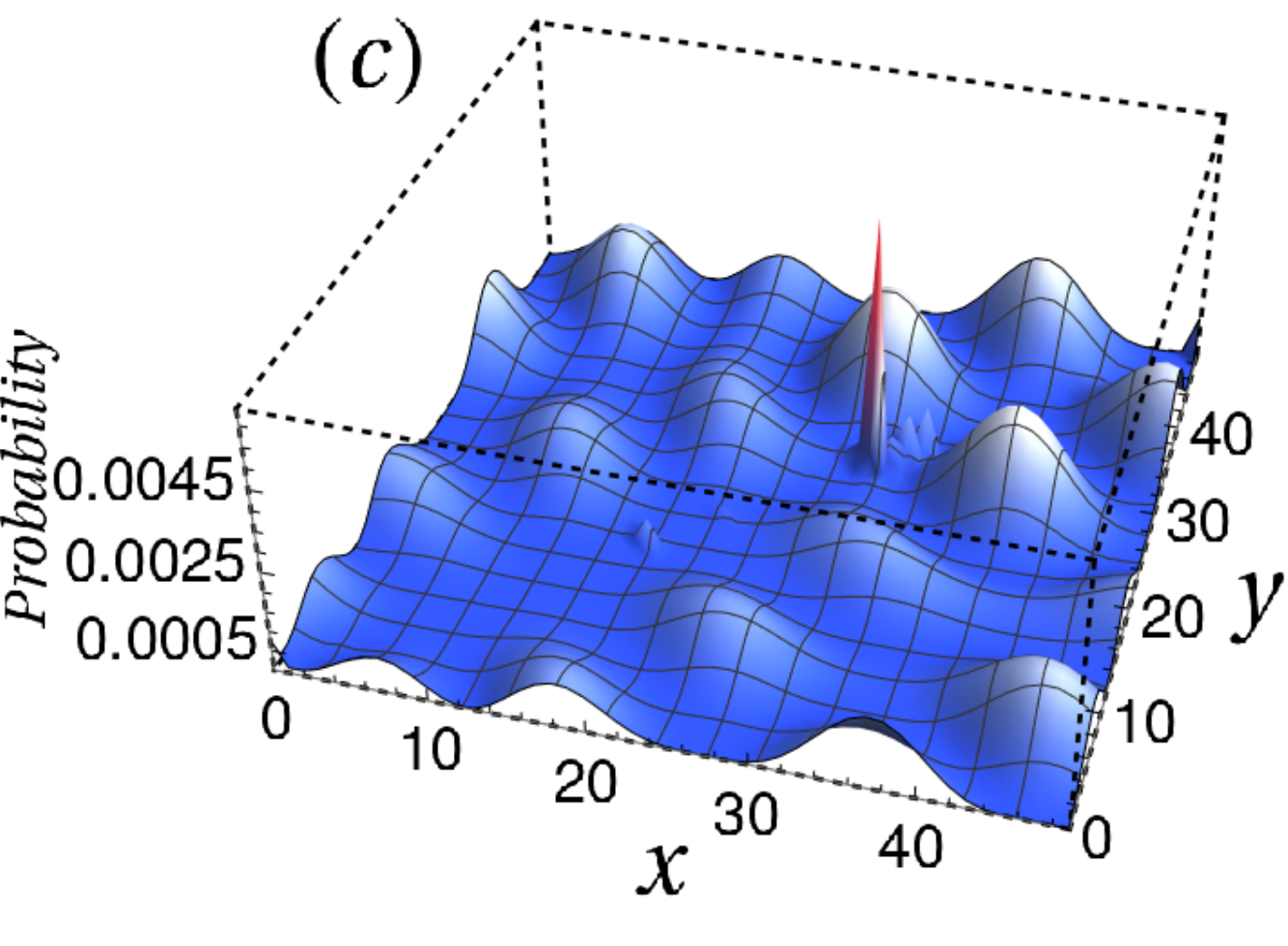}      \\
         \includegraphics[width=0.8\linewidth]{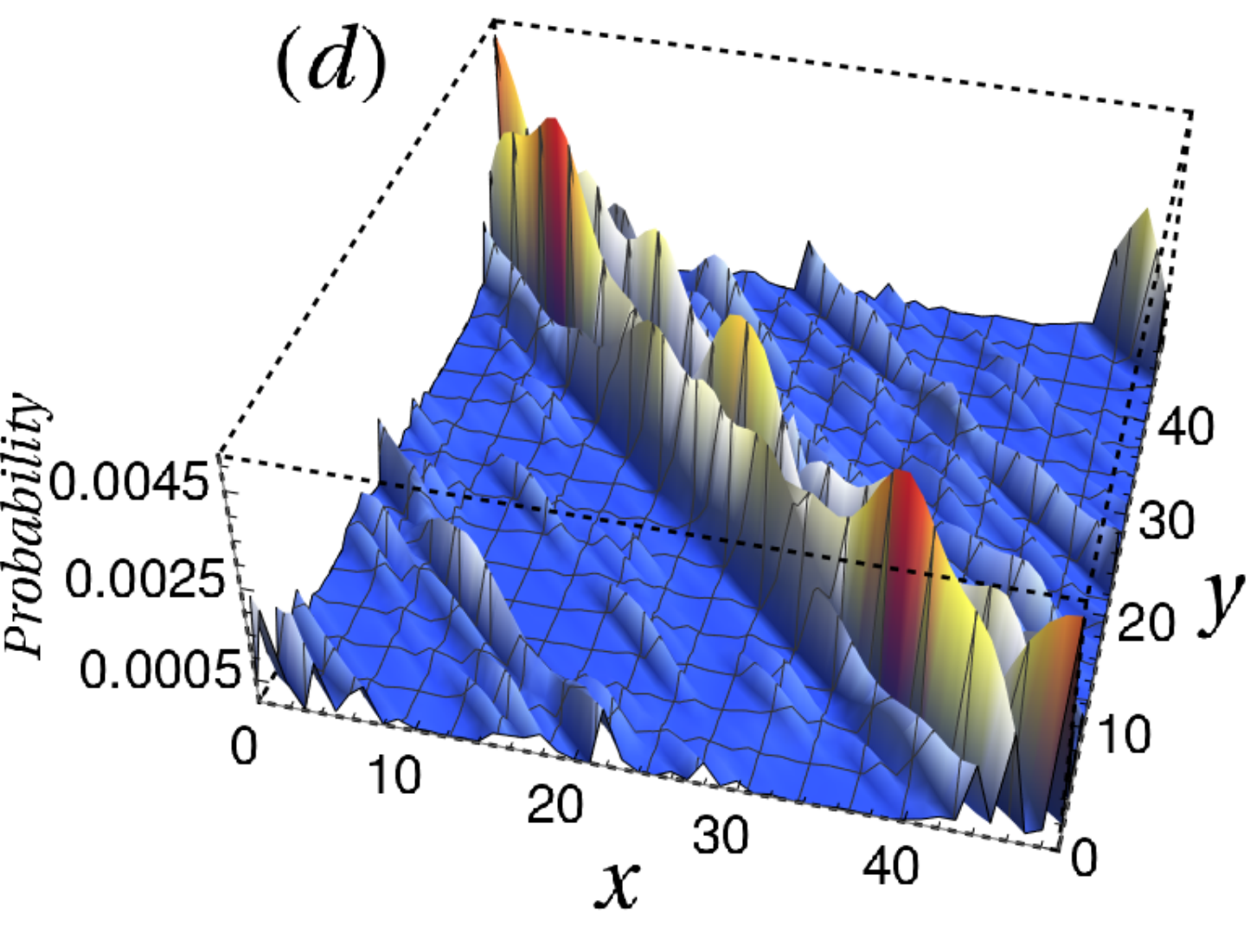}  
    
\end{tabular}

  \caption{Typical atomic eigenstates for a disordered interaction on a chain of $N=50$ sites: (a) quasi-ideal, (b) chaotic, (c) resonant, (d) separatrix. }
   \label{states_band}
\end{figure}

When weak disorder is turned on, we observe modifications in the eigenfunctions' IPR of the atomic band [see Fig.~\ref{spectrum}(b)]. In most cases, the IPR increases upon introducing finite disorder (this is discussed below under the name ``chaotic states''). However, mainly at the atomic band edges [see Fig.~\ref{spectrum}(c)], some states remain very delocalized. They are called quasi-ideal by~\cite{Cuevas1996}, because their weight on the disordered diagonal is small and they are almost like scattering states unaffected by the disorder [see Fig.~\ref{states_band}(a)]. In~\cite{Cuevas1996}, where $\bar{U}=0$, the authors argue that quasi-ideal states only exist due to finite-size effects in the presence of disorder and that their number is expected to vanish when $N\to\infty$. In Appendix~\ref{ap:energypertu}, we provide a generalized proof, valid also for $ \bar{U}\neq0$, that quasi-ideal states actually exist in the thermodynamic limit. However, they still can be considered as finite-size effects, because they form a vanishing measure set, the ratio of their number over that of chaotic states tending to 0 when $N\to \infty$. 
\medskip

\par (b) Most states inside the atomic band are ``chaotic" [see Fig.~\ref{spectrum}(b) and (c)]: their IPR slightly increases and scales as $3/N^2$, so that the participation ratio is lower ($\simeq 33\%$) but wavefunctions are still delocalized [see Fig.~\ref{states_band}(b)]. What we call ``chaotic states'' are similar to the ones observed by~\cite{Cuevas1996} and are typical of chaotic billiards. Such quantum states are discussed in detail in chapter 15 of Gutzwiller's book~\cite{Gutzwiller1990} (see in particular Figs.~44-46). They are delocalized, have a random character (but are not speckle) and have the same participation ratio ($1/3$) as eigenvectors of random matrices in the Gaussian orthogonal ensemble (GOE), see e.g.~\cite{Kaplan1999}. In addition, they feature filaments due to a preferred wavelength related to their energy content (see Fig.~2 in~\cite{Cuevas1996}). But filaments are not captured by eigenvectors of random matrices. Similarly to what was done in the continuum in~\cite{OConnor1987} following a conjecture by Berry~\cite{Berry1983}, in ``tight-binding billiards'', filaments can also be reproduced by building random superpositions of Bloch waves of a given energy. Fixing the energy is what selects a given wavelength that defines the width of the filaments. These filaments should be clearly distinguished from quantum scars~\cite{Heller1984,Kaplan1999}. The latter are enhanced probability in an eigenstate's wavefunction due to an underlying unstable periodic orbit of the corresponding classical billiard.

\medskip

\par (c) When the bottom of the molecular band at energy $\bar{U}-W$ becomes smaller than the top of the atomic band at $4$, there is band overlap. In the overlapping energy range, we observe states with an IPR in between that of typical atomic states and that of Anderson localized molecular states [see Fig.~\ref{spectrum}(d)]. These states are mainly localized along the diagonal and their wavefunction is close to that of Anderson localized molecular states. However, because their energy matches that of scattering states, they hybridize with them, which creates weight away from the diagonal [see Fig.~\ref{states_band}(c)]. They represent molecules that are coupled to atomic states and are partially dissociated. We call them ``resonant states''. They could also be called virtual bound states in analogy with the well-known phenomena occurring with impurities in metals discovered by Friedel (see e.g.~\cite{Georges2016}). We expect these states to become negligible in the thermodynamic limit as their number is at most $N$ (which is the number of bound states). They are discussed in more detail in Sec.~\ref{sec_overlap}.

\medskip

\par (d) The last type of atomic states that we observe are found in the middle of the atomic band near zero energy [see Fig.~\ref{spectrum}(d)]. Their IPR is quite large compared to the rest of the band ($I_2 \sim 1/N$). In fact, they are states which are localized only along the $x_+$ direction but extended into the relative motion direction $x_-$ [see Fig.~\ref{states_band}(d)]. These ``separatrix states'' are a consequence of the separatrix (iso-energy $E=0$) line in the dispersion relation when there is no interaction. They share some properties of scarred states~\cite{Heller1984,Kaplan1999} familiar in the context of quantum billiards but are clearly distinct (actually, we do not see scarred states in the present model). Since we could not find a description of these states elsewhere, we devote a complete section to them (see Sec.~\ref{sepstates}). As the number of these states is $N/2\ll N^2$, they are also expected to become negligible in the thermodynamic limit.

\subsection{Level-spacing statistics}

\par The localization of eigenfunctions can also be observed by the level-spacing statistics (for a review, see e.g.~\cite{Akkermans2007}). For a given energy $E$, an ensemble of normalized level spacings $s$ is obtained from an ensemble of the Hamiltonians, $H^{(r)}$, where $r$ is an integer index for the disorder realization, by  
\begin{align}
    \Delta E_n^{(r)} &= E_{n+1}^{(r)} - E_n^{(r)}, \quad \mathrm{with} \quad E \approx E_n^{(r)}, E_{n+1}^{(r)};\nn \\
    s_n^{(r)} &= \frac{\Delta E_n^{(r)}}{\overline{\Delta E}},
\end{align}
where $\overline{\Delta E}$ denotes the mean of the values of $\Delta E_n^{(r)}$. This procedure is known as spectrum unfolding in the literature on level-spacing statistics.

In the absence of disorder, for an integrable system with more than one degree of freedom, we generically expect that the normalized  level spacings $s$ have an exponential distribution (in this context also called Poisson distribution)~\cite{Berry1977}:
\begin{align}
    p(s) = e^{-s}.
\end{align}

The level spacing statistics is altered by disorder. 
Weak disorder usually breaks integrability and couples nearly degenerate eigenstates, leading to level repulsion and a universal behavior of the level spacing distribution, close to the Wigner surmise for the GOE:
\begin{align}
    p(s)=\frac{\pi}{2}s\, \exp \left(-\frac{\pi}{4}s^2\right).
\end{align} 
We will refer to this universal type of behavior as Wigner-Dyson or GOE distribution. 
Strong disorder, however, leads to Anderson localization of energy eigenstates. Thus nearly degenerate eigenstates can have wavefunctions localized to distant parts of the system, preventing hybridization between them. In this case the level-spacing statistics is again expected to be Poissonian, if the system size is considerably larger than the localization length. 
There are therefore two quite different situations in which Poisson statistics is obtained: clean integrable system (localized in momentum space) or strongly disordered Anderson localized system (localized in real space).

\par
\begin{figure}[h]
\begin{tabular}{cc}
  \includegraphics[width=0.5\linewidth]{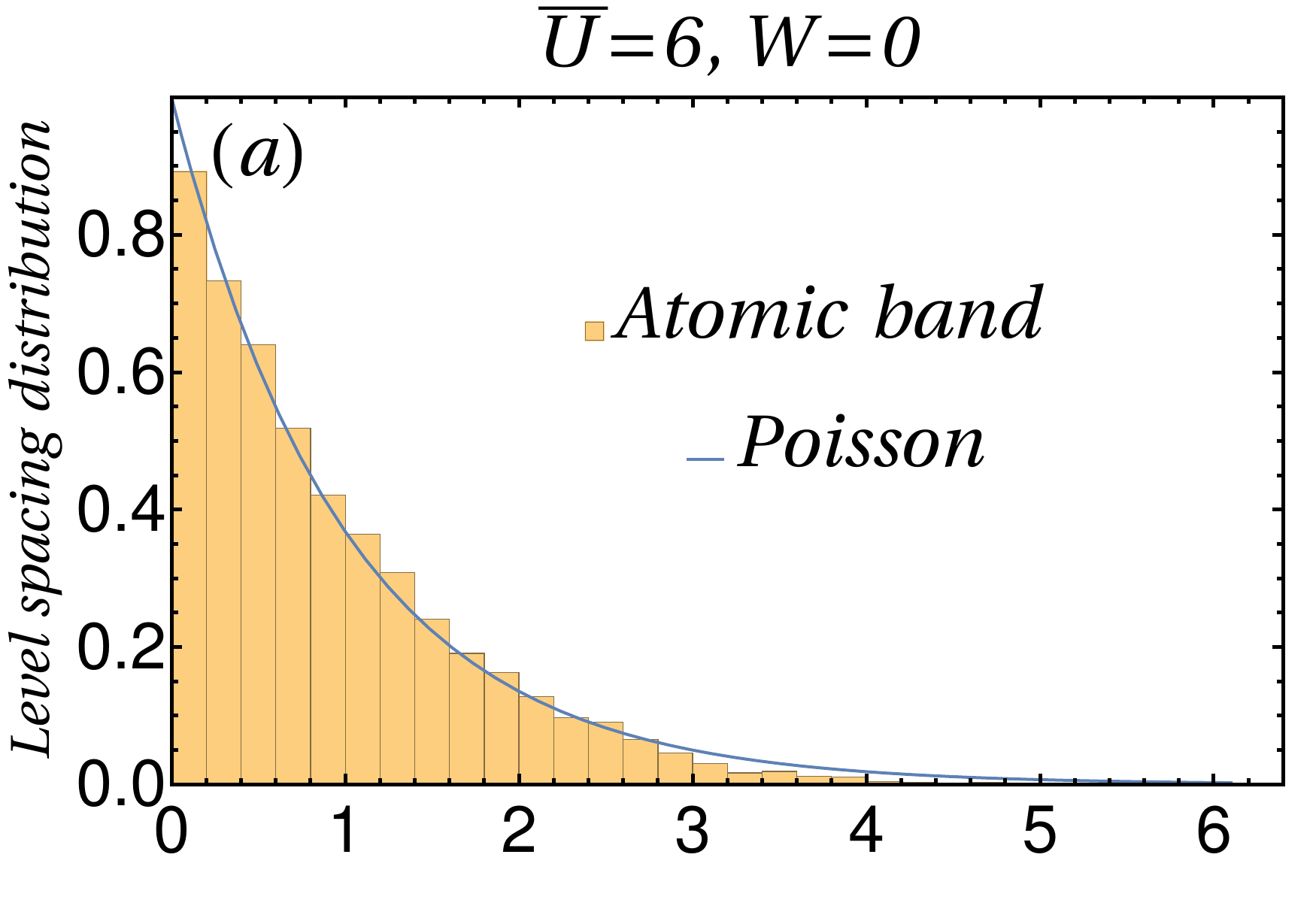}  & \includegraphics[width=0.5\linewidth]{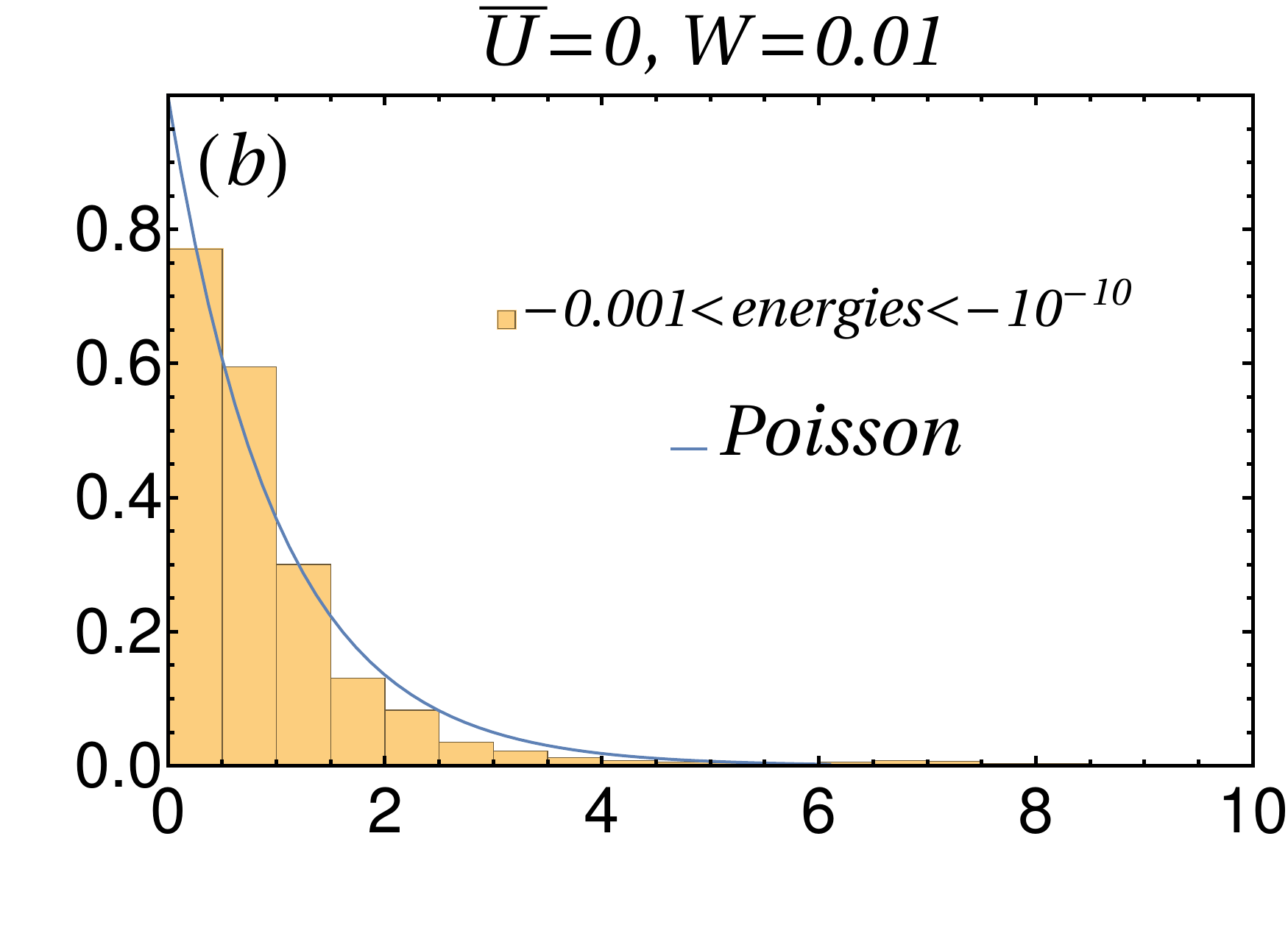}\\
   \includegraphics[width=0.5\linewidth]{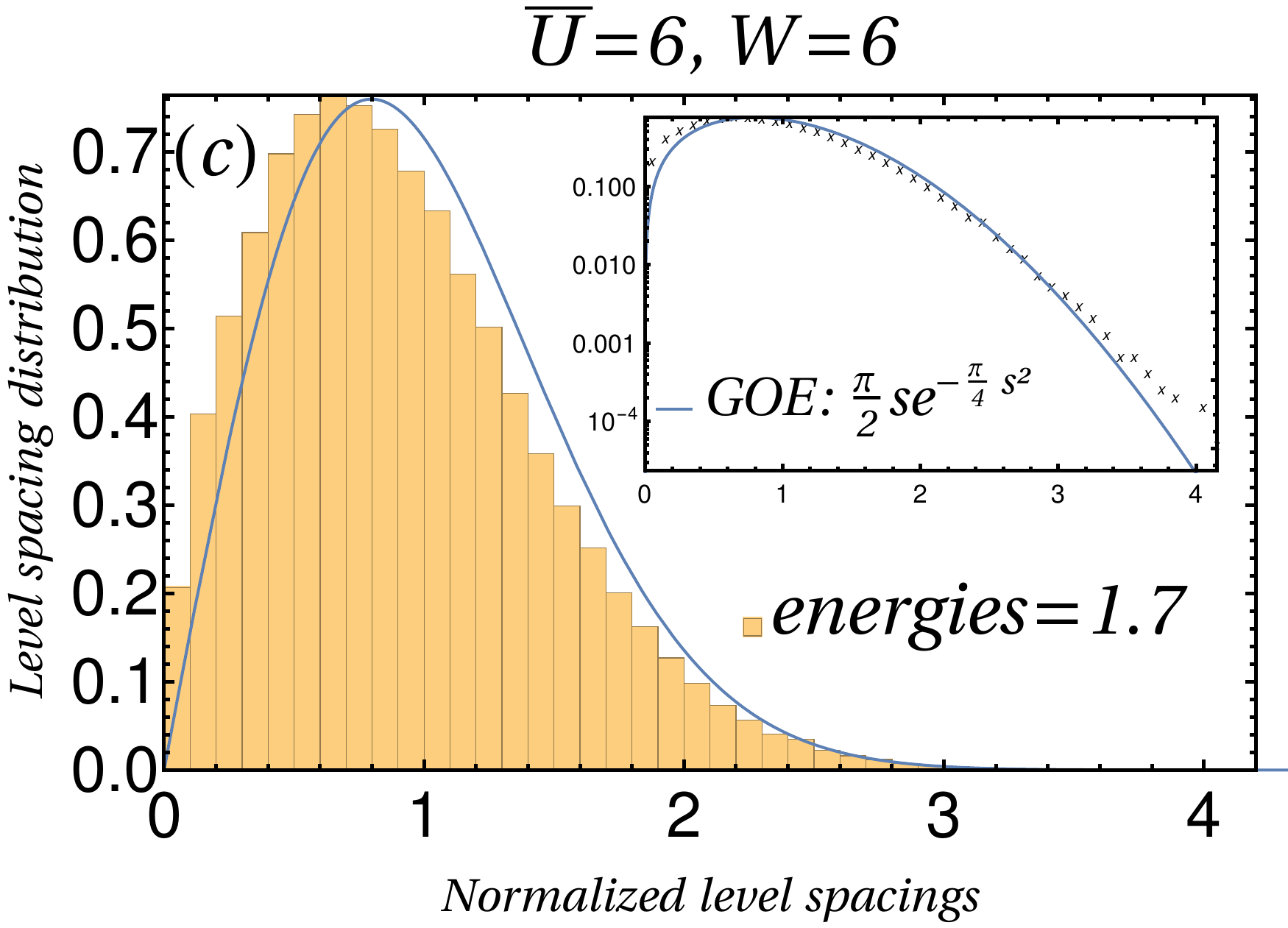}   &  \includegraphics[width=0.5\linewidth]{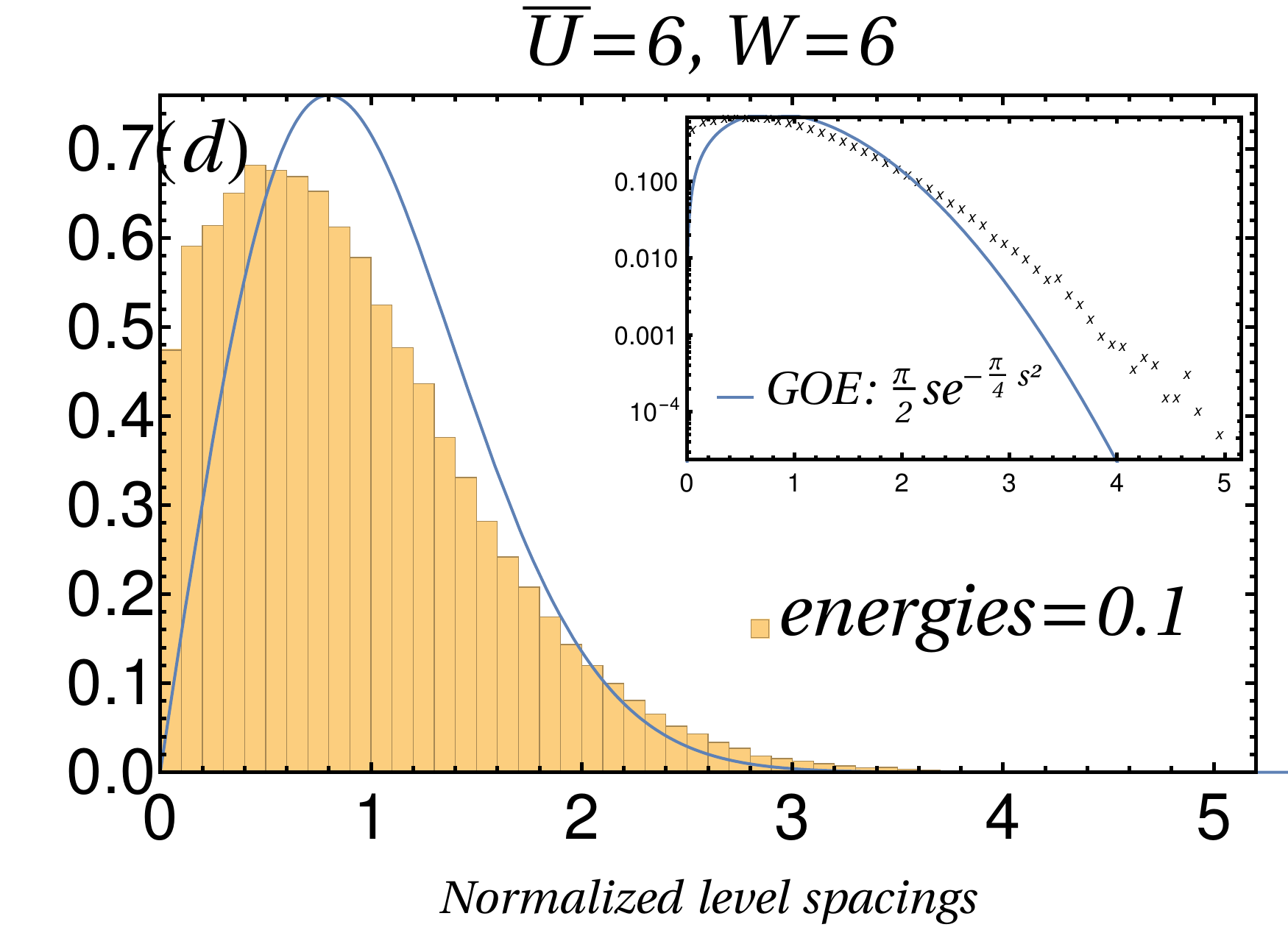} 
\end{tabular}
  
  \caption{Level-spacing distribution of atomic states for different regimes of energy $E$, mean interaction $\bar{U}$, strength of disorder $W$, chain size $N$ and number of disorder realizations $n_d$. (a): $E \in]-4,+4[$, $\bar{U}=6$, $W=0$, $N=70$ and $n_d=100$. Twisted boundary conditions were used to eliminate possible degeneracies. (b): $E$ between $-10^{-3}$ and $-10^{-10}$ (in order to isolate separatrix states), $\bar{U}=0$, $W=0.01$, $N=200$ and $n_d=100$. (c) $E \sim 1.7$, $\bar{U}=6$ and $W=6$, $N=900$ and  $n_d=1000$. (d) Same as (c) except that $E\simeq 0.1$. In (c) and (d), the Lanczos method was used to diagonalize the Hamiltonian. Both histogram and crosses in the inset represent the numerical data either with a linear or a log scale. Blue curves are the expected Poisson distribution for (a) and (b) and the GOE distribution for (c) and (d).}
    \label{spacingbulk}
\end{figure}

 Without disorder ($W=0$), for the band of atomic states, as expected for a 2D integrable system, we have a Poisson distribution [see Figure \ref{spacingbulk}(a)].

\par At $ \bar{U}\to\infty$ or $W\to\infty$, we expect the band of atomic states to be integrable and get a Poisson distribution as the  wavefunctions are expelled from the diagonal because of the on-site potential on the diagonal that tends to be infinite. 

\par The question of the level-spacing statistics at finite $ \bar{U}$ and $W$ is more subtle. For a finite $ \bar{U}$ and a small $W$, in Appendix~\ref{ap:energypertu}, we show that a perturbation in energy of the atomic scattering states due to $U$ scales as $W/( \bar{U}^2N^{3/2})\cos^2 (k_+/2)\sin^2\kappa_-$ where $k_+$ is the center of mass momentum and $\kappa_-$ one of the $N$-1 real solutions of the Bethe ansatz equation. They have to be compared with the mean level spacing which scales as $1/N^2$. If the perturbation is larger than the mean level spacing, atomic wavefunctions, because they are extended, lead to level repulsion and GOE statistics. Below this mean level spacing, we still have a Poisson distribution. Eventually, we conclude from Appendix~\ref{ap:energypertu} that, in the thermodynamic limit, we should obtain a distribution of level-spacings which converges slowly towards GOE.  The computation of Appendix~~\ref{ap:energypertu} does not take the contribution of the molecular band into account. However, as we will see in Sec.~\ref{sec_mol}, this convergence is speeded up when the molecular band overlaps the atomic band, the coupling between disorder and atomic states being much larger.  This is why, in practice, we could obtain the GOE statistics for reasonable $N$ but for $ \bar{U}-W<4$ (see Figure \ref{spacingbulk}-c). Indeed, in order to study the level-spacing statistics of the majority of states in the atomic band, we need to reach sample sizes much larger than in~\cite{Cuevas1996} in order to suppress the contribution of minority states (quasi-ideal states, resonant states and separatrix states).

\par At the center of the energy band near $E=0$, we do not obtain a universal level-spacing distribution because chaotic and separatrix states are mixed [see Figure \ref{spacingbulk}(d)]. At small disorder, separatrix states are localized along the direction of the center of mass and delocalized in the relative motion direction. Their IPR scales as $1/N$ (see Appendix~\ref{ap:iprsize}). They do not overlap and their level-spacing statistics agrees with the Poisson distribution [Fig.~\ref{spacingbulk}(b)]. For higher disorder, some of these separatrix states couple with atomic states and delocalize (therefore leading to level repulsion), while the others remain localized, which leads to the non-universal distribution seen in Fig.~\ref{spacingbulk}(d).

\section{Molecular states}
\label{sec_mol}
\par 
In this section, we focus on the molecular band made of bound pairs of particles that is analogous to defect or surface states in a 2D billiard. We consider the large-interaction ($\bar{U}-4>0$) and  weak-disorder ($W<\bar{U}-4$) regime, where these molecular states are clearly separated in energy from the atomic states and discuss the effect of disorder on this molecular band.

\subsection{Energy spectrum and level-spacing statistics}
\par The density of states (DoS) of the molecular band without disorder ($W=0$) is similar to the one of a clean 1D system with van Hove singularities at the edges (see Fig.~\ref{fig:dos}). When disorder is small, Lifshitz tails~\cite{Lifshitz1964} appear at the edges of the band between $ \bar{U}-W$ and $ \bar{U}$  for the bottom and $\sqrt{ \bar{U}^2+16}$ and $\sqrt{( \bar{U}+W)^2+16}$ for the top. They correspond to very rare events where the potential is approximately constant and minimal (for the bottom) or maximal (for the top) over a certain region, resulting in a box-like potential. The corresponding states are localized by disorder, but like a particle-in-a-box rather than due to interferences as in Anderson localization. 

When the disorder increases the DoS of the molecular band becomes flat and structureless. For weak disorder $W<\bar{U}-4$, the energy of the molecular band is in the range $[\bar{U}-W,\sqrt{( \bar{U}+W)^2+16}]$ and lies outside the atomic band (whose energy is in the range $[-4,+4]$). All such states are found to be localized around the diagonal contact interaction.

\par
\begin{figure}[h]
\begin{tabular}{cc}
  \includegraphics[width=0.5\linewidth]{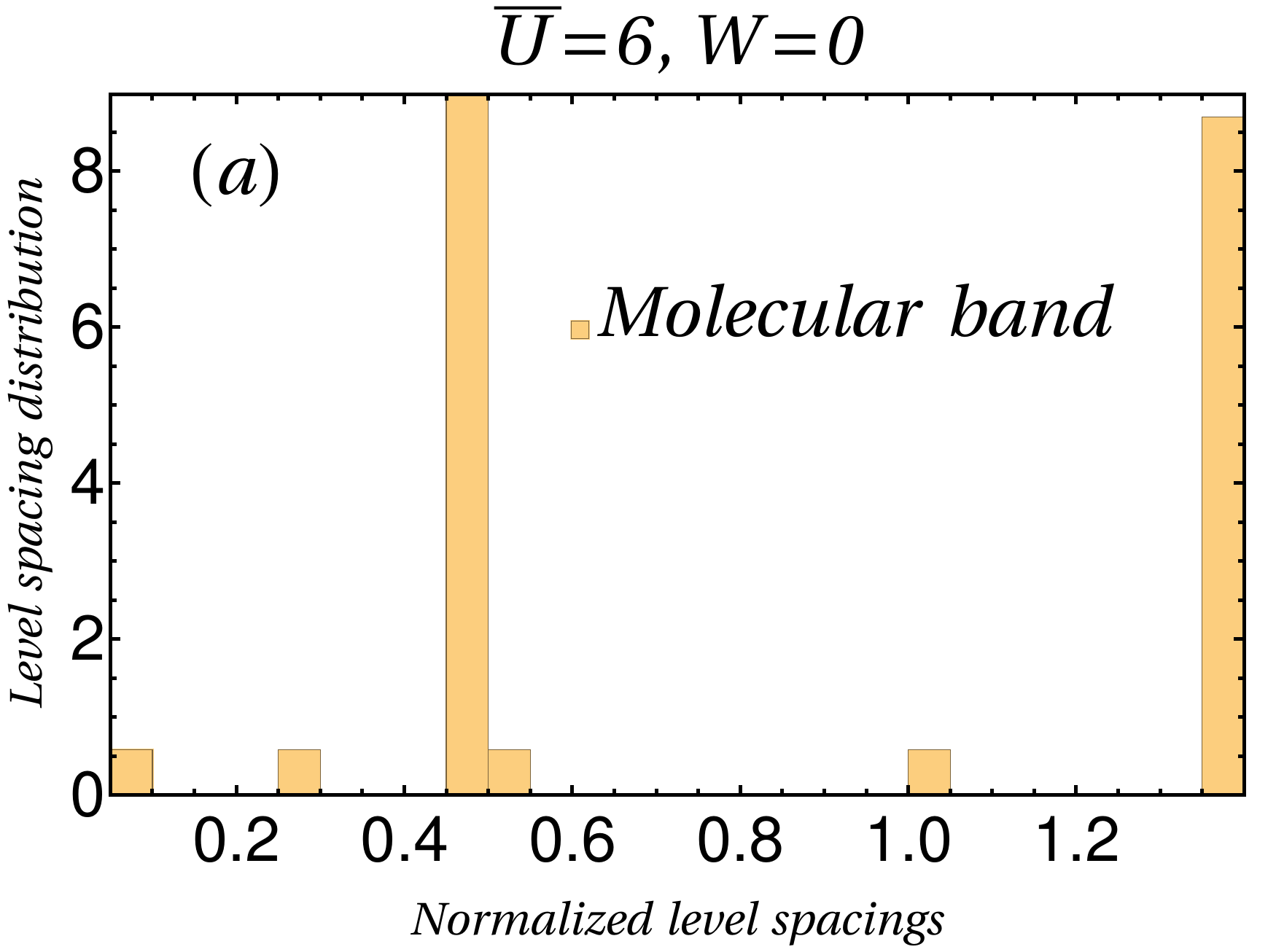}    & \includegraphics[width=0.5\linewidth]{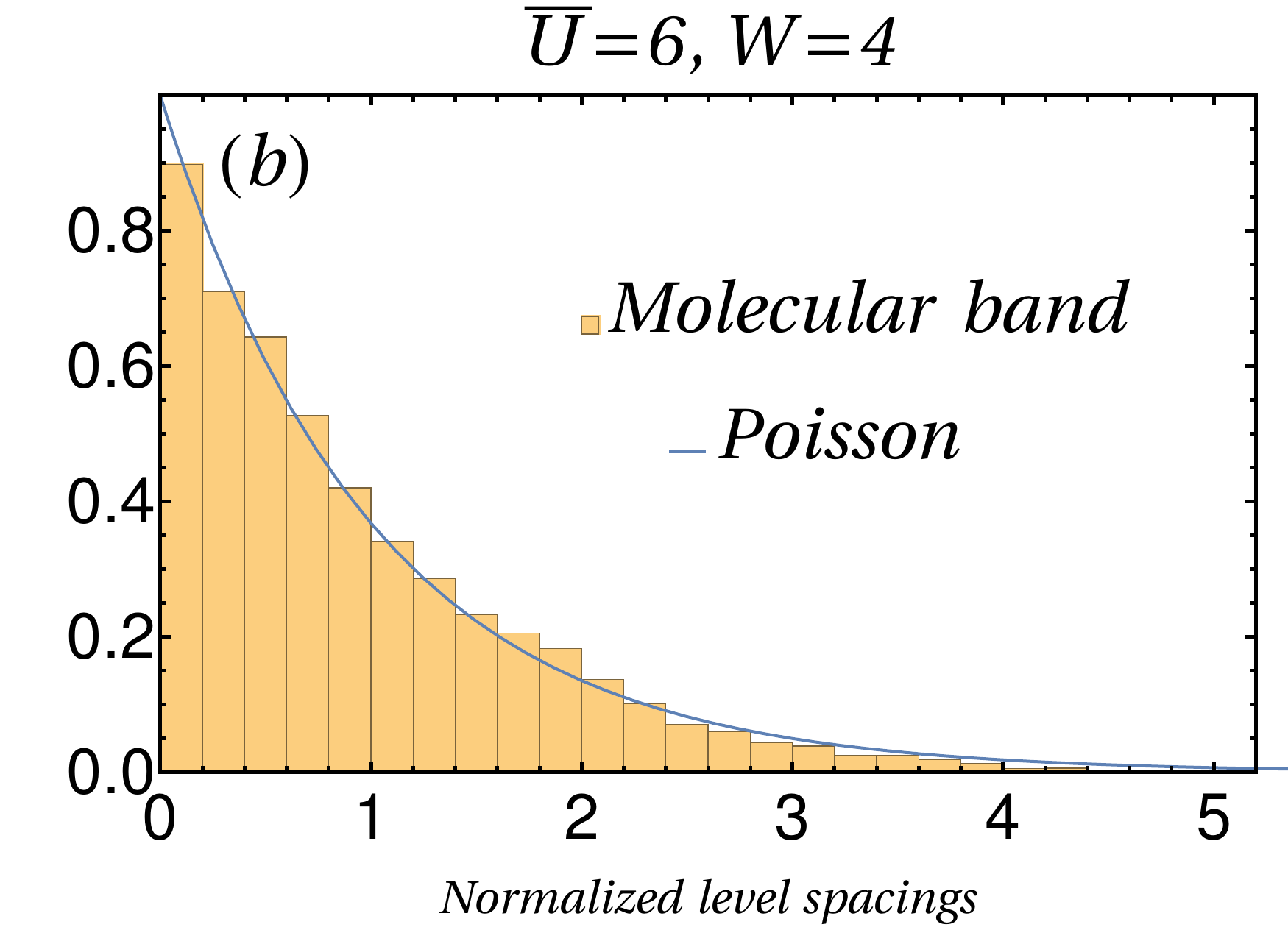}  \\
\end{tabular}

  \caption{Level-spacing distribution for the molecular band (energies $E$ such that $|E|>4.4$ in order to atomic states) computed for a chain of length $N=70$, a mean interaction $\bar{U}=6$ and $n_d=100$ disorder realizations. (a): $W=0$. (b): $W=4$.}
    \label{spacingmol}
\end{figure}

 At vanishing disorder, the system is integrable. The level-spacing statistics of the molecular band is not of universal type (i.e. neither Poisson nor GOE), see Fig.~\ref{spacingmol}(a), as is well known for a 1D integrable system~\cite{Berry1977}. Disorder  ($W>0$) leads to Anderson localization of the molecular states and   we find a Poisson distribution of level spacings (see Fig.~\ref{spacingmol}(b)).

\subsection{Molecular eigenfunctions}
\par For the molecular band, without disorder ($W=0$), quasi-momentum conservation ensures extended states in the center-of-mass $x_+$ direction. For interaction strength $\bar{U}=6$ and $N=50$, we find
an IPR of $I_2 \simeq 10^{-2}$, corresponding to exponentially localized states in the direction of relative motion, $x_-$, see Fig.~\ref{molecule}. The bound-state wavefunction is known analytically in that case, see Eq.~(\ref{eq:mol2}). Along $x_-$, it has an exponential decay over a typical length scale $\xi_\text{mol}=1/(2\kappa)\sim 0.3$, where $\sinh \kappa = \bar{U}/[4\cos(k_+/2)]\sim \bar{U}/(2\sqrt{2})$, as per Eq.~\eqref{eq:kappa}. We can check that the theoretical IPR is indeed $I_2 \approx \frac{1}{4N\xi_\text{mol}^2}\int d x_- e^{-2|x_-|/\xi_\text{mol}}\simeq 1.5\times 10^{-2}$.

\par
\begin{figure}[!h]
\begin{tabular}{cc}
      \includegraphics[width=0.5\linewidth]{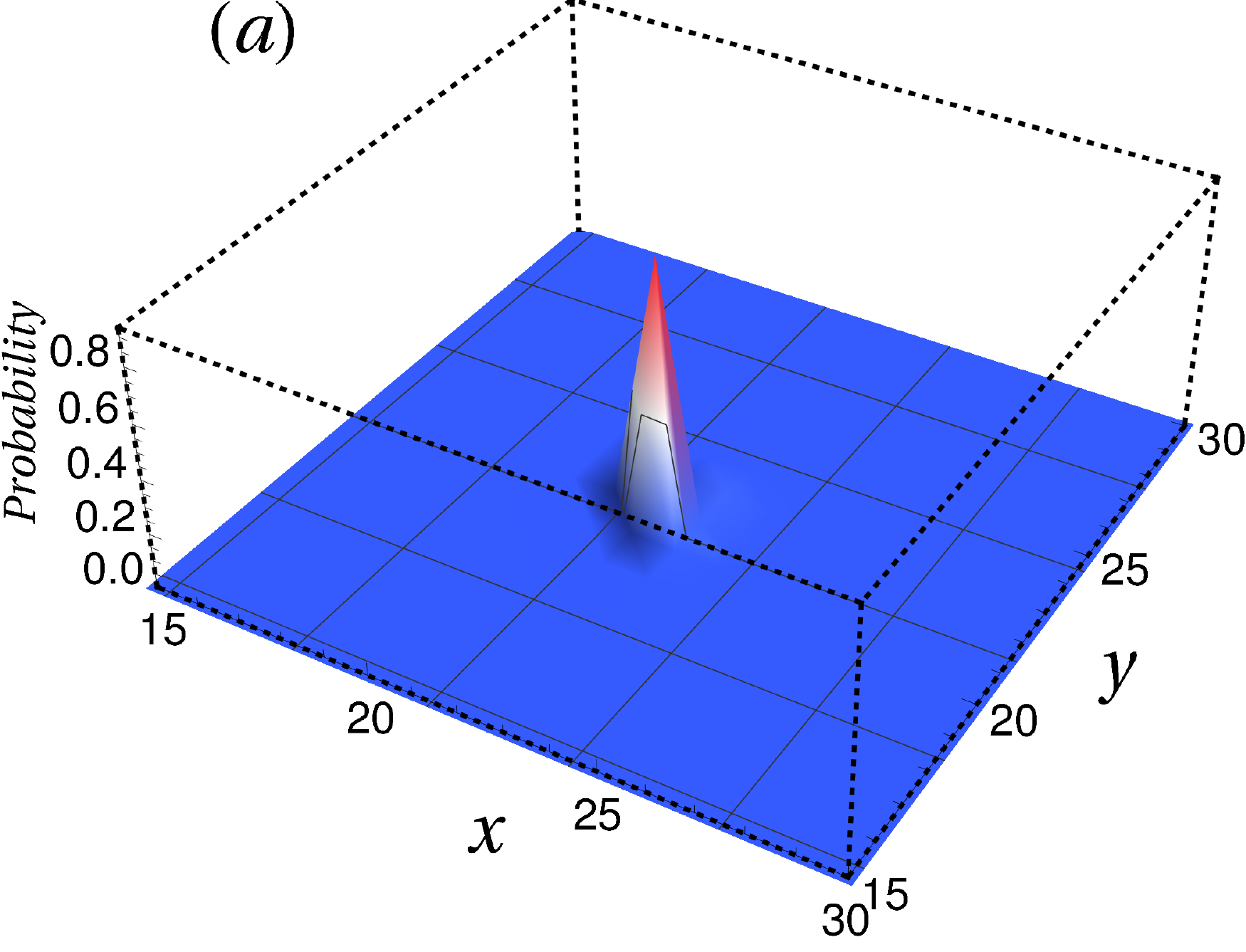}  & \includegraphics[width=0.5\linewidth]{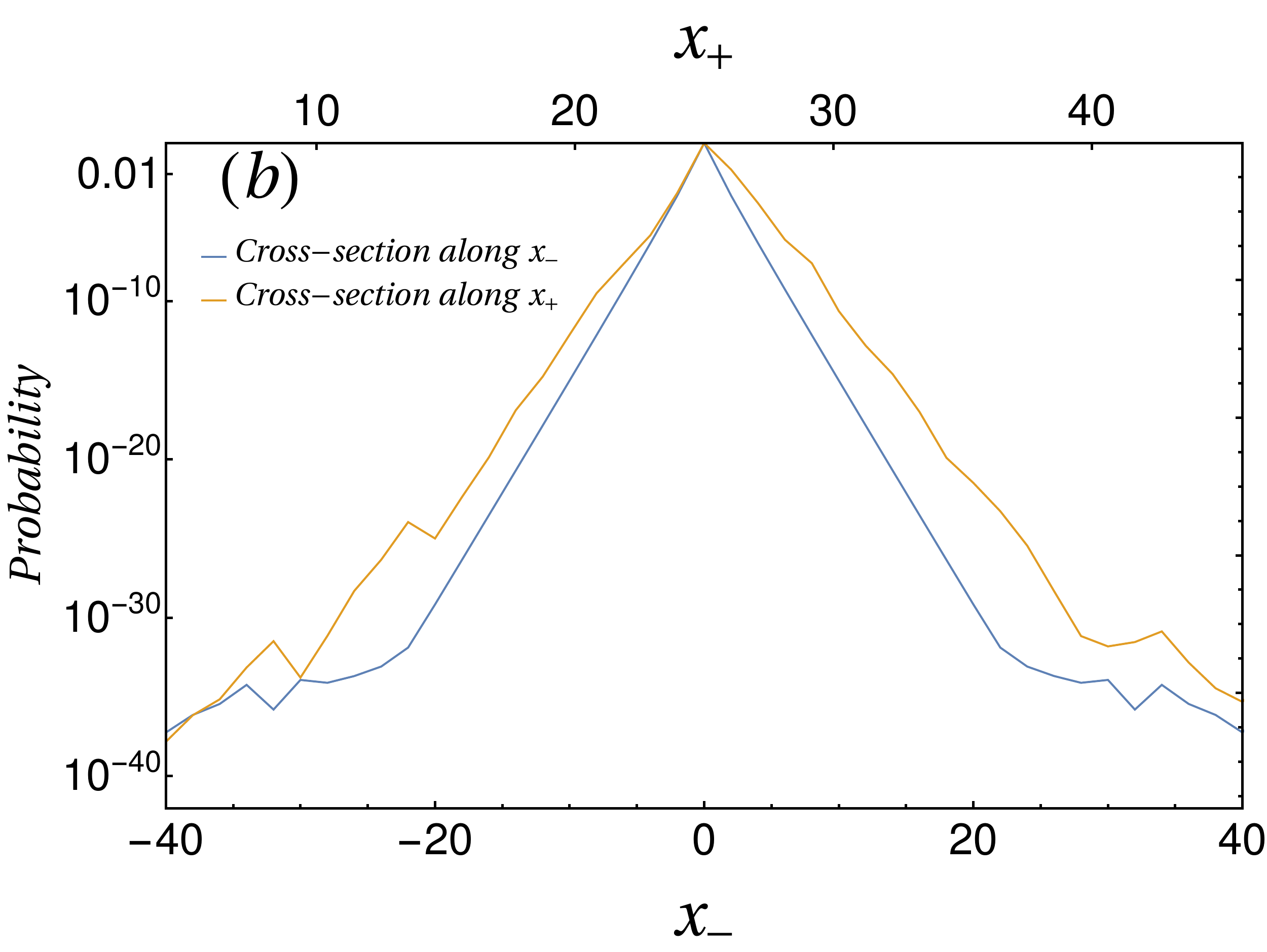} \\
 
\end{tabular}

  \caption{Anderson localized molecular state for a disorder $W=8$ and $\bar{U}=6$ on a chain of $N=50$ sites. (a): Probability of an eigenmode in the $(x,y)$ plane. (b): Cross-section of this wavefunction in the anti-diagonal direction $x_-$ (blue curve)  showing the extension of the molecule $\sim \xi_\text{mol}$. Cross-section in the diagonal direction $x_+$ (yellow curve) showing the localization length $\sim \xi_\text{loc}$ of the center of mass. A logarithmic vertical scale is used to emphasize the exponential decay.}
   \label{and_loc}
\end{figure}

 When disorder is switched on, the IPR of the molecular states increases, reaching approximately $I_2 \simeq 0.3$ when $W=2$ and $\bar{U}=6$, as shown in Fig.~\ref{spectrum}. We observe Anderson localization in the center of mass direction for the molecular states, see Fig.~\ref{and_loc}. Considering that the size of the molecules, $\xi_\text{mol}$ mostly depends on the mean interaction, $\bar{U}$, the characteristic Anderson localization length can be computed in the same manner as $\xi_\text{mol}$ and we find $\xi_\text{loc}\simeq 0.625$. Under the influence of disorder, the molecular band spreads and flattens. Its lowest energy is $\bar{U}-W=4$. When $W$ reaches $\bar{U}-4$, the two bands start to overlap and the molecular and atomic bands are no longer well separated. The case of overlap, corresponding to small interaction $\bar{U}-4<0$ or strong disorder $W>\bar{U}-4$, is analysed in Sec.~\ref{sec_overlap}.

\section{Overlapping bands}
\label{sec_overlap}
 If the interaction is weak ($\bar{U}-4<0$) or if the disorder strength is large ($W>\bar{U}-4$), there is an overlap in energy between the low lying states of the molecular band and the higher lying states of the atomic band. Let us consider the case where the overlap is due to $\bar{U}<4$ and the disorder is weak $W\ll1$. Because of disorder, bound states and scattering states are coupled. As a result, on the one hand, bound states dissociate and delocalize to become resonant (or virtual bound) states [see Fig.~\ref{states_band}(c)]. On the other hand, most scattering states couple weakly to molecular states and therefore remain almost unaffected, which we call quasi-ideal states [see Fig.~\ref{states_band}(a)].
 
 For even larger disorder, $W> \bar{U}$, the interaction $U_x$ at some positions can become negative and some molecular states with energy below the atomic band emerge. The molecular band then spreads in the range $[-\sqrt{( \bar{U}-W)^2+16},\sqrt{( \bar{U}+W)^2+16}]$ and all the atomic band is fully overlapped by molecular energies. When $W\to\infty$, we expect that the distribution of the molecular energies of the system tends to that of the on-site potential energies, i.e. a uniform distribution between $ \bar{U}-W$ and $ \bar{U}+W$.

 In the following, we discuss some properties of the states in the overlapping-band region.

\subsection{Eigenfunction analysis}

\par The regime of overlapping bands opens new channels of decay for some of the molecular states by removing the translation invariance of the center-of-mass-mode. Those whose energies are shifted into the band of atomic states can hybridize, dissociate and become delocalized. Instead of true bound states, they become resonances with finite lifetime (see Appendix~\ref{ap:lifetime} for a definition of this lifetime). This is a kind of re-entrance effect of disorder: delocalized molecular states at $W=0$ are localized by weak disorder $0<W<\bar{U}-4$ and then dissociate and delocalize due to hybridization with atomic states when the disorder further increases $W>\bar{U}-4$. If the whole energy spectrum is viewed as a single band, then a mobility edge at energy 4 separates high energy states that are localized from low energy states that are delocalized.

\par In Fig.~\ref{spectrum}, we see that the DoS is very weak at the edges of the disordered molecular band. These are the exponentially small Lifshitz tails~\cite{Lifshitz1964}. Therefore, we do not expect a large overlap of these tails with the atomic band for finite size system and, in practice, we observe resonances only when a macroscopic fraction of molecular states overlaps the atomic band, i.e., $W-\bar{U}+4\gtrsim 1$. Moreover, the IPR increases, e.g., $I_2\approx 0.5$, for $N=50$, $W=4$ and $ \bar{U}=6$, corresponding to $\xi_\text{loc}\sim0.42$. Eventually, at $W\geq 8$, the disorder is so strong that some previous resonant states leave the atomic band to spread below it and become again Anderson-localized molecules.\\

\par Atomic eigenfunctions are marginally affected by disorder when the molecular band is well separated from the atomic band, i.e., $W<\bar{U}-4$. Indeed, molecular eigenfunctions have an exponential decay over a typical length $\xi_\text{mol}$, and a coupling with the diagonal for atomic functions is then allowed but marginal. In the opposite limit of $\bar{W}\to\infty$, atomic wavefunctions have vanishing weight on the diagonal and are thus practically unaffected by the randomness. However, when the molecular band overlaps with the atomic band, i.e.,  $W>\bar{U}-4$ and $W$ finite, we expect a rather different behavior. Molecular wavefunctions with energies in the atomic range hybridize with atomic wavefuntions and delocalize, bringing a stronger coupling of the atomic wavefunctions to the disordered potential. We can understand that in this regime the coupling affects the atomic band much more intensively than in the case of $W<\bar{U}-4$. It is in this regime, that we find GOE statistics for the chaotic states, as shown in  Fig.~\ref{spacingbulk}(c).

\subsection{Molecular probability}
\par By definition, a molecular state is an eigenstate with energy outside the range $[-4,+4]$. In this section, we are interested in the probability for two particles initially on the same site $j$ to remain bounded at long times. 

We define the molecular (or survival) probability $P_\text{mol}$ as the square of the overlap between such a highly localized initial state and the molecular eigenstates:
\begin{equation}
P_\text{mol}(j) = \sum_{|E_\text{mol}|>4} |\langle \psi_\text{mol}|x=j,y=j\rangle|^2,
\end{equation}
where $j\in\mathbb{Z}$ is the initial position. This overlap is the long-time limit of the probability that two particles started initially from the same site $j$ do not dissociate and their distance remains bounded. Since its value depends on the initial position $j$ for any realization of disorder, to any set of parameters we can associate a probability distribution of $P_\text{mol}$ values, as we discuss below. For the moment, we concentrate on the long-time behavior and do not discuss the dynamics (see Appendix~\ref{ap:lifetime} for a discussion of the lifetime).   

\begin{figure}[h]
   \includegraphics[width=\linewidth]{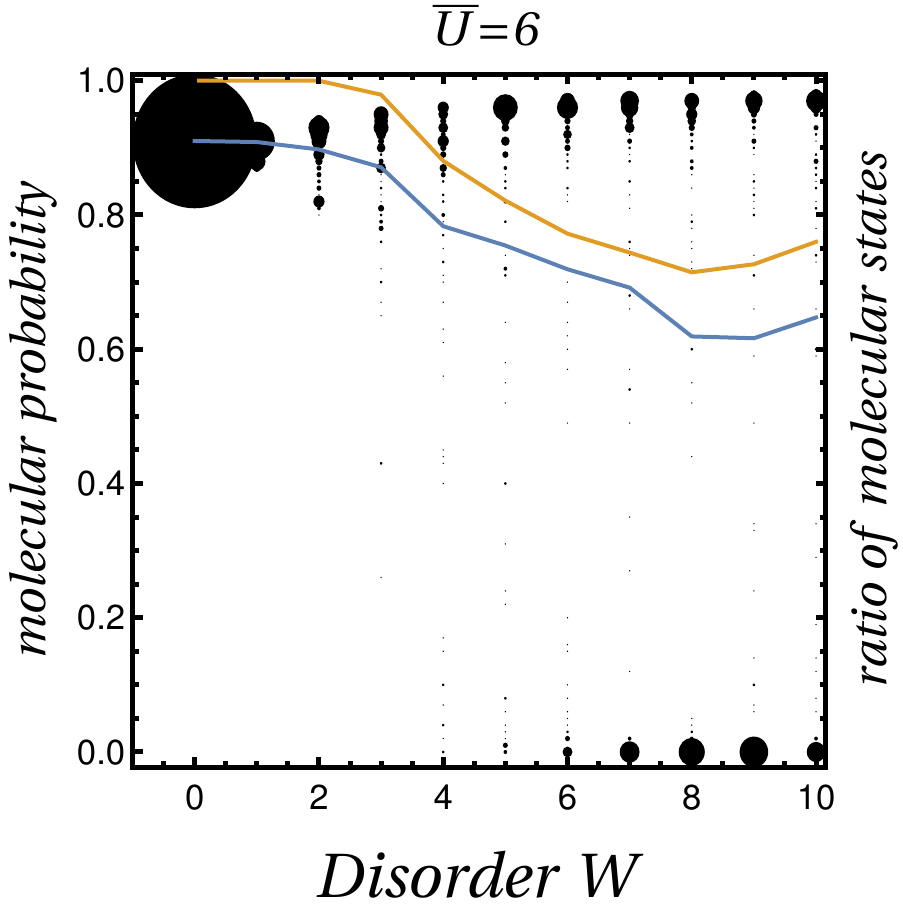}
  \caption{Molecular probability versus disorder $W$ for an average potential $\bar{U}=6$ on a chain of size $N=50$ and $n_d=10$ disorder realizations (for each, 10 randomly-chosen initial conditions on the diagonal are taken). The radius of a black dot is proportional to the number of occurrences. The blue curve is the mean of the distribution. The orange curve is the average on $n_d=100$ disorder realizations of the ratio between the number of molecular states (i.e energies outside [-4,+4]) and the number $N$ of bound states.}
    \label{probamoldis}
\end{figure}

We first consider how the distribution of the molecular probability $P_\text{mol}$ depends on the disorder $W$ at fixed $\bar{U}=6$, shown in Fig.~\ref{probamoldis}. 
At low disorder, the molecular band is separated from the atomic band by a gap. Here the molecular probability is  $91\%$ -- computed exactly as $2 K(-16/\bar{U}^2)/\pi$ using Eqs.~\eqref{eq:mol2} and \eqref{eq:kappa}, where $K$ is the complete elliptic integral of the first kind. When $W>2$, a fraction of the molecular states is coupled with the atomic states, and we thus find a bimodal distribution of molecular probabilities: either the initial state has a large overlap with an Anderson localized molecular state, and $P_\text{mol}\approx1$; or the initial state is mostly supported by resonant states with finite lifetimes, $P_\text{mol}\approx 0$. As the disorder increases, at first more and more molecular states are coupled with the atomic states and therefore the average $P_\text{mol}$ decreases. 

For $W>8$, however, we observe a re-entrant increase of $P_\text{mol}$: here increasing $W$ pushes some molecular states below the atomic band, and thus the mean $P_\text{mol}$ increases as a function of $W$. Thus, at high disorder, the mean $P_\text{mol}$ directly gives the fraction of molecular states with energy outside the atomic band. Qualitatively, the ratio between the number of molecular states (states that do not overlap in energy with the atomic band) and the number $N$ of bound states reproduces the trend of the molecular probability (see orange curve in Fig.~\ref{probamoldis}). As molecular states also have weight away from the diagonal sites, the latter ratio gives an upper bound on the molecular probability.

\begin{figure}[!!h]
\begin{tabular}{cc}
    \includegraphics[width=0.5\linewidth]{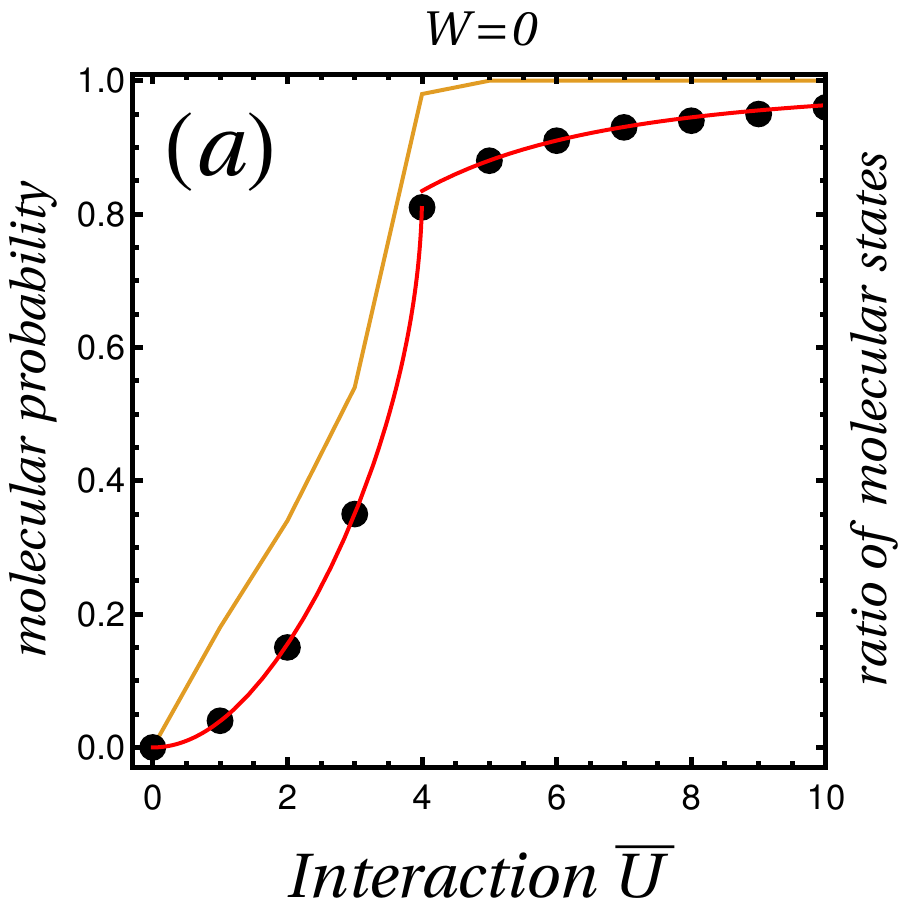}    &    \includegraphics[width=0.5\linewidth]{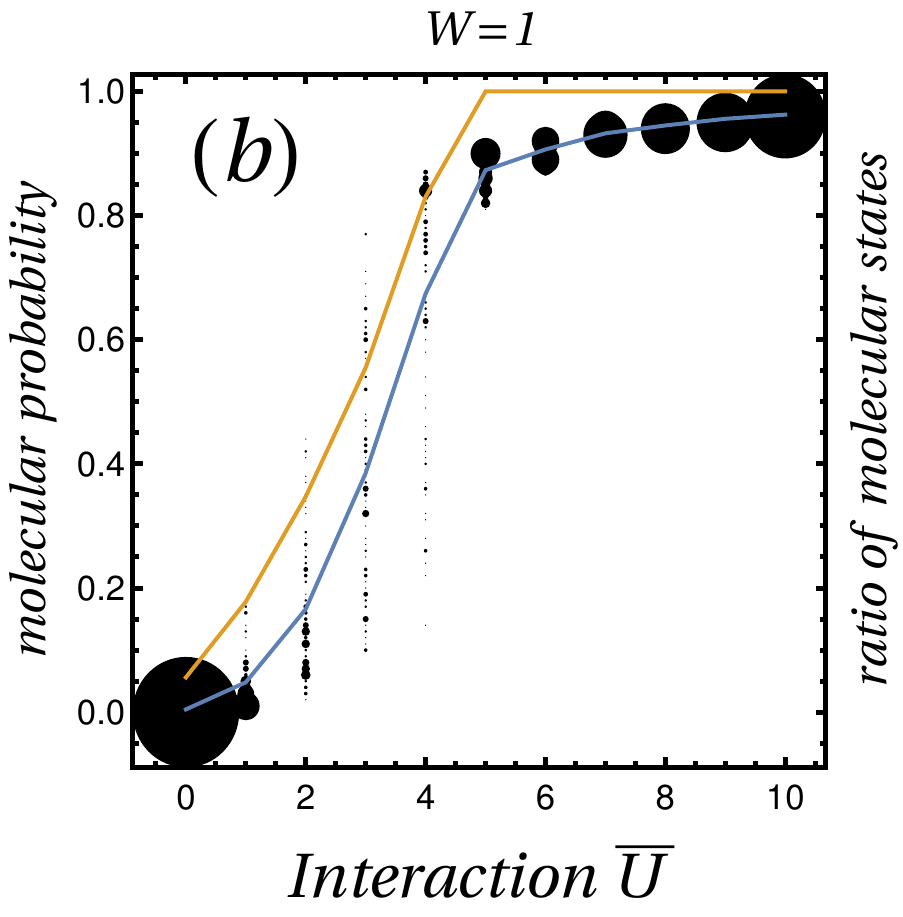}\\
      \includegraphics[width=0.5\linewidth]{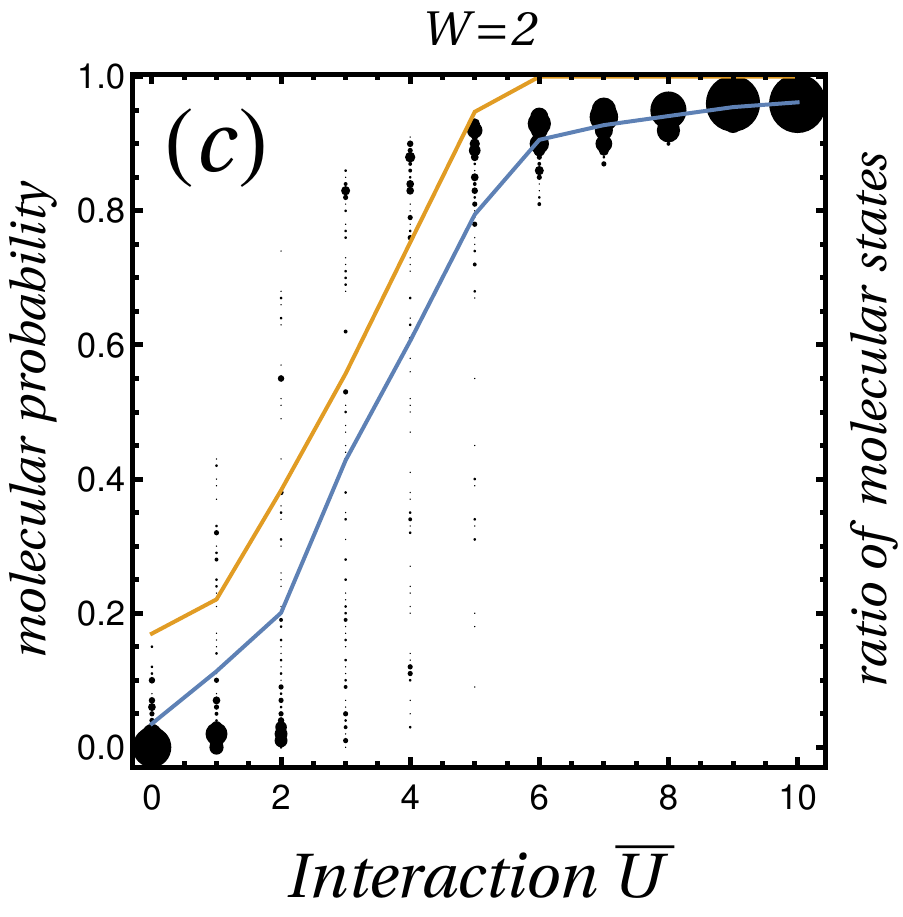} &   \includegraphics[width=0.5\linewidth]{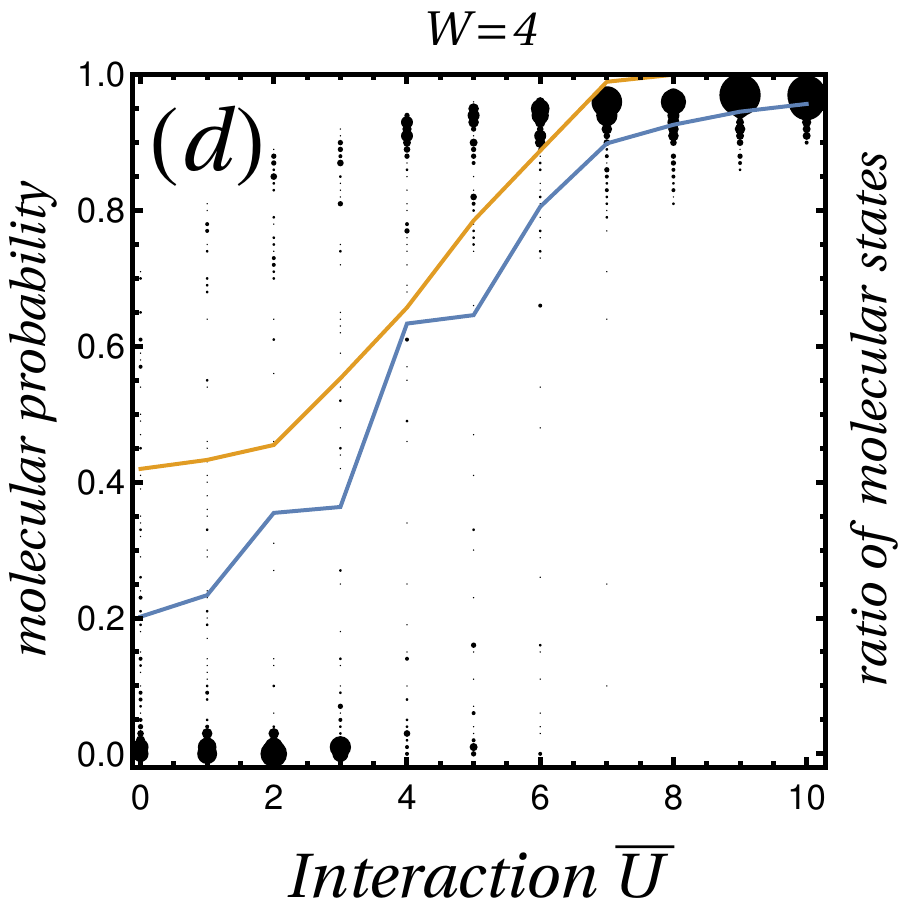}
\end{tabular}
  \caption{Molecular probability versus the average interaction $\bar{U}$ for different disorders $W$ on a chain of size $N=50$ and $n_d=10$ disorder realizations (for each, 10 randomly-chosen initial conditions on the diagonal are taken). (a): $W=0$. The red curve is the theoretical prediction of Eq.~\ref{eq:molproba}.
  (b): $W=1$, (c): $W=2$, (d): $W=4$. Black dots are the numerical computation of the molecular (survival) probability. The radius of a black dot is proportional to the number of occurrences. Blue curves are the mean of distributions. Orange curves give the average on $n_d=100$ disorder realizations of the ratio between the number of molecular states (i.e energies outside [-4,+4]) and the number $N$ of bound states.}
    \label{probamolint}
\end{figure}

\par One can also look at the molecular probability with respect to the average interaction $ \bar{U}$ for different strength of disorder $W$, see Fig.~\ref{probamolint}. At vanishing disorder, it is possible to compute this probability analytically. In Fig.~\ref{probamolint}(a), the solid red line is the theoretical prediction made using the definition of Sec.~\ref{sec_hubbard} and Eqs.~\eqref{eq:mol2} and \eqref{eq:kappa}: 
\begin{equation}
P_\text{mol}=\int_{-k_c}^{k_c} \frac{d k_+}{2\pi} \tanh \kappa=\frac{2 \bar{U}}{\pi} \frac{F\left[\tfrac{k_c}{2}, \tfrac{16}{16 + \bar{U}^2}\right]}{\sqrt{16 +\bar{U}^2}},
\label{eq:molproba}
    \end{equation}
where $F$ is the elliptic integral of the first kind and $k_c=\text{Re } 2 \arccos \sqrt{1-(\bar{U}/4)^2}$ is a cutoff that varies between $0$ and $\pi$ as a function of $\bar{U}$. The origin of this cutoff is the definition that molecular states have energies $E$ such that $|E|>4$.

\par For high disorder [$W=4$, Fig.~\ref{probamolint}(d)], we recover a bimodal distribution due to Anderson localization. There is a regime of rapid growth of the molecular probability when $ \bar{U}$ increases, which corresponds to the decoupling of molecular states leaving the atomic band. When the molecular band is completely outside the atomic band ($\bar{U}>W+4$), the initial state on the diagonal is mostly overlapped by few Anderson localized molecular eigenstates. However, their exponential decay over a length $\xi_\text{mol}$ allows a marginal coupling between the initial state and the atomic functions. This coupling vanishes when $\bar{U}\to\infty$. The greater $\bar{U}$ is, the  lower $\xi_\text{mol}$ is, the greater the molecular probability is.

\section{Separatrix states}
\label{sepstates}

Having treated the broad classes of molecular and atomic states and also the band overlap, we now focus on a special class of states at $E\approx 0$ that we call ``separatrix states''. These states are best understood by starting with the model without interaction (free time evolution of two particles hopping on a chain), and then considering how the disordered interaction affects the 0-energy eigenstates of the free model. Throughout this section we assume the number of sites $N$ to be even for simplicity.

\subsection{The 0-energy eigenspace of the noninteracting case }

\begin{figure}[!!h]
   \includegraphics[width=\linewidth]{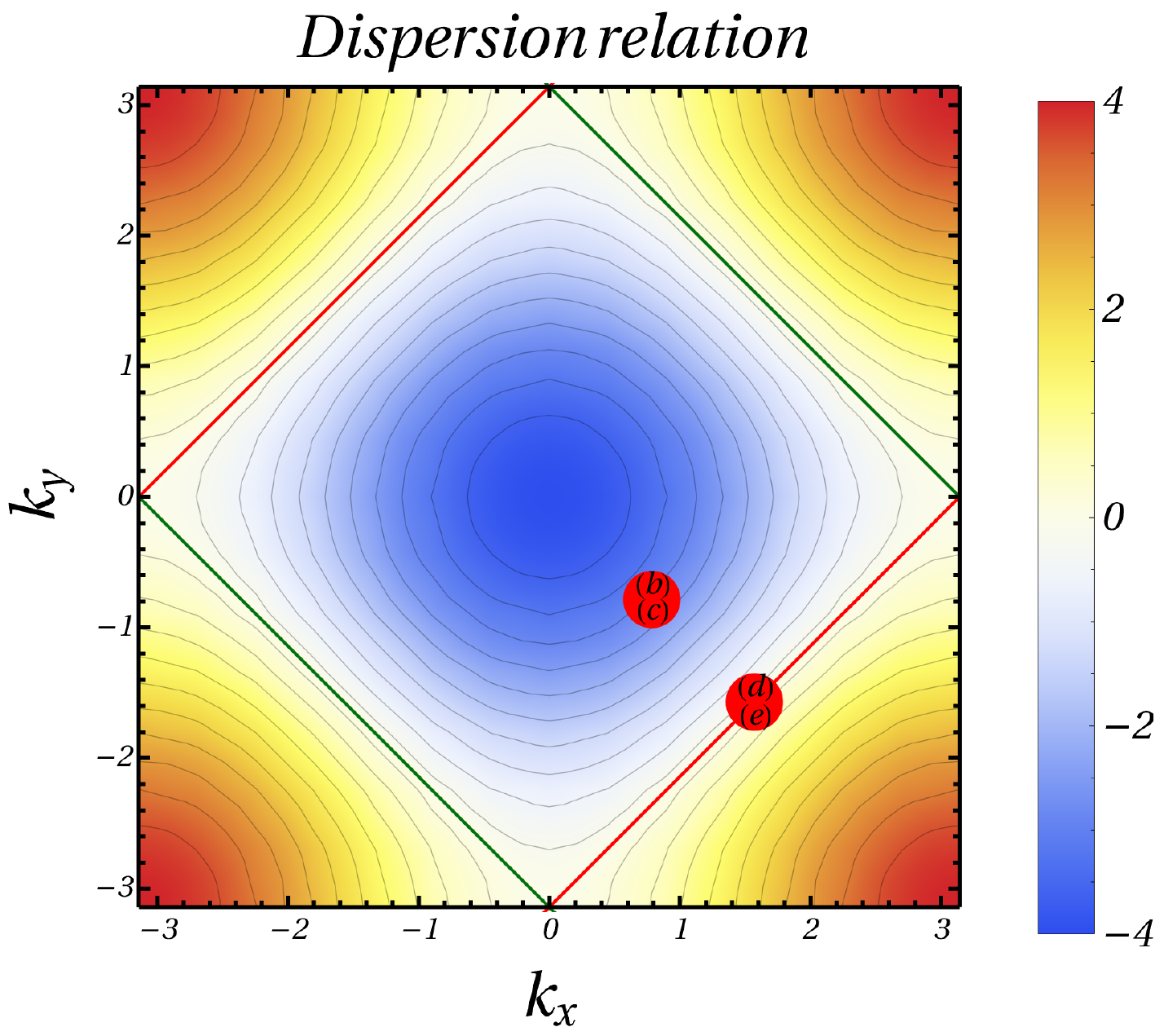}
  \caption{Dispersion relation at $\bar{U}=0$ and $W=0$. The red and green lines are the $E=0$ iso-energy lines. Linear combinations of the plane waves on green lines give unperturbed states when disorder is turned on, whereas those on red lines give rise to separatrix states. Red points $(b)-(c)$ and $(d)-(e)$ correspond respectively to initial wavepackets of Fig.~\ref{fig:wp}-b,c and Fig.~\ref{fig:wp}-d,e}
    \label{fig:E=0}
\end{figure}

The 0-energy subspace of the noninteracting problem, i.e., two particles on a chain of $N$ sites ($\bar{U}=W=0$), is spanned by $2N-2$ plane waves.  
The corresponding quasimomenta lie on the separatrix in the two-dimensional Brillouin zone ($k_x, k_y$). 
The separatrix is the iso-energy line at zero energy (separating particle-like and hole-like states in the band structure of the square lattice) with the shape of a rotated square, as shown in Fig.~\ref{fig:E=0}, with 
\begin{align}
&\text{red lines: } k_y=\pm\pi+ k_x  \Rightarrow k_-=\frac{k_x-k_y}{2}=\frac{\pm\pi}{2};
  \label{eq:redline}\\
&\text{green lines: } k_y=\pm\pi- k_x \Rightarrow k_+=k_x+k_y=\pm\pi.
  \label{eq:greenline}
\end{align} 

We note that using the alternative Brillouin zone defined in Appendix \ref{ap:comframe}, the equations for green lines become $k_+=-\pi$ with $-\pi\leq k_-<\pi$. 

Out of the plane-wave zero-energy modes, we can form $3N/2 -2$ linear combinations having wavefunctions that vanish for $x=y$, as we show below. These will be eigenstates of the system even if the on-site interaction between the particles is switched on. These come from two groups of 0-energy states, which we call diagonal and antidiagonal states. 

The $N$ diagonal states are linear combinations of the plane wave modes along the green lines in the Brillouin zone, chosen so that the distance between the two particles is fixed. The value of the distance $j=-N/2, -N/2+1, \ldots, N/2-1$ specifies the state,    
\begin{align}
    \ket{\psi_{d}(j)}  &= \frac{1}{\sqrt{N}} \sum_{x=1}^N 
    \ket{x, y=x+j \mod N}. 
\end{align}
Out of these $N$ mutually orthogonal states, $N-1$ have the property that their wavefunctions vanish for $x=y$. Only the state $\ket{\psi_d(0)}$ is affected by the on-site interaction.

The antidiagonal states are $N-2$ linear combinations of the plane wave modes along the red lines in the Brillouin zone. 
We only take plane quasimomenta with $k_x\neq 0$ and $k_y \neq 0$, since those plane wave modes were included in the construction of the diagonal states.

We can form $N/2-1$ linear combinations of the antidiagonal plane wave states that have 0 weight on $x=y$. These can be labeled by the quasimomentum component $k_x$, which is $0< k_x< \pi$, takes on $N/2-1$ different values. The states read,
\begin{align}
    \ket{\psi_{\text{ad}-}(k_x)} = \frac{1}{\sqrt{2N}} \sum_{x,y} \big( 
    e^{ik_x x} e^{i(k_x-\pi) y}  \nonumber \\ 
    -  e^{i(k_x-\pi) x} e^{i k_x y} \big)\ket{x, y}. 
\end{align}

The remaining $N/2-1$ linear combinations of antidiagonal plane wave states will be affected by the interaction. Their wavefunctions can be written as 
\begin{align}
    \ket{\psi_{\text{ad}+}(k_x)} = \frac{1}{\sqrt{2N}} 
    \sum_{x,y} \big( 
    e^{ik_x x} e^{i(k_x-\pi) y}  \nonumber \\ 
    +  e^{i(k_x-\pi) x} e^{i k_x y} \big)\ket{x, y}. 
\end{align}

\subsection{Perturbative effect of interaction on the separatrix states}

We now consider the effect of a weak interaction potential on the eigenstates, perturbatively up to first order in $U$. For simplicity, we restrict to the case $\bar{U}=0$ and finite $W$. We will refer to the $N/2$ states affected by the disorder as ``separatrix states".

\begin{figure}[!!h]
   \includegraphics[width=\linewidth]{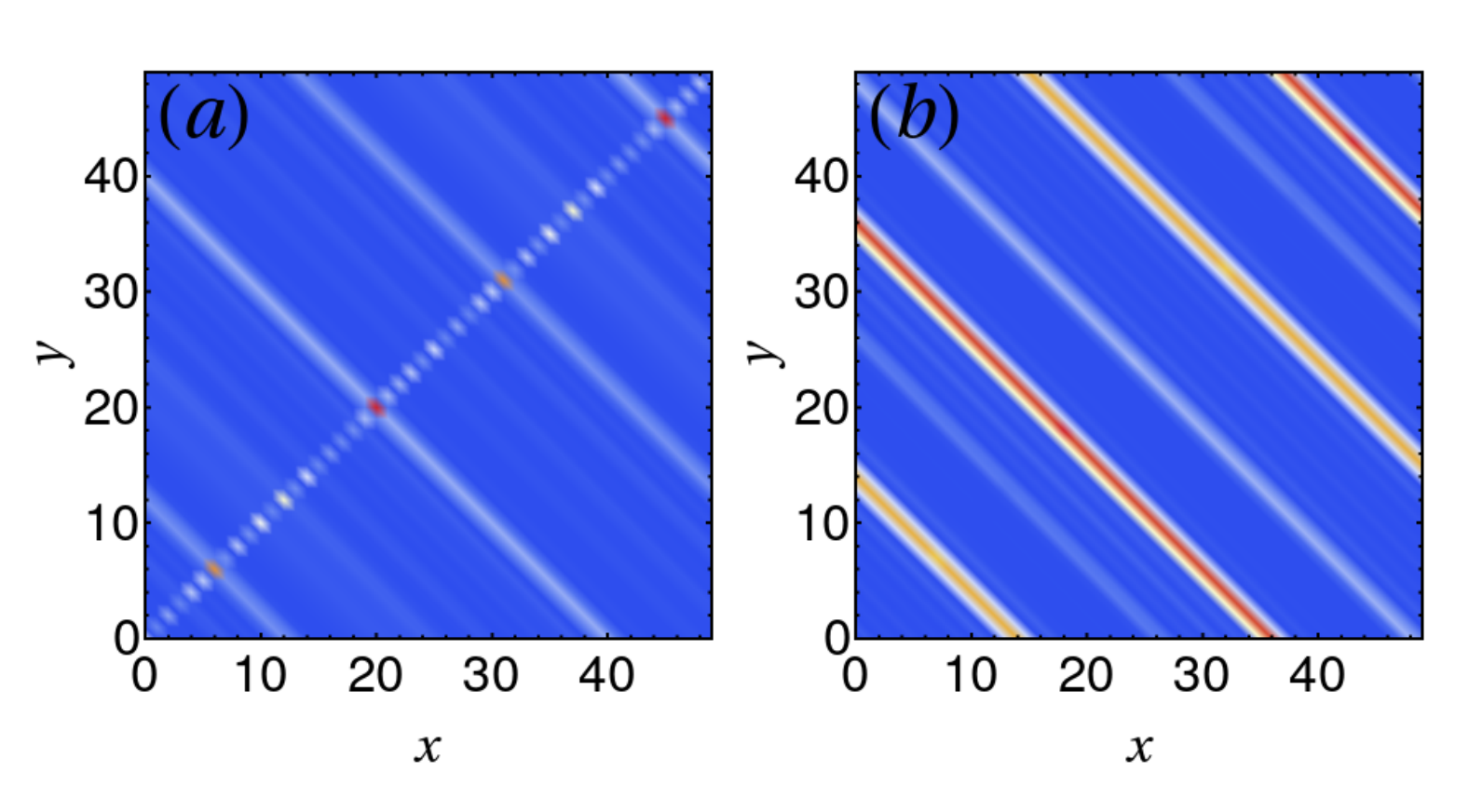}
  \caption{ Separatrix states. (a): An eigenstate of the perturbation $U$ having a non-negligible overlap with the state $\ket{\psi_d(0)}$. It represents a minority of eigenstates. (b): A typical eigenstate of the perturbation $U$ having a negligible overlap with the state $\ket{\psi_d(0)}$. 
  }
    \label{pertustate}
\end{figure}

In order to study separatrix states, we are therefore led to diagonalize a $N/2\cross N/2$ matrix.  We can numerically diagonalize this matrix to obtain eigenvectors that typically look like that shown in Fig.~\ref{pertustate}. The right panel  [see Fig.~\ref{pertustate}(b)] reproduces the main feature of the separatrix states, which is the localization along the center of mass $x_+$ direction together with delocalization along $x_-$. Further details on this disorder-induced localization are given in Appendix~\ref{ap:Upertu}. The perturbative analysis is valid for energy lower than the first non-zero energy (scaling as $1/N^2$). The eigenvalues of $U$ typically scale as $W/N$, thus the disorder should be very small ($W\ll 1/N$) for perturbation theory to hold.

\subsection{Wavepacket dynamics}
\begin{figure}[!!h]
   \includegraphics[width=\linewidth]{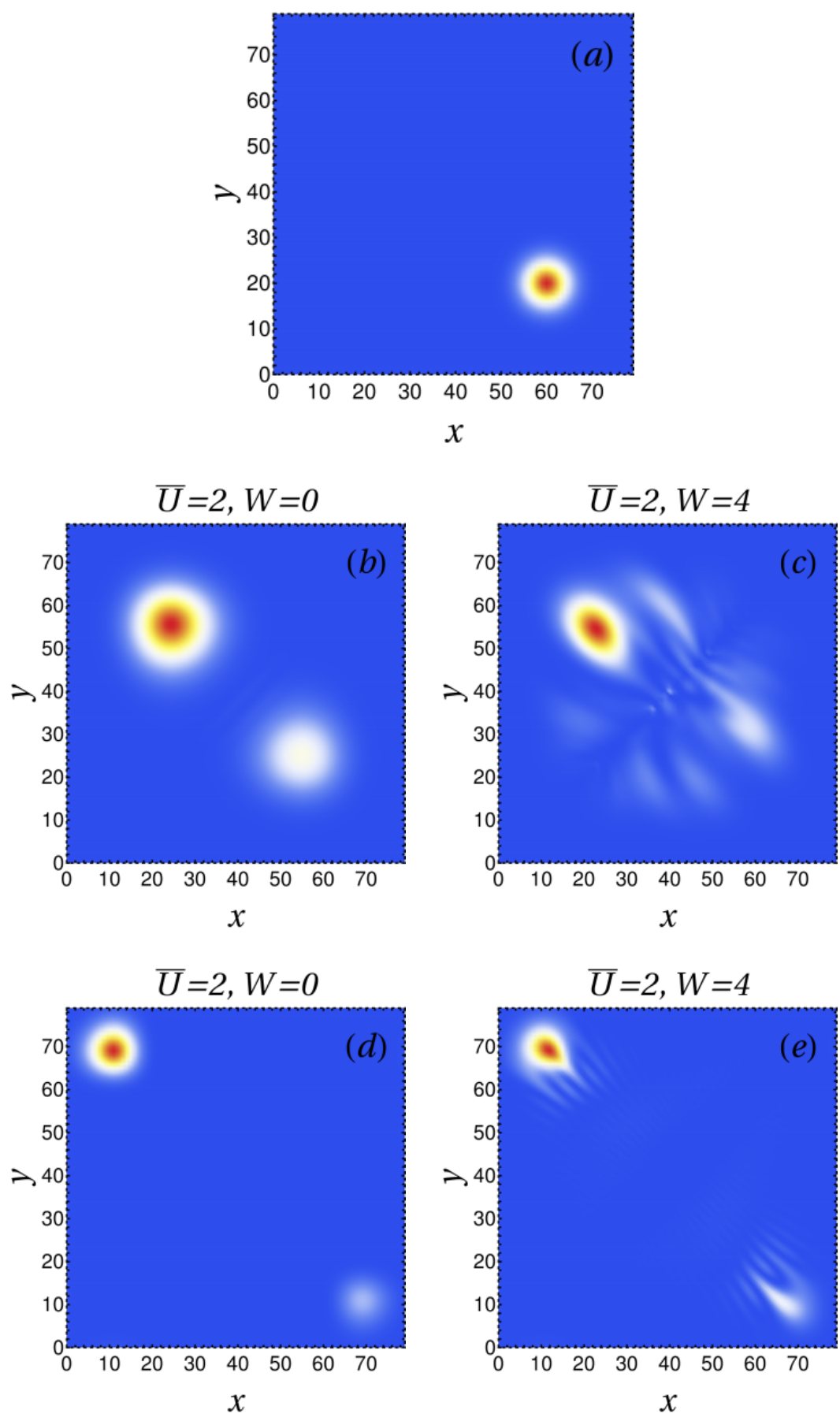}
  \caption{Position distributions of the two particles before (a) and after (time $t=25$) (b-e) a scattering of two Gaussian wavepackets on each other, for different initial momenta and with a translation invariant (b,d) or disordered (c,e) contact interaction $\bar{U}=2$, in a system of size $N=80$. (a) Initial wavepackets (time $t=0$) both have a width of 5 sites, prepared at $x_0=20$, $y_0=60$ with wavevector $k_0^+=0$ (i.e. vanishing group velocity along $x_+$) and either total energy $E\neq 0$, momentum $k_0^-=\pi/4$ [case (b) and (c) in Fig.~\ref{fig:E=0}] or $E=0$, $k_0^-=\pi/2$ [case (d) and (e) in Fig.~\ref{fig:E=0}]. (b, d) Without disorder, the post-collision wavefunction has two wavepackets, somewhat broadened, but centered on the same $x_+$ coordinates, due to the conservation of center-of-mass momentum - irrespective of the initial momenta. (c) When disorder in the contact interaction breaks translation invariance, and the total energy $E\neq 0$, we see a broader distribution of post-collision $x_+$ and $x_-$ coordinates. (e) When the contact interaction is disordered, and $E=0$, we see a broadening of the post-collision $x_-$, but only slight broadening of the $x_+$ distribution. This is due to the dominance of separatrix states in the initial wavepacket, as explained in the main text.}
    \label{fig:wp}
\end{figure}

The localization of separatrix states along the center of mass $x_+$ direction leads to an interesting effect for the scattering of the two particles, that we discuss below. 
Because of the disorder in the interaction potential, the center-of-mass momentum is not conserved. However, as we show below, if the two  particles have wavepackets with equal and opposite energies,  $+E$ and $-E$, then during the scattering the center-of-mass momentum does stay approximately conserved (in fact, the center of mass is stationary). We note that this is the same problem as that of a single particle moving on a two-dimensional lattice with a disordered diagonal potential barrier.

To set up a generic scattering problem, we prepare both particles in  Gaussian wavepackets, far from each other, centered around positions $x$ and $y$ with $x<y$. The velocities of the particles, $v_{x,y} = \partial E(k_{x,y})/\partial k_{x,y}$ should be such that $v_x>v_y$ so that a scattering event does occur. An example for initial-state quasimomenta $k_{x0}, k_{y0}$ representing this condition is indicated by the symbols ``(b)'' and ``(c)'' in Fig.~\ref{fig:E=0}. The corresponding distribution of positions is shown in Fig.~\ref{fig:wp} (a). In case of a translation-invariant contact interaction, during the collision both the energy $E$ and the center-of-mass momentum $k_+$ are conserved. In that case, after the collision, transmitted and reflected parts of the wavefunction will be wavepackets with $k_x\approx k_{x0} , k_y\approx k_{y0}$ and  $k_x\approx k_{y0} , k_y\approx k_{x0}$, respectively. The corresponding position distribution, obtained numerically, is shown in Fig.~\ref{fig:wp} (b).  If the interaction is disordered, the center-of-mass momentum is no longer conserved, only the total energy is. Thus after the collision we expect to see a broad distribution of quasimomentum values $k_x, k_y$ both for the reflected and the transmitted parts. Thus the distribution of post-collision velocities is broader, and as a result -- as shown for a concrete example in Fig.~\ref{fig:wp} (c) -- the post-collision position distributions are broader. 

A special case of the scattering problem is if the two incident wavepackets have opposite energies, so that the total energy is approximately zero. Thus $k_{x0}, k_{y0}$ lie somewhere on the red separatrix line of Fig.~\ref{fig:E=0}, an example indicated in the Figure by the symbols ``(d)" and ``(e)''. Then the initial state has a significant overlap with antidiagonal separatrix states, which, as explained in the previous section, are extended along $x_-$ but localized along $x_+$. Moreover, it has practically no overlap with diagonal states (eigenstates formed by linear combinations of plane waves from the green parts of the separatrix).
Thus the post-collision state should be composed of mostly plane wave-modes at or near the red parts of the separatrix, with velocities $v_x\approx -v_y$. This explains why the post-collision position distribution in this case can be broad along $x_-$, but should be not significantly broadened along $x_+$: the center-of-mass is approximately conserved. This is confirmed by a numerical example in Fig.~\ref{fig:wp}(e). For comparison, the post-collision position distribution with the same parameters, but without disorder in the interaction, is shown in Fig.~\ref{fig:wp} (d).
We note that disorder in the interaction also leads to a comb-like interference pattern of the position distribution of both the reflected and transmitted parts, which would merit further investigation. For the same time evolution and because of the non-parabolic dispersion relation, the natural wavepacket spreading is much smaller in Fig.~\ref{fig:wp} (d) than in Fig.~\ref{fig:wp} (b).

\subsection{Summary}
Separatrix states have a small participation ratio ($P_2 \sim N/N^2=1/N\ll 1$ when $N\gg1$) and as such could be mistaken for scarred states, well-known in the quantum chaos context~\cite{Heller1984,Kaplan1999}. However, they are markedly different. Indeed, they are due to separatrix iso-energy lines, that do not exist in a continuum billiard. 
In addition, they are not related to unstable and periodic classical orbits. 
Our understanding is that they are related to stable classical orbits in a peculiar billiard with a particular kind of kinetic energy $H(k_x,k_y)=-2\cos(k_x)-2\cos(k_y)$ instead of $H(k_x,k_y)=(k_x^2+k_y^2)/(2m)$. We leave it to future work to study these unusual classical billiards. We have not found scarred states in the present model.

\section{Conclusion and discussion \label{sec_conclusion}}
In the absence of disorder, two interacting particles on a chain can have a coherent dynamic as a bound state or independent motion as scattering states. When the interaction becomes spatially disordered, two very different effects are expected for the molecular bound state.

\par On the one hand, if the energy of the initial molecular state does not overlap with the atomic band, the molecule becomes Anderson-localized due to disorder. 

\par On the other hand, because the disorder in the interaction breaks the conservation of the center-of-mass quasi-momentum and spreads the molecular band in energy, an initially bound state with an energy that overlaps with the atomic band becomes a resonance with a finite lifetime and delocalizes over the whole system. The disorder breaks both the molecular bound-state and the Anderson localization.

\par Likewise, a few scattering states in the atomic band persist even when disorder is turned on. These are called quasi-ideal. Both resonant states and quasi-ideal states are not very different from bound states and scattering states that exist in the absence of disorder. There existence is due to the fact that the disorder is only a ``surface effect'' in our model. They are expected to become negligible in the thermodynamic limit as their number increases with $N$ but not as fast as $N^2$.

\par Near the center of the atomic band, we observe unusual states due to the disordered interaction and related to the presence of a separatrix zero-energy line in the square lattice dispersion relation. These separatrix states do not exist in standard quantum billiards (defined in the continuum rather than in a tight-binding model). They feature disorder-induced localization in real space but not of the Anderson type (not an interference effect). A remarkable consequence is that a wavepacket with zero average energy, i.e. built on these separatrix states, can not be laterally scattered when hitting a disordered barrier. The number of separatrix states ($N/2$) makes them negligible in the thermodynamic limit. 

\par Apart from resonant, quasi-ideal and separatrix states, which are all finite-size effects, most states in the atomic band are chaotic states. They are the typical states of a peculiar toric billiard possessing a disordered barrier along a closed loop that winds around the torus. These states are delocalized (with a participation ratio of 1/3), have a random character and feature filaments.

\par We also changed the disorder distribution (either Gaussian or binary, i.e., Bernoulli) and checked that the main effect is still valid: when a molecular state resonates with the atomic band it delocalizes.
In the Gaussian case, the main difference comes from the tails of the distribution, that no longer allows one to clearly separate a regime in which the molecular and atomic bands either overlap or do not overlap. In the binary (Bernoulli) case, we draw with a probability $p$ an on-site interaction strength $U_1$ and with probability $1-p$ an on-site interaction strength $U_2>U_1$. When $p\neq0,1$, the molecular energies are between  $U_1$ and $\sqrt{U_2^2+16}$ and this does not depend on the disorder parameter $p$. In summary, the main results are robust to changing the disorder model, but in the details, there are some differences.
\par We now discuss possible experimental realizations of the random $U$ Hubbard model. The type of disordered interaction we have considered could be realized with cold atoms trapped in an optical lattice. There the interaction between the trapped atoms can be magnetically tuned using Feshbach resonances~\cite{chin2010feshbach}. In a variant of this technique, optical Feshbach resonances~\cite{enomoto2008optical}, the resonance condition between the states is fulfilled with the help of an extra laser field (or pair of laser fields). Here the interaction strength can be made position-dependent if the spatial form of the lasers is modulated by optical speckle patterns.

\par An alternative experimental route would use the analogy to quantum billiards, i.e., realize the system as a single particle moving in two dimensions with a line of potential defects. This could be realized with photonic waveguides fabricated using femtosecond laser inscription, as in a recent experiment by Mukherjee et al\cite{Mukherjee2016}. There the 2-body 1D Hubbard model was mapped to a square tight-binding model in the presence of a barrier along the diagonal. Similarly, Di Liberto et al.~\cite{DiLiberto2016} have suggested this approach to simulate the effects of interaction in the Su-Schrieffer-Heeger model. In such an experimental setup, disorder could simply be included by varying the parameters of the waveguides also along the diagonal. This seems to be possible using the level of control over the parameters of the waveguides already demonstrated in the experiment~\cite{Mukherjee2016}. 

\acknowledgements
J.A. would like to thank LPTMC at Sorbonne university and CNRS for the hospitality in Paris where this project was started.
This work was supported by the National Research, Development  and Innovation Office of Hungary (NK-FIH) within the Quantum Technology National Excellence Program (Project No. 2017-1.2.1-NKP-2017-00001), projects FK124723 and K124351, and the Quantum Information National Laboratory of Hungary.

\appendix
\section{Center-of-mass frame \label{ap:comframe}}

In the main text, we use a mapping from the two-particle problem on a chain onto the dynamics of a single particle on a square lattice. 
In this appendix, we describe the  basis change from the canonical frame $(\vec{e_x},\vec{e_y})$ to the center of mass frame $(\vec{e_+},\vec{e_-})$, which reads
$$
\left\{
\begin{array}{ll}
  \vec{e_+}=\vec{e_x}+\vec{e_y},    &  ||\vec{e_+}||=\sqrt{2}; \\
  \vec{e_-}=\frac{\vec{e_x}-\vec{e_y}}{2},    & ||\vec{e_-}||=1/\sqrt{2}.
\end{array}
\right.
$$

In this orthogonal, but not normed, center-of-mass frame, the site coordinates $(x,y) \in \mathbb{Z}^2$ are replaced by the center of mass  $x_+=\frac{x+y}{2}$ and the relative coordinate $x_-=x-y$.

The vectors  $(\vec{e_+},\vec{e_-})$ generate a rectangular lattice which shares only half of the square lattice sites. This leads to integer and half integer coordinates for the square lattice sites: $x_+$ takes integer values when $x_-$ is even and half-integer values when $x_-$ is odd.

\begin{figure}
    \centering
\begin{tabular}{cc}
    \includegraphics[width=0.5\linewidth]{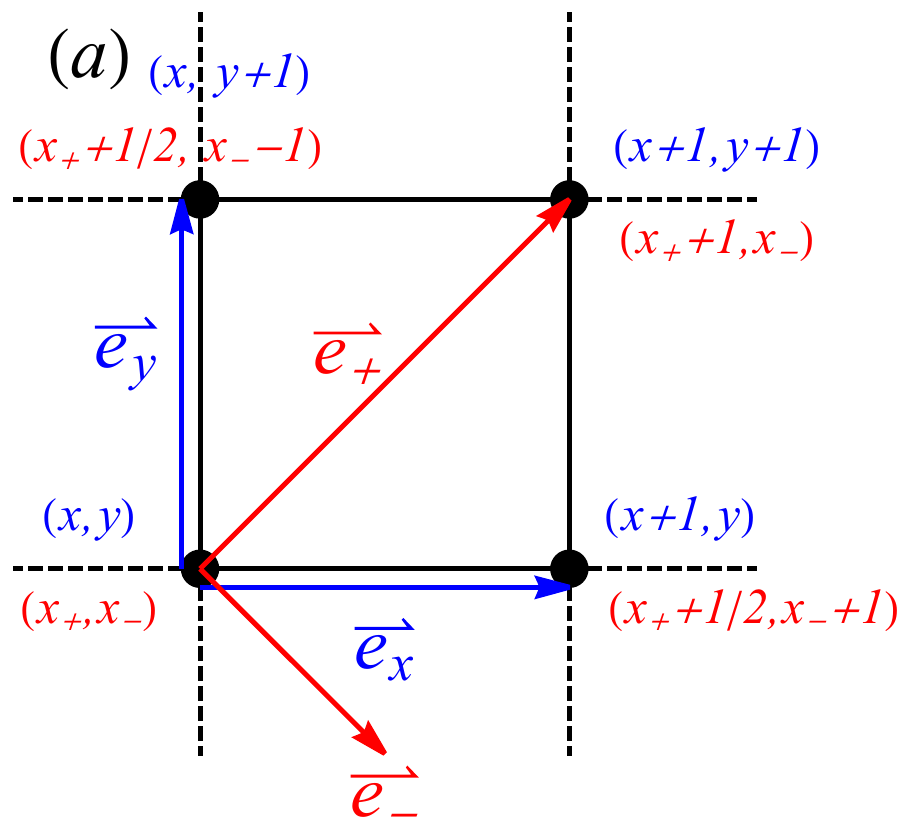}&
    \includegraphics[width=0.5\linewidth]{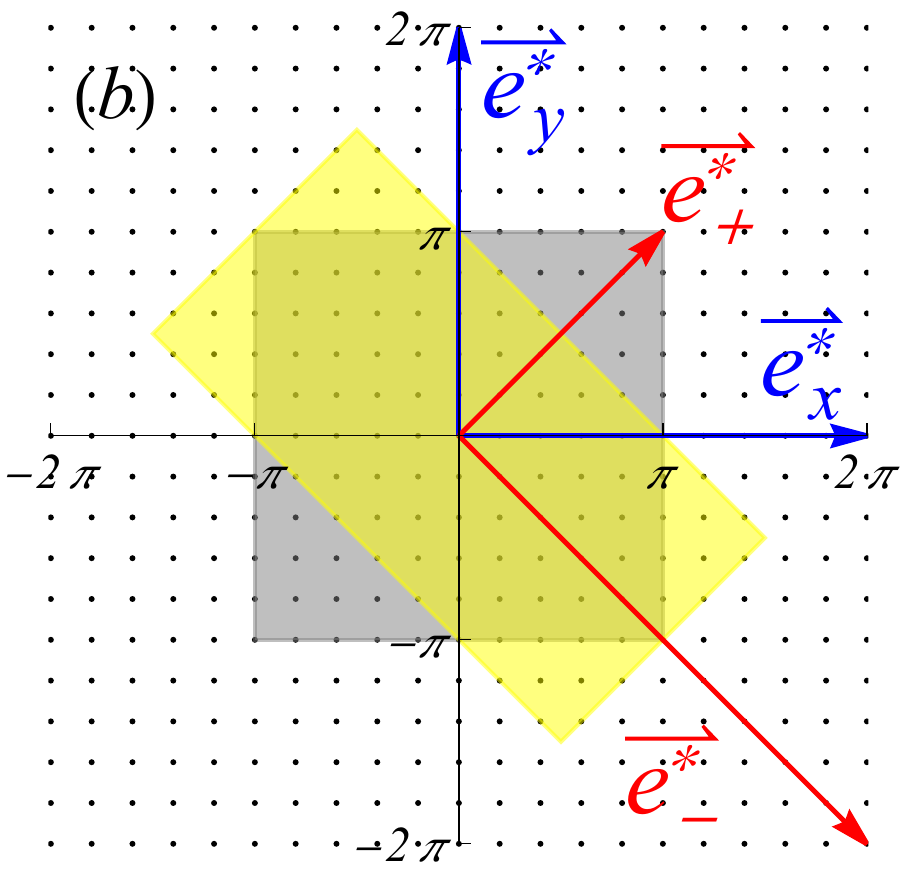}\\
    \end{tabular}
       \caption{(a) Portion of the square lattice with the coordinate and basis vectors for the natural frame $(x,y)$ in blue and the center of mass frame $(x_+,x_-)$ in red. (b) First Brillouin zone of the reciprocal lattice in gray. The dual vectors of each frame are indicated, in blue (red) for the natural (center of mass) frame. The dots represent the allowed values taken by the wavevector for a finite size system with $N=10$. An alternative Brillouin zone better adapted to $k_+$ and $k_-$ is indicated in yellow.}
    \label{comframe}
\end{figure}

Turning to reciprocal space, we have the canonical reciprocal frame  $\vec{e_x^*}=2\pi\vec{e_x}$ and $\vec{e_y^*}=2\pi\vec{e_y}$, wavevectors  $\vec{k}=k^*_x\vec{e^*_x}+k^*_y\vec{e^*_y}=k_x\vec{e_x}+k_y\vec{e_y}$, and the associated first Brillouin zone defined by $(k^*_x,k^*_y)$ both in the range $[-1/2,1/2[$ or equivalently,  $(k_x,k_y)=2\pi (k_x^*,k_y^*)$ both in the range $[-\pi,\pi[$.  The reciprocal frame associated with the center of mass coordinates reads : 

$$
\left\{
\begin{array}{ll}
  \vec{e_+^*}=\pi\vec{e_+}=\frac{\vec{e^*_x}+\vec{e^*_y}}{2},    &  ||\vec{e_+^*}||=\pi\sqrt{2}\\
  \vec{e_-^*}=4\pi\vec{e_-}=\vec{e^*_x}+\vec{e^*_y},    & ||\vec{e_-^*}||=2\sqrt{2}\pi
\end{array}
\right.
$$

Working with a finite $N \times N$ patch with periodic boundary conditions amounts to select a finite set of allowed $\vec{k}$ vectors, whose coordinates $(k_x,k_y)$ inside the first Brillouin zone [gray region in Fig.~\ref{comframe}(b)] read $2\pi j_{x,y} /N$ with $(j_x,j_y)$ in the range $[-N/2,N/2-1]$ for $N$ even and  $[-(N-1)/2,(N-1)/2]$ for $N$ odd.

In the center of mass reciprocal frame, these allowed $\vec{k}$ vectors read
$\vec{k}=k^*_+\vec{e^*_+}+k^*_-\vec{e^*_-}=k_+\vec{e_+}/2+2k_-\vec{e_-}$ with
$$
\begin{array}{lr}

\left\{
\begin{array}{ll}
 k_+=2\pi(j_x+j_y)/N\\
k_-=\pi(j_x-j_y)/N
\end{array}\right.
\end{array}
$$
with $(j_x,j_y)$ running in the same range as above.

The two direct space frames $(\vec{e_x},\vec{e_y})$ and $(\vec{e_+},\vec{e_-})$ are shown in Fig.~\ref{comframe}(a) , and the two reciprocal frames $\{\vec{e^*_x},\vec{e^*_y}\}$ and $\{\vec{e^*_+},\vec{e^*_-}\}$ together with allowed  $\vec{k}$ vectors (with $N=10$) in the first Brillouin zone are displayed in Fig.~\ref{comframe}(b). 

\par The extremal values taken by one coordinate of the reciprocal center of mass frame when ranging over the first Brillouin zone [gray region in Fig.~\ref{comframe}(b)] depends on the other coordinate. In practice, this makes computation harder. However, one can define equivalently an alternative Brillouin zone [yellow region in Fig.~\ref{comframe}(b)], in which both coordinates range over $[-\pi,\pi[$. The allowed $(k_+,k_-)$ coordinates now read:
$$\left\{
\begin{array}{l}
 k_+=2\pi K/N\\
k_-=\pi q/N
\end{array}\right. ,
$$
where $K$ takes integer values in $ [-N/2,N/2-1]$ and $q\in [-N,N-1]$ takes even (resp. odd) values when $K$ is even (resp. odd).\\

\section{Scaling of the inverse participation ratio}
\label{ap:iprsize}
\par The localization of eigenfunctions can be measured using the inverse participation ratio (IPR) defined in Eq.~\eqref{eq:ipr}. The scaling of the eigenfunctions' IPR with respect to the system size $N$ shows different behaviors according to the degree and nature of localization. As  we  increase  the  system  size,  wavefunctions completely delocalized over the whole system should have an IPR $\sim 1/N^2$;  those  localized  along  only one  direction are expected to have IPR $\sim 1/N$; and completely localized wavefunctions should have an IPR $\sim N^0$ for large $N$. Table~\ref{tab:iprscale} associates the different type of states defined throughout the article with their corresponding IPR scaling.

\par
\begin{figure}
\begin{tabular}{c c}
       \includegraphics[width=0.5\linewidth]{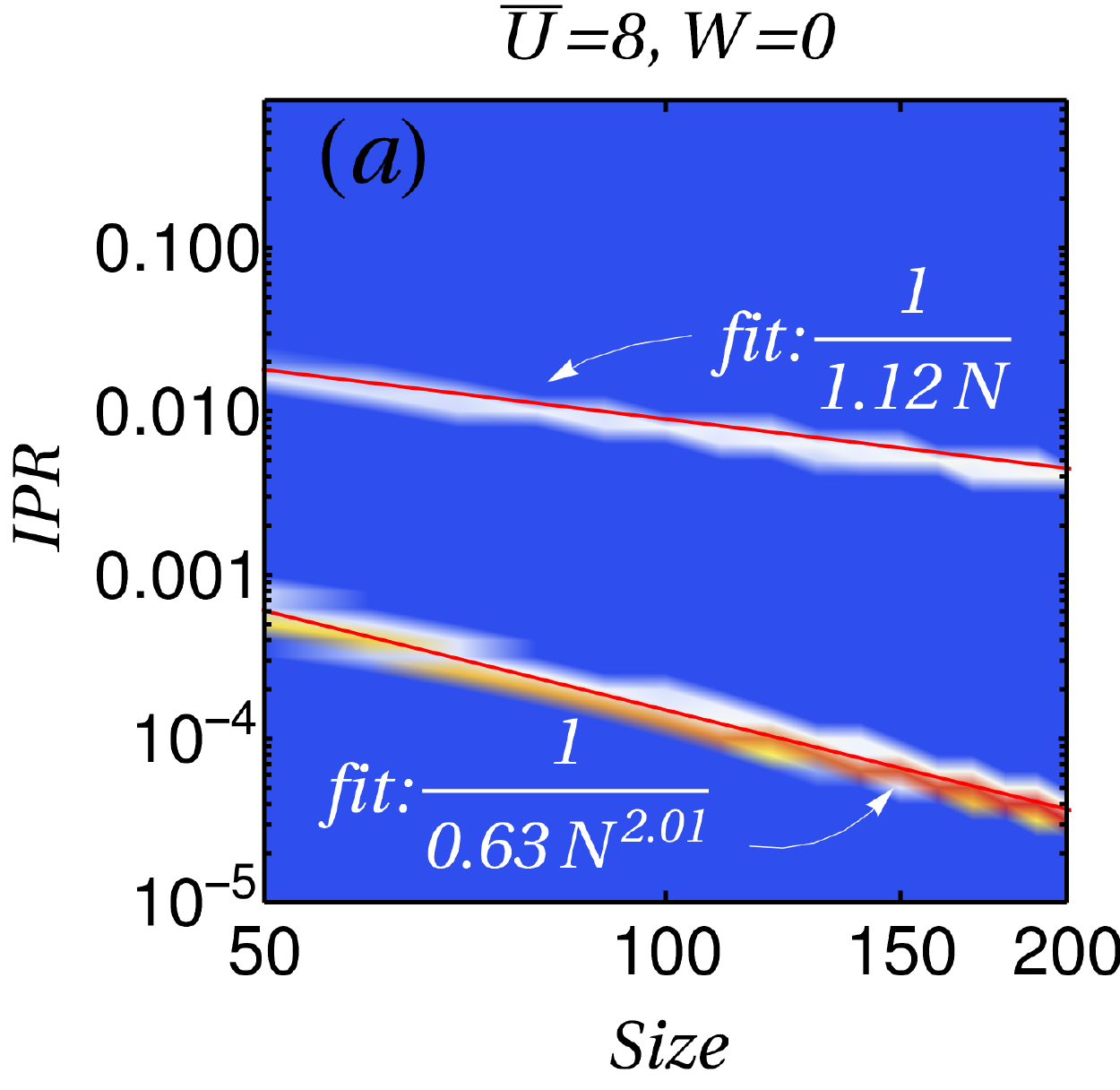}  &     \includegraphics[width=0.5\linewidth]{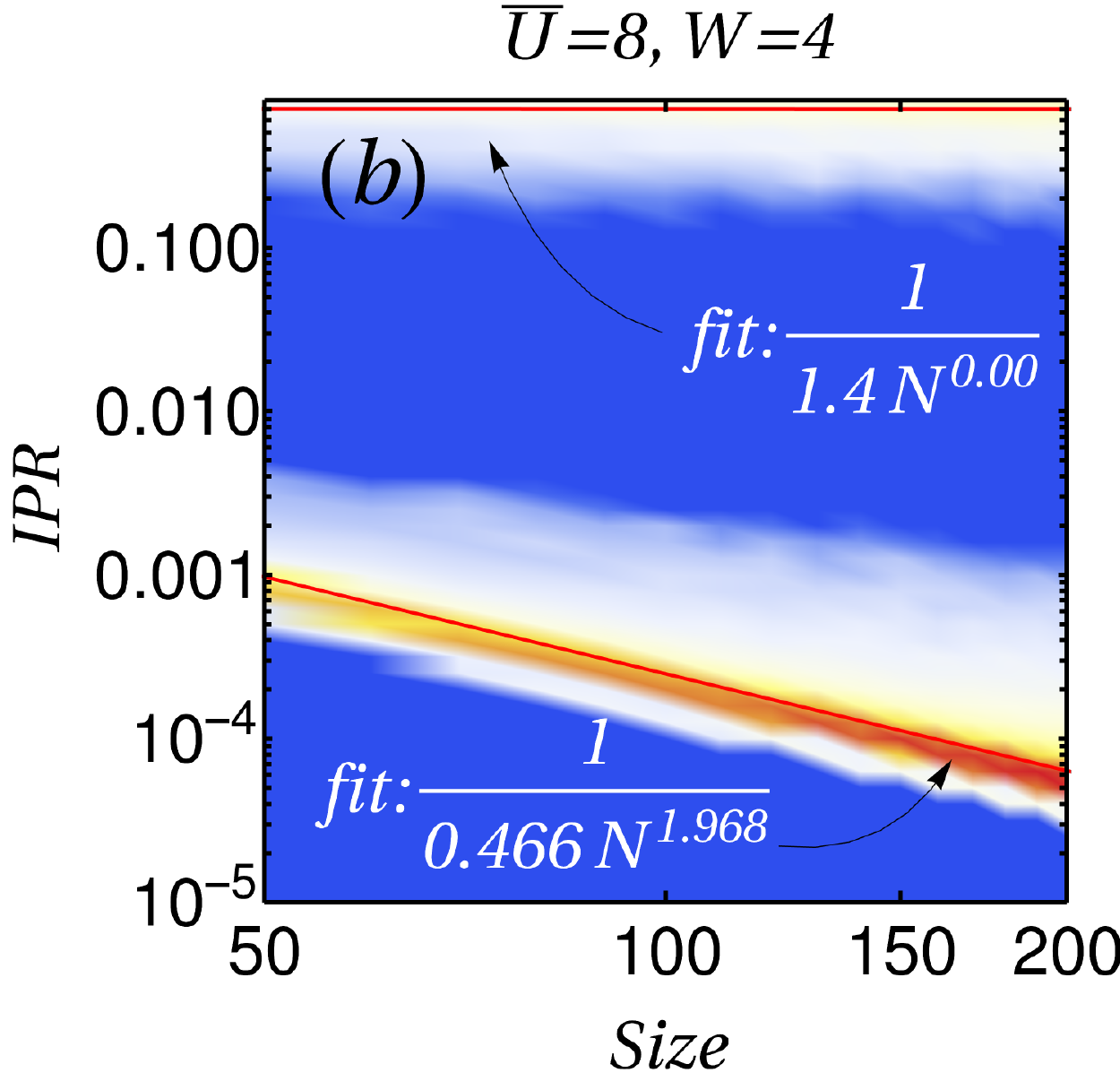} \\
     \includegraphics[width=0.5\linewidth]{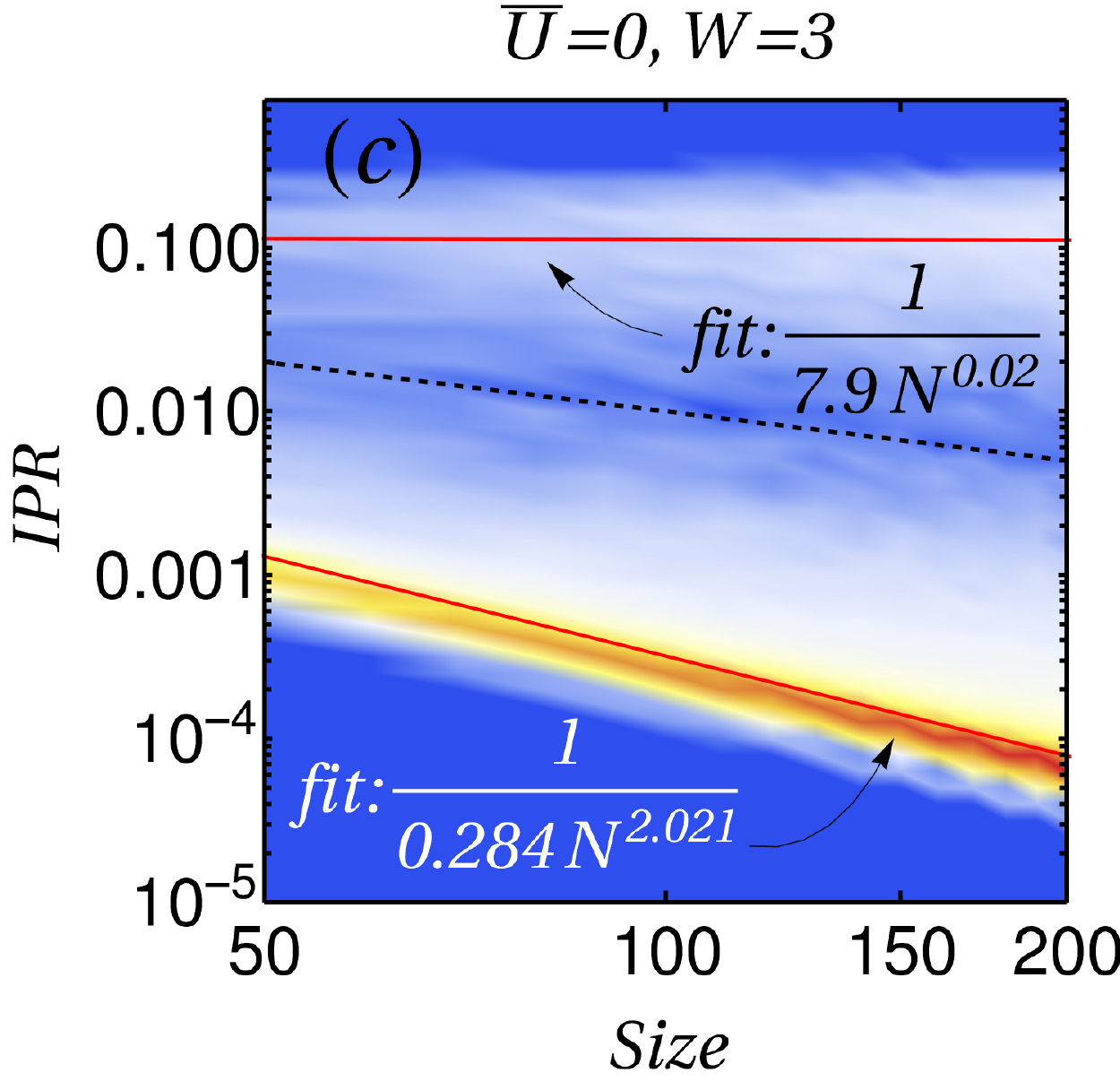}    &     \includegraphics[width=0.5\linewidth]{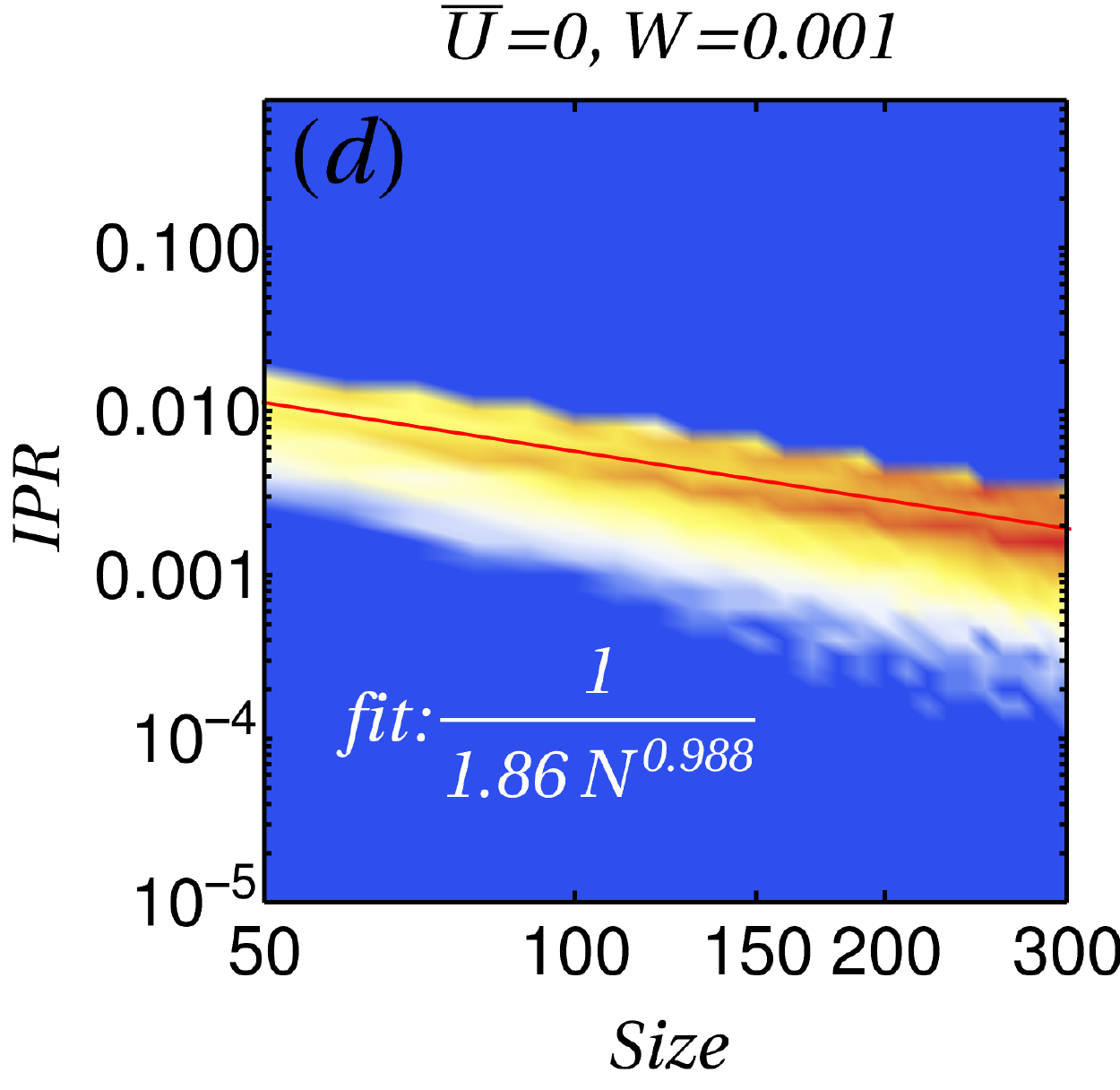}
\end{tabular}

    \caption{Log-log plot of the inverse participation (IPR) of eigenfunctions with respect to the size of the system $N$ for different average interaction strength $\bar{U}$ and disorder $W$. (a), (b) and (c): Complete energy spectrum for system sizes from $N=50$ to $N=200$ by increment of $10$. (a) $\bar{U}=8$, $W=0$; (b) $\bar{U}=8$, $W=4$; and (c) $\bar{U}=0$, $W=3$. (d): Only separatrix states at $\bar{U}=0$ and $W=0.001$ between $N=50$ and $N=300$ by increment of $10$. Linear fits in red highlight three types of behavior: $I_2 = 1/(P_2 N^2)$ (2D-delocalized), $\sim 1/N$ (1D-delocalized) and $\sim N^0$ (localized). Black dotted line in (c) delimits the region between localized and 2D-delocalized states.}
    \label{fig:iprsize}
\end{figure}

 Without disorder, eigenfunctions of the atomic band are the scattering wavefunctions which scale as $1/N^2$, the fit shows a participation ratio $P_2\simeq63\%$ [see Fig.~\ref{fig:iprsize}(a)]( we employ twisted boundary conditions in order to avoid degeneracy). This numerical result matches the analytical computation of the participation ratio, using Eqs.~\ref{eq:bethe} for $\bar{U}\to\infty$, which gives $2/3\simeq 67 \%$.  The second fit in Fig.~\ref{fig:iprsize}(a) shows a dependence in $1/N$ which corresponds to the molecular states delocalized along the center of mass direction and localized along the relative motion direction.

\par With disorder, molecular states become Anderson localized, so that their IPR does not depend on $N$ anymore [see Fig.~\ref{fig:iprsize}(b)]. The majority of atomic states remain delocalized and their IPR scales as $1/N^2$. In this regime (both bands do not overlap), the latter are either quasi-ideal states ($P_2 \sim67\%$) either chaotic states ($P_2 \sim33\%$). One notices that the distribution of IPR [see Fig.~\ref{fig:iprsize}(b)] is larger than in the free-disorder case.  The average value of the participation ratio is $P_2\sim 46\%$. The participation ratio of chaotic states ($1/F\simeq 33\%$) can be obtained analytically from random matrix theory (with $F=3$ for the GOE), see  e.g.~\cite{Kaplan1999}. However, we observe that the distribution of atomic states' IPR widens as $N$ increases which is due to a minority of states: the separatrix states. At finite $\bar{U}$ and $W$, those states do not show a well defined IPR scaling but rather form a continuous transition between the 1D localized regime (IPR$\sim1/N$) and the 2D completely delocalized regime (IPR$\sim1/N^2$).

\par When both bands overlap ($\bar{U}-W<4$), the disorder is strongly felt by atomic states (see Sec.~\ref{sec_overlap}). Figure~\ref{fig:iprsize}(c) shows an IPR for atomic states scaling as $N^{-2}$, i.e., a participation ratio $P_2 \sim 28\%$ not far from $33\%$ expected for chaotic states. The difference probably comes from the fact that not all atomic states are chaotic states: there are also quasi-ideal, resonant and separatrix states. Their effect should disappear in the thermodynamic limit. Similarly to the separatrix states, the resonant states do not have a proper scaling and smoothly connect the completely localized with the 2D completely delocalized IPR's distribution.

\par However, it is possible for separatrix states to exhibit a well defined IPR scaling when they do not couple with other atomic states. They should present a $1/N$ scaling for the IPR because they are localized in the center of mass direction and delocalized in the relative motion direction. We know that they are related to states at energy $E=0$ when $\bar{U}=W=0$. To keep track of them, we perturb slightly the system ($W=0.001$ and $\bar{U}=0$) to stay in the regime where separatrix and atomic states do not resonate [see Fig.~\ref{fig:iprsize}(d)] and indeed find the expected $1/N$ behaviour.

\begin{table}
\centerline{
\begin{tabularx}{\linewidth}{|X|X|}
\hline
 \textbf{Type of states}    &  \textbf{IPR scaling with system size $\bm{N}$}\\
 \hline
 \hline
  scattering ($W=0$) or quasi-ideal ($W\neq 0$)  & $\frac{1}{P_2 N^2}$ with $P_2\geq67\% $\\
   \hline
  chaotic  &$ \frac{1}{P_2 N^2}$ with $P_2\simeq33\% $\\
   \hline
  separatrix & $\sim 1/N$ when $W\ll1$, $\bar{U}=0$\\
  &no proper scaling for finite $W$ or $\bar{U}$\\
   \hline
  resonant  &no proper scaling\\
   \hline
  delocalized bound& $\sim 1/N$ when $W=0$  \\
  \hline
  Anderson localized molecular& $\sim 1/N^0$\\
  \hline
\end{tabularx}}
\caption{Scaling of IPR with respect to system size $N$ for the different type of states and sorted in increasing order (from most delocalized to most localized).}
\label{tab:iprscale}
\end{table}

\section{First-order perturbation of the Bethe ansatz solutions}
\label{ap:energypertu}

In this appendix, we use perturbation theory in the disorder strength $W$ to study the effect of the interaction potential $U$ (with average $\bar{U}$ and fluctuations $W$) on the exact eigenstates of the two-particle problem known at nonzero $\bar{U}$ but $W=0$. For convenience, we describe the system as the dynamics of a single particle on a two-dimensional grid (with periodic boundary conditions), with a disordered onsite potential on the diagonal.

\par The cases $\bar{U}=0$ and $\bar{U}\neq0$ are expected to lead to different quantitative results. Indeed, the disordered diagonal perturbs eigenfunctions of $\bar{U}=0$ (plane waves) much more intensively than eigenfunctions of $\bar{U}\neq 0$ (scattering states) because their weight on the diagonal is greater.

\subsection{$\bar{U}=0$}
We start by considering $\bar{U}=0$ and follow the steps of~\cite{Cuevas1996}. The first order perturbation in energy is given by the matrix element of the perturbation operator $U$ between unperturbed eigenvectors, i.e. plane waves obtained at $W=0$. In our case, the potential is on a diagonal, whereas in~\cite{Cuevas1996}, it is along the four edges of a square billiard. The weight of a plane wave on a site is $1/N^2$. The typical deviation of the energy is the standard deviation $\sigma$ of the  variable $\mathcal{U}=\sum_{x=1}^N U_x$ where $U_x$'s are the random diagonal potential uniformly chosen in $[-W,+W]$. We have $\sigma(\mathcal{U})=W\sqrt{N}/\sqrt{3}$ where $N$ is the length of the disordered barrier. We therefore obtain a typical perturbation in energy scaling as $W.N^{-3/2}$, which has to be compared to the mean level spacing $\sim N^{-2}$. In the thermodynamic limit $N\to\infty$, as $W.N^{-3/2}\gg N^{-2}$, every state will be sensitive to the neighbouring energy levels and we will eventually lead to GOE statistics. In addition, taking into account that the DoS is not flat but has a maximum at $E=0$ and minima at band edges $E=\pm4$ (see Fig.~\ref{fig:dos}), the last states to be affected by the disorder are those near band edges. Indeed, quasi-ideal states (i.e. states almost unaffected by the disorder) are found mostly near band edges.

\subsection{$\bar{U}\neq 0$}
Next we consider $ \bar{U}\neq0$. When $W=0$ and with periodic boundary conditions, scattering states at finite $\bar{U}$ are given by the Bethe ansatz~\cite{Lieb1968,Caffarel1998} and read:
\begin{eqnarray}
\psi_\text{sc}(x,y)&=&Ce^{ik_+x_+}[\sin (\kappa_-x_-)\nn \\ 
&-&\frac{4}{ \bar{U}}\cos \frac{k_+}{2}\sin \kappa_-\cos(\kappa_-x_-)],
\end{eqnarray}
where $C=\sqrt{2}/(N\sqrt{1+16/ \bar{U}^2\cos^2 \frac{k_+}{2}\sin^2 \kappa_-})$ is the normalization constant, $k_+=k_x+k_y=2\pi K/N$ and $K \in \llbracket -N/2,N/2-1\rrbracket$. 
The quantity $\kappa_-$ is the solution of the Bethe ansatz equation: $\kappa_-N=2\pi\lambda+\theta$ where $\lambda \in \llbracket -N/2,N/2-1\rrbracket$ and $\theta$ is such that 
\begin{align}
e^{i\theta} &=-\frac{1+4i/ \bar{U}\cos(k_+/2)\sin\kappa_-}{1-4i/ \bar{U}\cos(k_+/2)\sin\kappa_-}.
\end{align}
At zeroth order in $\bar{U}$, one has $\kappa_-\simeq 2\pi \lambda/N =k_-$ and the normalization constant $C\simeq \sqrt{2}/N$. 

The weight of the eigenvectors on a disordered site $(x_-=0)$ is $C^2 16 \bar{U}^{-2}\cos^2\frac{k_+}{2}\sin^2\kappa_-$. By a similar computation as the one done in the case $\bar{U}=0$, the first order perturbation scales as $W \bar{U}^{-2}\cos^2\frac{k_+}{2}\sin^2\kappa_-N^{-3/2}$. The number of Bethe solutions for which the first order perturbation is lower than the mean level spacing $\propto1/N^2$ depends on the value of $k_+$. The latter satisfy the condition:
\begin{align}
\cos^2\frac{k_+}{2}\sin^2\kappa_-\ll \bar{U}^2W^{-1} N^{-1/2} .
\end{align}

\par We distinguish two regimes: (1) when $k_+$ is far from $\pi$, (2) when $k_+$ is close to $\pi$. For the first case, we find that among the $N-1$ scattering states, a number of levels proportional to $ N^{3/4}$ does not hybridize with other levels. For the second case, the interval is split in two. In the first interval, the number of Bethe solutions unaffected by other levels is sublinear $\propto N^{1-p}$, with $p>0$. In the second interval, the latter grows linearly, however, the interval shrinks to $0$ when $N\to\infty$.

\par (1): If $k_+$ is far enough from $\pi$ such that $\cos^2 k_+/2=a\sim1$ not too close from 0 then $|\sin\kappa_-|\ll \bar{U}W^{-1/2} N^{-1/4} $. As a consequence the sine being small for $N\to\infty$ and $\kappa_-\sim 2\pi\lambda/N$, among the $N$ Bethe solutions of a fixed value $k_+$, only a number growing as $\propto \bar{U}W^{-1/2} N^{3/4} $ does not hybridize with other levels.

\par (2): If $k_+=\pi-2\pi r/N$ with $r$ an integer such that $0\leq r\leq r_1$ with $2\pi r_1/N\ll1$ and $r_1$ is the upper bound of $r$. Below it, the following approximation is valid: $\cos(k_+/2)\sim r^2/N^2$ and $|\sin\kappa_-|\ll \bar{U}W^{-1/2} N^{7/4}r^{-2} $. When $r>r_1$, we are then back to the first case (1). We want to estimate $r_0$, the number of $k_+$ for which the overall $N$ corresponding Bethe solutions do not hybridize with neighbouring levels. It is fulfilled when $ \bar{U}W^{-1/2} N^{7/4}r^{-2}>1\Rightarrow r<\sqrt{ \bar{U}}W^{-1/4}N^{7/8}=r_0 $.  In the following, it will be necessary to get an upper bound $r'_0$ on this  $r_0$  so we relax the constraint and ask for $ \bar{U}W^{-1/2} N^{7/4}r^{-2}> \bar{U}W^{-1/2}N^{-p}$ where $p>0$, the upper bound is $r'_0=N^{7/8+p/2}$. For $r_1\geq r\geq r'_0$, $|\sin\kappa_-|\ll \bar{U}W^{-1/2} N^{7/4}r^{-2}\leq \bar{U}W^{-1/2}N^{-p}\Rightarrow\lambda\ll N^{1-p}$. We expect that the number of Bethe solutions which does not hybridize scales as $N^{1-p}$. We choose $p<1/4$ and obtain at the thermodynamic limit that $r_1>r'_0$.

\par In summary, we want to estimate the total number of Bethe solutions $S=\sum_{j=-N/2}^{N/2-1} f(2\pi j/N)$ for which the first perturbation energy does not cross neighbouring energy levels, where $f$ is the function which counts these Bethe solutions at fixed $k_+$. We obtain 3 typical different behaviors for $f$ depending on the value of $k_+=2\pi j/N=\pi-2\pi r/N \Rightarrow j=N/2-r$. We define $j_0=N/2-r_0$ and $j_1=N/2-r_1$. $f_0$ is the behavior of $f(2\pi j/N)$ for $0\leq|j|<j_1$ computed in the first case (1). $f_1$ and $f_2$ described the behavior of $f(2\pi j/N)$ respectively in the interval $j_1\leq|j|\leq j_0$ and $j_1<|j|\leq N/2$ computed in the second regime (2). Then
\begin{align}
S<\sum_{|j|=0}^{j_1-1}f_0+\sum_{|j|=j_1}^{j_0}f_1+\sum_{|j|=j_0+1}^{N/2}f_2,
\end{align}
where $f_0\propto \bar{U}W^{-1/2}N^{3/4}$, $f_1\propto N^{1-p}$ and $f_2=N$. The number of terms in the first and second sum is proportional to $N$. In the third sum, the number of terms is proportional to $r'_0=N^{7/8+p/2}$. For $0<p<1/2$, $S$ scales as $N^{\alpha}$ where $\alpha<2$. Therefore, in the thermodynamic limit, these states, which do not see neighbouring levels, won't be the majority.

\par This result is quite different from the case $\bar{U}=0$. When $ \bar{U}=0$ and in the thermodynamic limit, every state will eventually hybridize and lead to GOE statistics. At $ \bar{U}\neq0$, some states will not hybridize. As their number scales as $N^\alpha$ with $\alpha<2$, they are a minority compared to the total number $N^2$ of eigenstates of the problem. 

\section{Localization of separatrix states along the center-of-mass direction}
\label{ap:Upertu}
\par This Appendix presents the typical form of eigenstates in the Fig.~\ref{pertustate}, in particular that of typical separatrix states shown in panel (b). We wish in particular to explain the mechanism of disorder-induced localization and to distinguish it from Anderson localization.

\par The interaction operator $U$ in the basis made of the diagonal state $\psi_d(x_-=0)$ plus the $N/2-1$ symmetric red states reads
\begin{equation}
    \begin{pmatrix}
    b&\sqrt{2} u\\
    \sqrt{2}u^\dagger& 2 U_|
    \end{pmatrix},
    \label{eq:Ures}
    \end{equation}
where $b=\f{1}{N}\sum_j U_j$, $u$ is a vector of length $N/2-1$ and $U_|$ is a matrix of size $N/2-1\times N/2-1$. It has components and matrix elements
\begin{align}
   u_{k_+}&=\f{1}{N\sqrt{N}}\sum_j U_j\e^{-i(k_+ +\pi) j} \label{eq:uvec}\\
   U_{|_{k_+,k_-}}&=\f{1}{N^2}\sum_j U_j\e^{-i (k_+-k'_+)j}\label{eq:umat}
\end{align}
 with $k_+=\f{2\pi K}{N}$, $k'_+=\f{2\pi K'}{N}$, $K$ and $K'$ being one of the $N/2-1$ even integers in $[-N/2+1,N/2[$. 

The interaction operator written in this basis looks almost like an on-site potential Hamiltonian $H_{\text{op}}$ of dimension $N/2$ but written in the plane wave basis:
\begin{equation}
H_\text{op}=\sum_j U_j\ket{j}\bra{j}=\sqrt{2/N}\sum_{k,k'} U_j\e^{2i\pi j(k-k')}\ket{k'}\bra{k}.
\label{eq:potham}
\end{equation}

\par If the interaction operator would be exactly proportional to $H_\text{op}$ by establishing the one to one correspondence between the quasi-momentum center-of-mass $k_+$ (see Eqs.~\eqref{eq:uvec} and~\eqref{eq:umat}) and the 1D quasi-momentum $k$ (see Eqs.~\eqref{eq:potham}), each eigenvectors would be localized exactly on one center of mass $x_+$ and delocalized in the relative motion direction $x_-$. Figure \ref{pertustate}-b shows a localisation of eigenvectors along the center of mass direction but we cannot assign precisely a center of mass position to an eigenvector. In the following, we make explicit the differences between the interaction operator $U$ and the on-site potential hamiltonian $H_\text{op}$.

\par Making the substitution $U_j\rightarrow U_j\sqrt{2}/(N\sqrt{N})$, and establishing a one to one correspondence between the quasi-momentum center-of-mass $k_+$ of Eq.~\eqref{eq:umat} and the 1D quasi-momentum $k$ of Eq.~\eqref{eq:potham}, $2U_|$ corresponds exactly to the restricted part of $H_{\text{op}}$ onto the $N/2-1$ dimensional subspace where we remove the plane wave of lowest quasi-momentum $k=-\pi$. However, if one wants $b$ of Eq.~\ref{eq:Ures} to match with the first matrix element of $H_{op}$, one has to make a different substitution $U_j\rightarrow U_j/(\sqrt{2N})$. Eventually, the correspondence between $\sqrt{2}u$ of Eq.~\eqref{eq:uvec} and $\bra{-\pi}H_{op}\ket{k\neq-\pi}$ is established by still another substitution $U_j\rightarrow U_j/N$. Therefore, numerical factors (and scaling with $N$) do not match between the different part of the matrix in Eq.~\ref{eq:Ures} and this constitutes one of the differences with the on-site potential Hamiltonian of Eq.~\ref{eq:potham}. 

The other difference is that the first vector in Eq.~\ref{eq:Ures} is the diagonal state $\psi_d(0)$ and is not a plane wave state as the other symmetric red states or the 1D plane wave of the on-site potential model. 

If one would have $H_\text{op}$ instead of $U$, the eigenvectors would be the symmetric antidiagonal states, well-localized on one center-of-mass position. The resulting interaction operator being quite similar to $H_\text{op}$, we do not expect very different eigenstates and we observe also localization along the center-of-mass direction. If this analogy with the potential model holds, the localization leading to separatrix states is a trivial localization by disorder potentials (indeed there is no kinetic energy in the potential model) and not an Anderson localization resulting from multiple scattering interferences.

\section{Lifetime of resonant states}
\label{ap:lifetime}
\begin{figure}[!h]
    \includegraphics[width=1\linewidth]{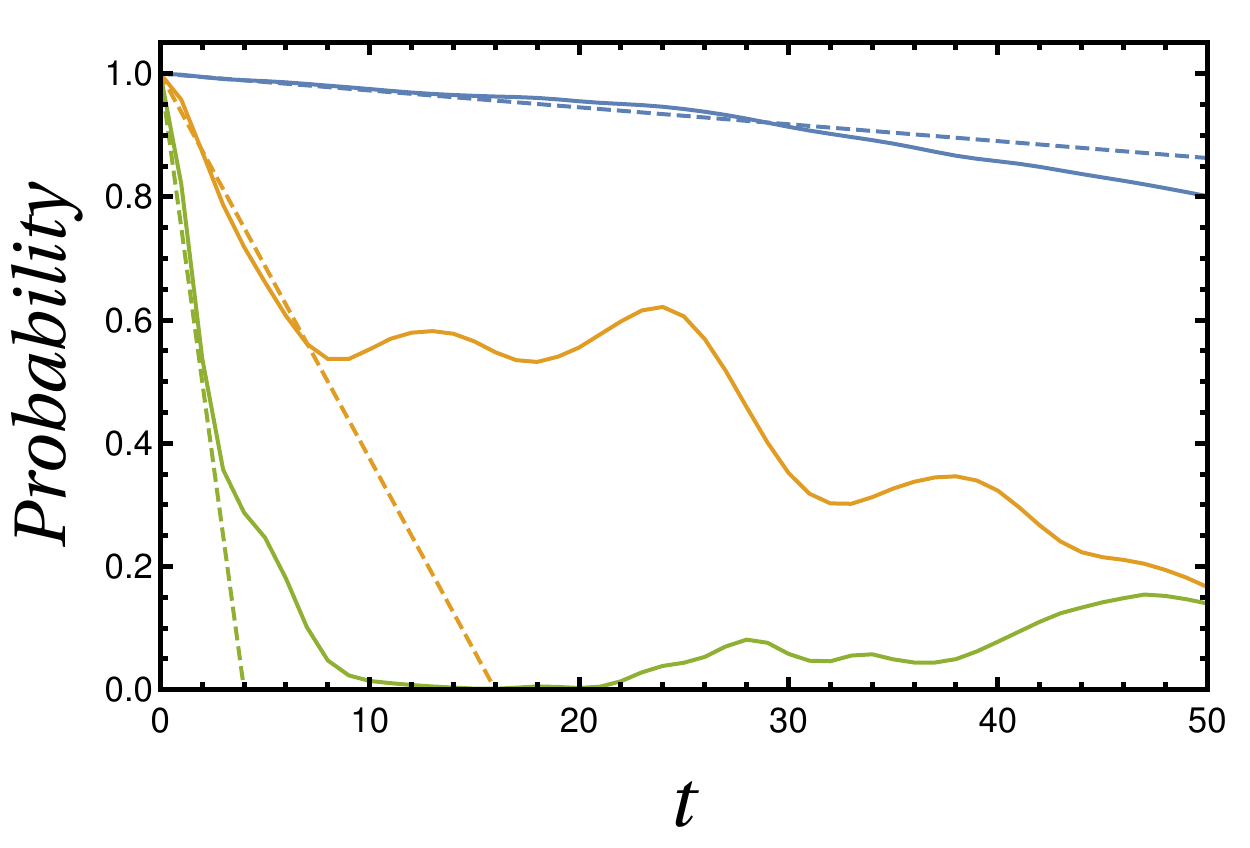}
  \caption{Time evolution of the probability $P(t)=|\langle \psi(0)|\psi(t)\rangle|^2$ for three different disorder strengths: $W=0.1$ (blue), $0.5$ (yellow) and $1$ (green). The initial bound state $|\psi(0)\rangle$ is an eigenstate at $W=0$ with energy $E_0\simeq 3.087$ for $N_x=30$ and $\bar{U}=2$. From $P(t)\simeq 1-t/\tau$ (dashed lines), the lifetime $\tau$ is (a) $365$, (b) $16$ and (c) $4$, which agrees with $\tau \sim 3.5/W^2$.}
    \label{fig:lifetime}
\end{figure}

For resonant states (or virtual bound states), one may define a lifetime. Resonant states can be seen as the result of the coupling, via the disorder, between bound states and scattering states that coincide in energy, i.e. in the region of band overlap. Because of this coupling, bound states are no longer eigenstates but acquire a finite lifetime. For example, we take $\bar{U}=2$ and $W=0$ and consider a bound state with energy $E_0$ in between $\bar{U}$ and $4$. Such a bound state will be taken as initial state $|\psi(0)\rangle$. It satisfies $H_0 |\psi(0)\rangle = E_0 |\psi(0)\rangle$. Now, we turn on a finite but weak disorder $1 \gg W>0$, and study the time evolution of $|\psi(t)\rangle=e^{-i H t}|\psi(0)\rangle$ by considering the probability $P(t)=|\langle \psi(0)|\psi(t)\rangle|^2$. At short time $t\ll \tau$, we expect that it decays as $e^{-t/\tau}\simeq 1-t/\tau$, where $\tau$ is the lifetime given by Fermi's golden rule $1/\tau \sim \rho(E_0) W^2$, where $\rho(E_0)$ is the density of states (per site) of the atomic band at energy $E_0$. The lifetime should therefore scale as $1/W^2$. This is indeed what we observe: for example, for $E_0\simeq 3.087$, we find $\tau \sim 3.5/W^2$, see Fig.~\ref{fig:lifetime}. At longer time, the evolution is more complicated.

\end{document}